\documentclass[11pt,A4paper]{article}
\usepackage{jcappub}
\usepackage{url,amsmath,wrapfig,multirow}
\usepackage{graphicx}
\usepackage{epsfig}
\usepackage{xcolor}
\usepackage{lineno}
\usepackage{longtable}
\usepackage{hyperref}
\usepackage{amsmath}
\usepackage{amssymb}

\usepackage{xspace}
\def\Xmax{$X_\mathrm{max}$~}

\arxivnumber{1612.07155}

\title{Combined fit of  spectrum and composition data as measured by the Pierre Auger Observatory}

\collaboration{The Pierre Auger Collaboration}

\author[63]{A.~Aab,}
\author[70]{P.~Abreu,}
\author[48,47]{M.~Aglietta,}
\author[29]{I.~Al Samarai,}
\author[16]{I.F.M.~Albuquerque,}
\author[1]{I.~Allekotte,}
\author[8,11]{A.~Almela,}
\author[62]{J.~Alvarez Castillo,}
\author[79]{J.~Alvarez-Mu\~niz,}
\author[38]{G.A.~Anastasi,}
% ...

\affiliation[1]{Centro At\'omico Bariloche and Instituto Balseiro (CNEA-UNCuyo-CONICET), Argentina}
% ...
\affiliation[8]{Instituto de Tecnolog\'\i{}as en Detecci\'on y Astropart\'\i{}culas (CNEA, CONICET, UNSAM), Centro At\'omico Constituyentes, Comisi\'on Nacional de Energ\'\i{}a At\'omica, Argentina}
% ...
\affiliation[11]{Universidad Tecnol\'ogica Nacional -- Facultad Regional Buenos Aires, Argentina}
% ...
\affiliation[16]{Universidade de S\~ao Paulo, Inst.\ de F\'\i{}sica, S\~ao Paulo, Brazil}
% ...
\affiliation[29]{Laboratoire de Physique Nucl\'eaire et de Hautes Energies (LPNHE), Universit\'es Paris 6 et Paris 7, CNRS-IN2P3, France}
% ...
\affiliation[38]{Gran Sasso Science Institute (INFN), L'Aquila, Italy}
% ...
\affiliation[47]{INFN, Sezione di Torino, Italy}
\affiliation[48]{Osservatorio Astrofisico di Torino (INAF), Torino, Italy}
% ...
\affiliation[62]{Universidad Nacional Aut\'onoma de M\'exico, M\'exico}
\affiliation[63]{Institute for Mathematics, Astrophysics and Particle Physics (IMAPP), Radboud Universiteit, Nijmegen, Netherlands}
% ...
\affiliation[70]{Laborat\'orio de Instrumenta\c{c}\~ao e F\'\i{}sica Experimental de Part\'\i{}culas -- LIP and Instituto Superior T\'ecnico -- IST, Universidade de Lisboa -- UL, Portugal}
% ...
\affiliation[79]{Universidad de Santiago de Compostela, Spain}
% ...
\affiliation[c]{also at Vrije Universiteit Brussels, Brussels, Belgium}
% ...

\emailAdd{auger\_spokespersons@fnal.gov}
\dedicated{This article is dedicated to Aurelio Grillo, 
who has been deeply involved in this study 
for many years and who has inspired several 
of the developments described in this paper. 
His legacy lives on.}
\abstract{We present a combined fit of a simple astrophysical model of UHECR sources to both the energy spectrum and mass
composition data measured by the Pierre Auger Observatory. The fit has been performed for energies above $5 \cdot 10^{18}$ eV, i.e.~the
region of the all-particle spectrum above the so-called ``ankle'' feature. The astrophysical model we adopted consists of
identical sources uniformly distributed in a comoving volume, where nuclei are accelerated through a rigidity-dependent
mechanism. The fit results suggest sources characterized by relatively low maximum injection energies, hard spectra and heavy chemical composition. 
 We also show that uncertainties about physical quantities relevant to UHECR propagation and shower development have a non-negligible impact on the fit results.
 }
\begin{document}
\maketitle
\flushbottom
\section{Introduction \label{sec:intro}}
Cosmic rays have been detected up to particle energies around $10^{20}$ eV and are the highest energy particles in the present Universe. In spite of a reasonable number of events already collected by large experiments around the world, their origin is still largely unknown. It is widely believed that the particles of energy above a few times $10^{18}$ eV (Ultra High Energy Cosmic Rays, UHECRs) are of extragalactic origin, since galactic magnetic fields cannot confine them, and the distribution of their arrival directions appears to be  nearly isotropic \cite{Auger:2012an,ThePierreAuger:2014nja}.  \\ 
With the aim of investigating the physical properties of UHECR sources, we here use cosmic ray measurements performed at the Pierre Auger Observatory, whose design, structure and operation are presented in detail in \cite{ThePierreAuger:2015rma}. 
The observables we use are the energy spectrum \cite{Valino:2015zdi}, which
is mainly provided by the surface detector (SD), and the shower depth ($X_\text{max}$)
distribution, provided by the fluorescence detector (FD) \cite{Aab:2014kda}, which gives information about  the nuclear mass of the cosmic particle hitting the Earth's atmosphere. \\
The observed energy spectrum of UHECRs is close to a power law with index $\gamma \approx 3$ \cite{Valino:2015zdi,TheTelescopeArray:2015mgw}, but there are two important features: the so-called ankle at $E \approx 5 \cdot 10^{18}$ eV where the spectrum becomes flatter (the spectral index decreasing from about $3.3$ to $2.6$) and a suppression above $4 \cdot 10^{19}$~eV after which the flux sharply decreases. The ankle can be interpreted  to reflect e.g.~the transition from a galactic to an extragalactic origin of cosmic rays, or the $\mathrm{e}^+\,\mathrm{e}^-$~pair-production dip resulting from cosmic ray proton interactions with the cosmic microwave background. The decrease at the highest energies, observed at a high degree of significance,  can be ascribed to interactions with background radiation \cite{Greisen:1966jv,Zatsepin:1966jv}  and/or to a maximum energy of cosmic rays at the sources. \\
Concerning the mass composition, the  average $X_\text{max}$ measured at the Pierre Auger Observatory (as discussed in \cite{Aab:2014kda,Aab:2014aea,Abreu:2013env}) indicates that UHECRs become heavier with increasing energy above $2 \cdot 10^{18} $~eV. Moreover, the measured fluctuations of $X_\text{max}$ indicate a small  mass-dispersion at energies above the ankle.
The average $X_\text{max}$ has also been measured by the Telescope Array collaboration~\cite{Abbasi:2014sfa}. Within the uncertainties this measurement is consistent with a variety of primary compositions and it agrees very well with the Auger results~\cite{Unger:2015rzh}. \\
In this paper we investigate the constraining power of the Auger measurements of spectrum and composition with respect to source properties. Therefore we compare our data  with simulations that are performed starting from rather simple astrophysical scenarios featuring only a small number of free parameters. We then constrain these astrophysical parameters taking into account experimental uncertainties on the measurements. 
To do so, we perform a detailed analysis of the possible processes that determine the experimental measurements from the sources to the detector (section \ref{sec:gen}). \\
We assume as a working hypothesis that the sources are of extragalactic origin. 
The sources inject nuclei, accelerated in electromagnetic processes, and therefore with a rigidity $(R=E/Z)$ dependent cutoff (section \ref{sec:astro}). \\
Injected nuclei propagate in extragalactic space and experience interactions with cosmic photon backgrounds. 
Their interaction rates depend on the cross sections of various nuclear processes and on the spectral density of background photons. 
We use two different publicly available Monte Carlo codes to simulate the UHECR propagation, CRPropa \cite{Armengaud:2006fx,Batista:2014xza,Batista:2016yrx} and SimProp \cite{Aloisio:2012wj,Aloisio:2015sga,Aloisio:2016tqp}, together with different choices of photo-disintegration cross sections and models for extragalactic radiation (section \ref{sec:simu}). \\
After propagation, nuclei interact with the atmosphere to produce the observed flux. These interactions happen at energies larger than those experienced at accelerators, and are modelled by several interaction codes. 
The interactions in the atmosphere  are discussed in section \ref{sec:shower}. \\
In section \ref{sec:fitting}, we describe the data we use in the fit and the simulations we compare them to, and in section  \ref{sec:fproc} we describe the fitting procedures we used.
The results we obtain (section \ref{sec:ref}) are in line with those already presented by several authors \cite{Aloisio:2013hya,Hooper:2009fd} and in \cite{bib:fitICRC15,Boncioli:2015pds} using a similar analysis approach to that of this work. The Auger composition data indicate relatively narrow $X_\text{max} $ distributions which imply little mixing of elemental fluxes and therefore limited production of secondaries.
This in turn implies  low maximum rigidities at the sources, and hard injection fluxes to reproduce the experimental all-particle spectrum \cite{Abreu:2013env}.\\ 
The novelty of the present paper is that we discuss in detail the effects of theoretical uncertainties on propagation and interactions in the atmosphere of UHECRs and we generally find that they are much larger than the statistical errors on fit parameters (which only depend on experimental errors). These uncertainties are the feature that limit the ability to constrain source models, as will be discussed in the conclusions. Moreover, we investigate the dependence of the fit parameters on the experimental systematic uncertainties. \\  
We present some modifications of the basic astrophysical model, in particular with respect to the homogeneity of the sources, in section \ref{sec:hom}. \\
In this paper we are mainly interested in the physics above $5 \cdot 10^{18}$ eV. However in section \ref{sec:proton} we briefly discuss how our fits can be extended below the ankle, where additional components (e.g.\ different astrophysical sources) would be required. \\
In section \ref{sec:discussion} we present a general discussion of our results and their implications. \\
Finally  appendix \ref{app-a} contains some details on the simulations used in this paper and appendix \ref{app} on the forward folding procedure used in the fits. 

\section{UHECRs from their sources to Earth \label{sec:gen}}
 \subsection{Acceleration in astrophysical sources \label{sec:astro}}
 The hypotheses of the origin of UHECRs can be broadly classified into two distinct scenarios, the ``bottom-up'' scenario and the ``top-down'' one.  The ``top-down'' source models generally assume the decay of super-heavy particles; they are disfavoured as the source of the bulk of UHECRs below $10^{20}$~eV by  upper limits on photon \cite{Abraham:2009qb,Aglietta:2007yx,Abraham:2006ar} and neutrino \cite{Abraham:2007rj,Abreu:2011zze,Abreu:2012zz,Abreu:2013zbq} fluxes that should be copiously produced at the highest energies, and will not be considered any further in this work. In the ``bottom-up'' processes charged particles are accelerated in astrophysical environments, generally via electromagnetic processes.\\
The distribution of UHECR sources and the underlying acceleration mechanisms are still subject of ongoing research. In particular the non-thermal processes relevant for acceleration to the highest energies constitute an important part of the theory of relativistic plasmas. Various acceleration mechanisms discussed in the literature include first order Fermi shock acceleration with and without back-reaction of the accelerated particles on the magnetized plasma~\cite{Bykov:2012ca}, plasma wakefield acceleration~\cite{Chen:2002nd} and reconnection~\cite{Guo:2014via}. 
Sufficiently below the maximal energy these mechanisms typically give
rise to power-law spectra $\mathrm{d}N/\mathrm{d}E\propto E^{-\gamma}$, with $\gamma\simeq2.2$ for relativistic shocks, and $\gamma$ ranging from $\simeq2.0$ to $\simeq1.0$ for  the other cases. Other processes can even result in `inverted' spectra, with $\gamma < 0$ \cite{Blasi:2000xm,Kotera:2015pya,Ptitsyna:2015nta,Winchen:2016koj}.\\
A common representation of the maximal energy of the sources makes use of an exponential cutoff; yet not all of these scenarios predict power law particle spectra with an exponential cut-off~\cite{Niemiec:2006zd}. 
Reconnection, in particular, can give rise to hard spectra up to some characteristic maximal energy.  This is also the case if cosmic rays are accelerated in an unipolar inductor that can arise in the polar caps of rotating magnetized neutron stars~\cite{Blasi:2000xm,Arons:2002yj,Fang:2013cba}, or black holes~\cite{Neronov:2007mh}. If interactions in the magnetospheres can be neglected this also gives rise to $\gamma \approx 1$. Also, second order Fermi acceleration has been proposed to give rise to even harder spectra \cite{Winchen:2016koj}. 
Close to the maximal energy, where interactions can become significant, the species dependent interaction rates can give rise to complex all-particle spectra and individual spectra whose maximal energies are not simply proportional to the charge $Z$~\cite{Kotera:2015pya}.\\
On the other hand, individual sources will have different characteristics and integrating over them can result in an effective spectrum that can differ significantly from a single source spectrum. For example, integrating over sources with power law acceleration spectra of different maximal energies or rigidities can result in an effective power law spectrum that is steeper than that of any of the individual sources~\cite{Kachelriess:2005xh,Blaksley:2011kw}.\\
Taking into account all these possibilities would render the fit unmanageable. 
In this work, the baseline astrophysical model we use assumes identical extragalactic UHECR sources uniform in comoving volume and isotropically distributed; source evolution effects are not considered. We also neglect possible effects of extragalactic magnetic fields and therefore the propagation  is considered one-dimensional. 
This model, although widely used, is certainly over-simplified. We briefly discuss some possible extensions  in section \ref{sec:improve}.\\ 
A description of the spectrum in terms of elementary fluxes injected from astrophysical sources, all the way from $\log_{10}(E/\text{eV})\approx 18$ up to the highest energies, with a single component, is only possible if UHECRs are protons, since protons  naturally exhibit the ankle feature as the electron-positron production dip; however this option  is at strong variance with Auger composition measurements \cite{Abreu:2013env,Aab:2016htd} and, to a lesser extent, with the measured HE neutrino fluxes  \cite{Aloisio:2015ega,Heinze:2015hhp,Aartsen:2016ngq}.
On the other hand, the ankle can also be interpreted as the transition between two (or more) different populations of sources. 
In this paper we will assume this to be the case; for this reason,  we will generally present results of fits for energies above $\log_{10}(E/\mathrm{eV})=18.7$. An attempt to extend the analysis in the whole energy range is discussed in section \ref{sec:proton}. Recently there have been attempts \cite{Unger:2015laa,Globus:2015xga} to unify the description of the spectrum down to energies of a fraction of EeV, below which galactic CRs are presumed to dominate, by considering the effects of the interactions of nuclei with photon fields in or surrounding the sources. We do not follow this strategy here, but concentrate only on the highest energies.\\
We assume that sources accelerate different amounts of 
nuclei; in principle all nuclei can be accelerated, however it is  reasonable to assume that considering only a representative subset of injected masses still produces approximately correct results.
We therefore assume that sources inject five representative stable nuclei: Hydrogen ($^1\text{H}$), Helium ($^4\text{He}$), Nitrogen ($^{14}\text{N}$), Silicon ($^{28}\text{Si}$) and Iron ($^{56}\text{Fe}$).  
Of course  particles other than those injected can be produced by photonuclear interactions during propagation.
These nuclei are injected with a power law of energy ($E=ZR$) up to some maximum rigidity $R_\text{cut}$, reflecting the idea that acceleration is electromagnetic in origin: 
\begin{equation}
\frac{\mathrm{d}N_A}{\mathrm{d}E}  = J_A(E)=f_A J_0 ~\left( \frac{E}{10^{18}~\mathrm{eV}} \right)^{-\gamma} \times f_\text{cut}(E,Z_AR_\text{cut}),
\label{equation:1}
\end{equation}
where $f_A$ is defined as the fraction  of the injected nucleus $A$ over the total. 
This fraction is defined, in our procedure, at fixed energy $E_0=10^{18}$ eV, below the minimum cutoff energy for protons. The power law spectrum is modified by the cutoff function, which  describes physical properties of the sources near the maximum acceleration energy. 
Here we (arbitrarily) adopt a purely instrumental point of view and use a broken exponential cutoff function
\begin{eqnarray}
 f_\text{cut}(E,Z_A R_\text{cut}) = \begin{cases}
  1 &(E<Z_A R_\text{cut})  \\
  \exp \left(1-\frac{E}{Z_A R_\mathrm{cut}} \right) & \left( E>Z_A R_\mathrm{cut} \right) \label{equation:2}
 \end{cases}
\end{eqnarray}
in order to improve the sensitivity to $\gamma$ in the rather limited range from the lowest energy in the fit to the cutoff.
The effect of this choice will be further discussed  in section~\ref{sec:cutoff}. \\
The free parameters of the fit are then the injection spectral index $\gamma$, the cutoff rigidity $R_\text{cut}$, the spectrum normalization $J_0$ and four of the mass fractions $f_A$, the fifth being fixed by $\sum_A f_A = 1$.

 \subsection{The propagation in the Universe \label{sec:simu}}
At the UHECR energies, accelerated particles travel through the extragalactic environment and interact with photon backgrounds, changing their energy and, in the case of nuclei, possibly splitting into daughter ones. The influence of these processes is discussed in detail in \cite{Batista:2015mea}.
In the energy range we are interested in, the photon energy spectrum includes the cosmic microwave background radiation (CMB), and the infrared, optical and ultra-violet photons (hereafter named extragalactic background light, EBL). The CMB has been extremely
well characterized and has been shown to be a very isotropic pure black-body spectrum, at least to the accuracy relevant for UHECR propagation. The EBL, which comprises the radiation produced in the Universe since the formation of the first stars,
is relatively less known: several models of EBL have been proposed \cite{Stecker:2005qs,Stecker:2006eh,Kneiske:2003tx, Gilmore:2011ks, Dominguez:2010bv, Franceschini:2008tp,Inoue:2012bk}, 
among which there are sizeable differences, especially in the far infrared and at high redshifts.\\ 
Concerning the interactions of protons  and nuclei in the extragalactic environment, 
the loss mechanisms and their relevance for the propagation were predicted  by Greisen \cite{Greisen:1966jv} and independently by Zatsepin and Kuzmin \cite{Zatsepin:1966jv} soon after the discovery of the CMB.
First, all particles produced at cosmological distances lose energy adiabatically by the expansion of the Universe, with an energy loss length $c/H_0 \sim 4$~Gpc. This is the dominant energy loss mechanism for protons with $E \lesssim 2 \cdot 10^{18}$~eV and nuclei with $E/A \lesssim 0.5\cdot 10^{18}$~eV. At higher energies, the main energy loss mechanisms are electron-positron pair production mainly due to CMB photons and, in the case of nuclei, photo-disintegration in which a nucleus is stripped by one or more nucleons or (more rarely) $\alpha$ particles, for which interactions both on the EBL and the CMB have a sizeable impact. At even higher energies ($E/A \gtrsim 6 \cdot 10^{19}$~eV), the dominant process is the photo-meson production on CMB photons.\\
%
The cross sections for pair production can be analytically computed via the Bethe-Heitler formula, and those for photo-meson production have been precisely measured in  accelerator-based experiments and have been accurately modelled by codes such as SOPHIA \cite{Mucke:1999yb}. On the other hand, the cross sections for photo-disintegration of nuclei, especially for exclusive channels in which charged fragments are ejected, have only been measured in a few cases; there are several phenomenological models that can be used to estimate them, but they are not always in agreement with the few experimental data available or with each other  \cite{Batista:2015mea}.\\ 
In order to interpret UHECR data within  astrophysical scenarios some modelling of the extragalactic propagation is needed. Several approaches have been used to follow the interactions in the extragalactic environment, both analytically and using  Monte Carlo codes. In this paper simulations based  on CRPropa \cite{Armengaud:2006fx,Batista:2014xza,Batista:2016yrx} and  SimProp \cite{Aloisio:2012wj,Aloisio:2015sga,Aloisio:2016tqp}  will be used, along with the Gilmore \cite{Gilmore:2011ks} and Dom\'inguez (fiducial)  \cite{Dominguez:2010bv} models
of EBL and the Puget, Stecker and Bredekamp (PSB) \cite{Puget:1976nz,Stecker:1998ib}, TALYS \cite{talys,Koning20122841,talys1.6-manual,Batista:2015mea} and Geant4 \cite{Allison:2006ve} models of photo-disintegration in various combinations.\\
A comparison between the  Monte Carlo codes used here is beyond the scope of this paper, and has been discussed  in \cite{Batista:2015mea}.

 \subsection{Extensive air showers and their detection \label{sec:shower}}
 Once a nucleus reaches the Earth it produces an extensive air shower by interacting with the atmosphere.
Such a shower can be detected by surface detectors (SD) and, during dark moonless nights, by fluorescence detectors (FD) \cite{ThePierreAuger:2015rma}. The FD measures the shower profile, i.e.~the energy deposited by the shower per unit atmospheric depth. The integral of the profile gives a measurement of the calorimetric energy of the shower, while the position of the maximum $X_{\max}$ provides information about the primary nucleus which initiated the cascade. The SD measures the density of shower particles at ground level, which can be used to estimate the shower energy, using the events simultaneously detected by both the SD and the FD for calibration.
There are several models available to simulate the hadronic interactions involved in the shower development. They can be used to estimate the distribution of $X_{\max}$ for showers with a given primary mass number $A$ and total energy $E$.
In this work, we use the Auger SD data for the energy spectrum and the Auger FD data for the $X_{\max}$ distributions. The simulated mass compositions from the propagation simulations are converted to $X_{\max}$ distributions assuming EPOS-LHC~\cite{Pierog:2013ria}, QGSJetII-04~\cite{Ostapchenko:2010vb} and Sibyll 2.1~\cite{Ahn:2009wx} as the hadronic interaction models.

\section{The data set and the simulations \label{sec:fitting}}
The data we fit in this work consist  of the SD event distribution in $15$ bins of $0.1$ of $\log_{10}(E/\text{eV})$, ($18.7 \leq \log_{10}(E/\text{eV}) \leq 20.2$) and  $X_\text{max}$ distributions \cite{Aab:2014kda} (in bins of $20$ g/cm$^2$) in the same bins of energy up to $ \log_{10}(E/\text{eV})=19.5$ and a final bin from $19.5$ to $20.0$, for a total of 110 non zero data points.
In total, we have $47767$ events in the part of the spectrum we use in the fit and $1446$ in the $X_\text{max}$ distributions.\\
In the Auger data the energy spectrum and the \Xmax distributions are independent measurements and the model likelihood is therefore given by $    L = L_J \cdot L_{X_\mathrm{max}}$.
The goodness-of-fit is assessed with a generalized $\chi^2$, (the {\it deviance, D}), defined as the negative log-likelihood ratio of a given model and the {\it saturated} model that perfectly describes the data:
\begin{eqnarray}
    D &=D(J) + D(X_\text{max})= -2 \ln \frac{L}{L^\text{sat}} = -2 \ln \frac{L_J}{L_J^\text{sat}} - 2 \ln \frac{L_{X_\mathrm{max}}}{L_{X_\mathrm{max}}^\text{sat}}
\end{eqnarray}
Details on the simulations used are given in appendix \ref{app-a}.
\subsection{Spectrum}
Measurements of UHECR energies are affected by uncertainties of the order of 10\%,
due to both shower-to-shower fluctuations and detector effects.
These can cause detected events to be reconstructed in the wrong energy bin.
As a consequence of the true spectrum being a decreasing function of energy,
more of the events with a given reconstructed energy~$E_\text{rec}$
have a true energy~$E_\text{true} < E_\text{rec}$ than~$E_\text{true} > E_\text{rec}$.
The net effect of this is that the reconstructed spectrum is shifted to higher energies and smoothed compared to the true spectrum.
\\
In ref.~\cite{Valino:2015zdi}, we adopted an unfolding procedure to correct the
measured spectrum for these effects, consisting in assuming a phenomenological ``true'' spectrum~$J_\text{unf}^\text{mod}$,
convolving it by the detector response function to obtain a folded spectrum~$J_\text{fold}^\text{mod}$,
computing the correction factors $c(E) = J_\text{unf}^\text{mod}(E)/J_\text{fold}^\text{mod}(E)$,
and obtaining a corrected measured spectrum as $J_\text{unf}^\text{obs}(E) = c(E)J_\text{fold}^\text{obs}(E)$,
where $J_\text{fold}^\text{obs}(E)$ is the raw event count
in a reconstructed energy bin divided by the bin width and the detector exposure.
This procedure is an approximation which does not take into account the dependence of the correction factors
on the assumed shape of~$J_\text{unf}^\text{mod}$, thereby potentially underestimating the total uncertainties
on~$J_\text{unf}^\text{obs}$.
\\
To avoid this problem, in this work we apply a forward-folding procedure 
to the simulated true spectrum~$J(E)$ obtained from each source model (spectral index, maximum rigidity and composition at the sources, section~\ref{sec:astro}) in order to compute the expected event count in each energy bin,
and directly compare them to the observed counts, so that the likelihood can be correctly modelled as Poissonian, without approximations, resulting in the deviance (in each bin $m$ of $\log_{10} (E/\mathrm{eV})$)
\begin{equation}
    D  = -2 \sum_m\left(\mu_m-n_m+n_m\ln \left(\frac{n_m}{\mu_m}\right)\right)
\label{eq:devsum}
\end{equation}
 where $n_m$ is the experimental count in  the $m$-th (logarithmic) energy bin, and $\mu_m$ are the corresponding expected numbers of events.
The details of the forward folding procedure, together with the experimental resolutions used, are described in appendix~\ref{app}.
\\
We decided to use only the experimental measurements from the surface detector since in the energy range we are interested in (above $\approx 5 \cdot 10^{18}$~eV), the contribution of the other Auger detectors is negligible and the SD detector efficiency is saturated \cite{Ghia:2015kfz}.
The spectrum reconstructed from the SD detector is produced differently depending on the zenith angle of primary particles, namely vertical  ($\theta_Z <60^{\circ}$), and inclined ($60^\circ < \theta_Z <80^{\circ}$) events, $\theta_Z$ being the shower zenith angle \cite{ThePierreAuger:2015rha}. For the present analysis, the vertical and inclined SD spectra are combined (i.e.~$\mu_m = \mu^\text{vert}_m + \mu^\text{incl}_m$, $n_m = n^\text{vert}_m + n^\text{incl}_m$) with the exposures rescaled as described in \cite{Valino:2015zdi}.
\subsection{Composition}
The $X_\mathrm{max}$ distribution at a given energy can be obtained using
standard shower propagation codes. The distributions depend on the mass
of the nucleus entering the atmosphere and the model of hadronic interactions. 
In this work we adopted a parametric model~ for the $X_\mathrm{max}$ distribution, which takes the form of a generalized Gumbel distribution $g(X_\mathrm{max} | E,A)$~\cite{DeDomenico:2013wwa}.
The Gumbel parameters have been determined   
with CONEX~\cite{Pierog:2004re} shower simulations using different hadronic
interaction models: we use here the parameterizations for EPOS-LHC~\cite{Pierog:2013ria}, QGSJetII-04~\cite{Ostapchenko:2010vb} and Sibyll 2.1~\cite{Ahn:2009wx}. The Gumbel parameterization provides
a reasonable description of the $X_\mathrm{max}$ distribution in a wide energy range with the resulting $\langle X_\mathrm{max} \rangle$ and $\sigma(X_\mathrm{max})$ differing by less than a few $\mathrm{g ~cm^{-2}}$ from the CONEX
simulated value~\cite{DeDomenico:2013wwa}. 
The advantage of using this parametric function is that it allows us to evaluate the model $X_\mathrm{max}$ distribution for any mixture of nuclei without the need to simulate showers for each primary.
In this analysis the distributions of mass numbers $A = 1 - 56$ are considered.\\
To compare with the measured $X_\mathrm{max}$ distributions, the Gumbel distributions  are corrected for detection effects to give the expected model probability $(G_{m}^\text{model})$, evaluated at the 
logarithmic average of the
energies of the observed events in the bin $m$, for a given mass distribution at detection (see appendix B). 
The last energy bin of the measured $X_\mathrm{max}$  distribution combines the 
energies $\log_{10} (E/\mathrm{eV}) \geq 19.5$,  having
$\langle \log_{10} (E/\mathrm{eV}) \rangle = 19.62$.  We, therefore,
combine the same bins in the simulated $X_\mathrm{max}^\mathrm{rec}$ 
distribution~(\ref{eq:GumbelMix}).
%
%
In the \Xmax measurement \cite{Aab:2014kda} the total number of events $n_m$ per energy bin $m$ is fixed, 
as the information about the spectral flux is already captured by the spectrum likelihood.
The probability of observing a $X_\text{max}$ distribution $\vec{k}_m = (k_{m1}, k_{m2} ...)$ then follows a multinomial distribution.
\begin{eqnarray}
    L_{X_\mathrm{max}} = \prod_m n_m! \prod_x \frac{1}{k_{mx}!} (G_{mx}^\text{model})^{k_{mx}}
    \label{eq:likelihood_xmax}
\end{eqnarray}
where $G_{mx}^\mathrm{model}$ is the probability to observe an event in the \Xmax bin $x$ given by Eq. \ref{eq:GumbelMix}.
\\
The reason for using the full $X_{\max}$ distributions rather than just their first two moments $\langle X_{\max} \rangle, \sigma(X_{\max})$ is that the former contain information not found in the latter, i.e.~two different compositions can result in the same $X_{\max}$ average and variance, but different distributions~\cite{Aab:2014aea}.
\section{Fitting procedures \label{sec:fproc}}
In order to derive the value of fitted parameters (and the associated errors) and cross-check the results, we follow two independent fitting procedures that are described below. 
\\
We use four parameters for the mass composition at injection, assuming that the sources inject only Hydrogen, Helium, Nitrogen, Silicon and Iron.  We do this because not including Silicon among the possible injected elements would result in a much worse fit to the measured spectrum ($D(J)=41.7$ instead of $13.3$ in the reference scenario), due to a gap in the simulated spectrum between the disintegration cutoffs of Nitrogen and Iron not present in the data, whereas including more elements  would only marginally improve the goodness of fit.  In either case, there are no sizeable changes in the best-fit values of $\gamma$, $R_\text{cut}$, or $D(X_{\max})$. In all the cases considered below, the best fit Iron fraction is zero.\footnote{In practice, after we noticed that the best-fit Iron fraction
was zero in both the reference scenario and in a few other cases,
we only used Hydrogen, Helium, Nitrogen and Silicon (three
parameters) in the remaining fits, in order to have a faster,
more reliable minimization.}
\subsection{Likelihood scanning \label{ssec:minuit}}
In this approach a uniform scan over $(\gamma,\log_{10}(R_\text{cut}))$ binned pairs is performed and for each pair the deviance is minimized as a function of the fractions ($f_A$) of masses ejected at the source.
The Minuit \cite{Minuit} package is used; since $\sum f_A =1$, the number of parameters $n$ in the minimization is equal to the number of masses at source minus one.
The elemental fractions are taken as the (squared) direction cosines in $n$ dimensions. 
The scan is performed in the $ \gamma=-1.5 \div 2.5, \log_{10}(R_\text{cut}/\text{V})=17.5\div 20.5 $  intervals, on a grid with $0.01$ spacing in $\gamma$ and $\log_{10}(R_\text{cut})$.
This range contains the predictions for Fermi acceleration ($\gamma \sim 2 - 2.2$), as well as possible alternative source models (see section \ref{sec:astro}).
\\
The best fit solution found after the  scan procedure is used for evaluating errors on fit parameters. In order to do so $n_\text{mock}$ simulated data sets are generated from the best solution found in the fit, with statistics equal to the real data set. With $n_\text{mock} = 10^4$ we found stability in the outcomes. The procedure is similar to the one described in \cite{Aab:2014aea}.
The quality of the fit, ``$p$-value'', is calculated as the fraction of mock datasets with $D_\text{min}$ worse than that obtained from the real data. 
%
%
%
The best fit solution  corresponding to each mock data set is found and the mean value of the distribution of each fit parameter is evaluated. One standard deviation statistical uncertainties are calculated within the limits containing $68 \%$ of the area of the corresponding distribution of $( \gamma, R_\text{cut}, f_A )$.
Concerning $\gamma$ and $R_\text{cut}$ we found that the errors so obtained are well approximated by those evaluated considering the intervals where $ D \leq D_\text{min}+1$ (the profile likelihood method \cite{Rolke:2004mj}) so in order to compute parameter uncertainties in most cases we only used the latter method, which is computationally much faster.

 \subsection{Posterior sampling \label{ssec:bayes}}
In the second approach we apply a fit constraining all the parameters, ($\gamma, R_\text{cut}$ and the four mass fractions) simultaneously, taking into account the statistical and correlated systematic uncertainties of the measured data.
For this we use the Bayesian formalism where the posterior probability of the astrophysical model parameters in light of the data is denoted as $P(\text{model}|\text{data})$. It is calculated from $P(\text{model}|\text{data})\propto L(\text{data}|\text{model}) P(\text{model})$, where $L(\text{data}|\text{model})$ is the likelihood function, i.e. probability of the data to follow from the model, and $P(\text{model})$ is a prior probability.
%
In this phenomenological work, we do not assign prior   probabilities $P(\text{model})$ based on astrophysical plausibility, but we use a uniform prior for $\gamma$ ranging from $-3$ to 3 and for $\log_{10}(R_\text{cut}/\mathrm{eV})$ ranging from $17.9$ to $20.5$. 
As for the elemental fractions we use uniform priors for the $(n-1)$ dimensional region given by $\sum_A f_A = 1$ (employing the method described in \cite{Onn2011}). For the experimental systematic uncertainties we use Gaussian priors with mean $0$ and standard deviation corresponding to the systematic uncertainty of the measurements.\\
The posterior probability is sampled using a Markov chain Monte Carlo (MCMC) algorithm \cite{Patil:2010}.
For each fit, we run the MCMC algorithm from multiple random starting positions in the parameter space, and require that the Gelman-Rubin statistic $\hat R$ \cite{gelman1992} be less than 1.04 for all parameters in order to assess the convergence of the fit.\\
To include the experimental systematic uncertainties the fit performs continuous shifts of the energy scale and shower maximum with a technique called template morphing \cite{Barlow:2003sg}.
This is done by interpolating between template distributions corresponding to discrete systematic shifts.
The nuisance parameters representing these shifts are fitted simultaneously with the parameters of the astrophysical model.
For each parameter we report the posterior mean and the shortest interval containing $68\%$ of the posterior probability, as well as the best fit solution.

\subsection{Systematic uncertainties}
The most important sources of experimental systematic uncertainties in the present analysis are on the energy scale and on  \Xmax. The systematic uncertainties can be treated as nuisance parameters to be determined simultaneously in the fitting procedure, starting from Gaussian prior distributions. We use this approach when performing the posterior sampling method of section \ref{ssec:bayes}.
Alternatively, all measured energy and/or \Xmax values can be shifted by a fixed amount corresponding to one systematic standard deviation in each direction; this is the approach used when performing the likelihood scanning method of
section \ref{ssec:minuit}.

\section{The fit results \label{sec:ref}}
In this section, we first present the fit results in a ``reference'' scenario; then we study the effect on the fit results of variations of this scenario, using different propagation simulations, air interaction models, shapes of the injection cutoff functions, or shifting all the measured energy or $X_{\max}$ data within their systematic uncertainty.
\subsection{The reference fit}
\begin{figure}[!t]
\centering
\includegraphics*[width=0.7\textwidth]{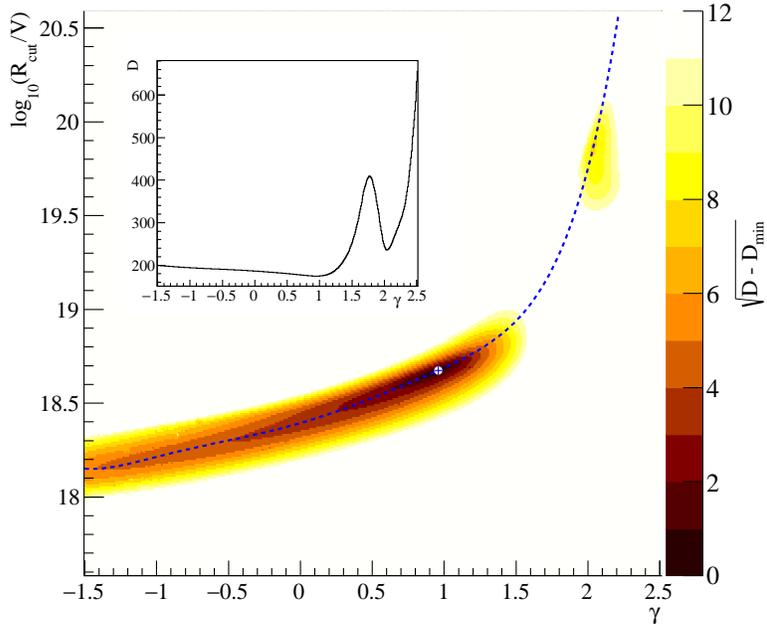}
\caption{Deviance $\sqrt{D-D_\text{min}}$, as function of $\gamma$ and $\log_{10}(R_\text{cut}/\text{V})$. The dot indicates the position of the best minimum, while the dashed line connects the relative minima of~$D$ (valley line). In the inset, the distribution of $D_\text{min}$ in function of $\gamma$ along this line.}
\label{fig:m3}
\end{figure}

\begin{figure}[!t]
\centering
\includegraphics*[width=0.7\textwidth]{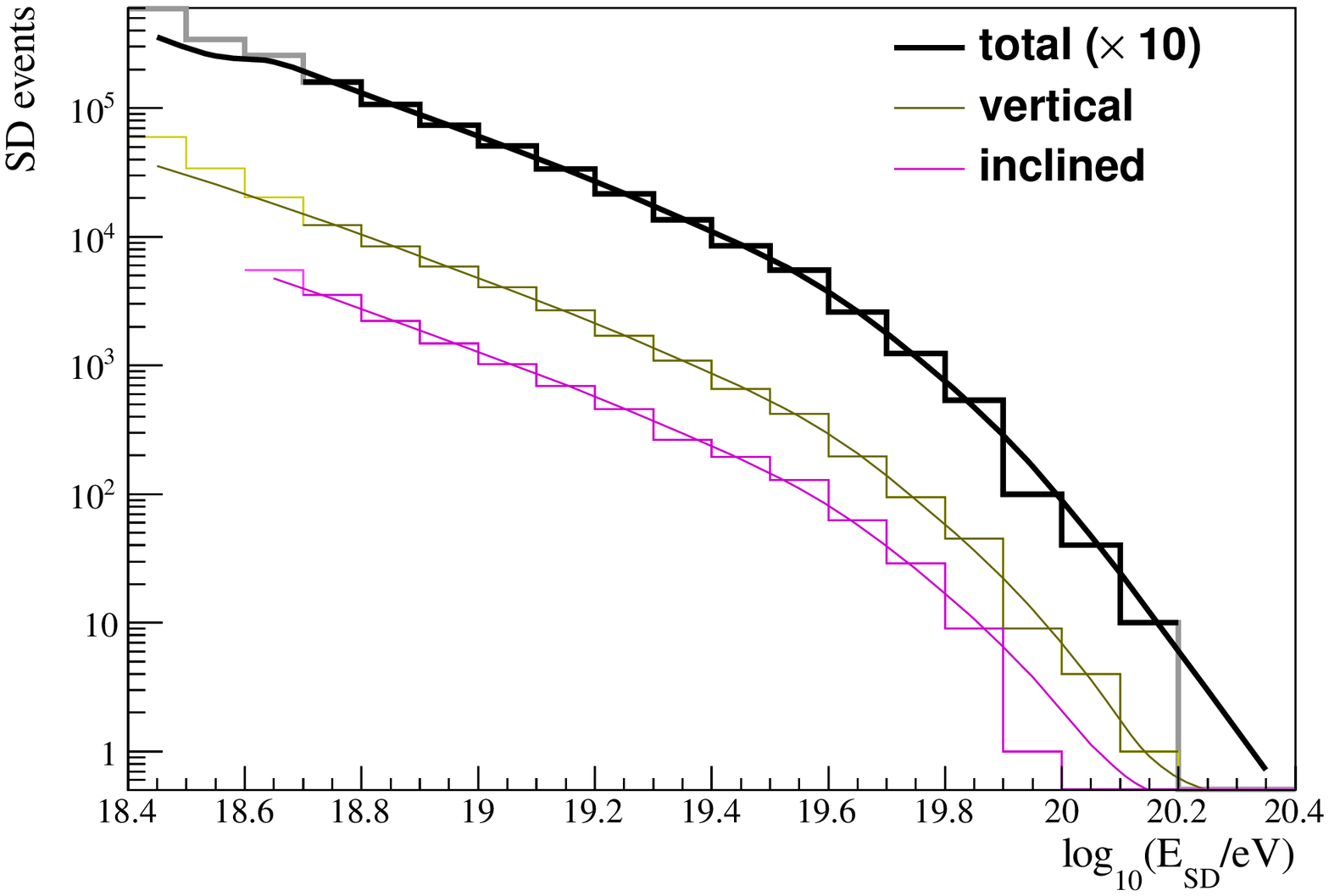}
\includegraphics[width=0.3\textwidth]{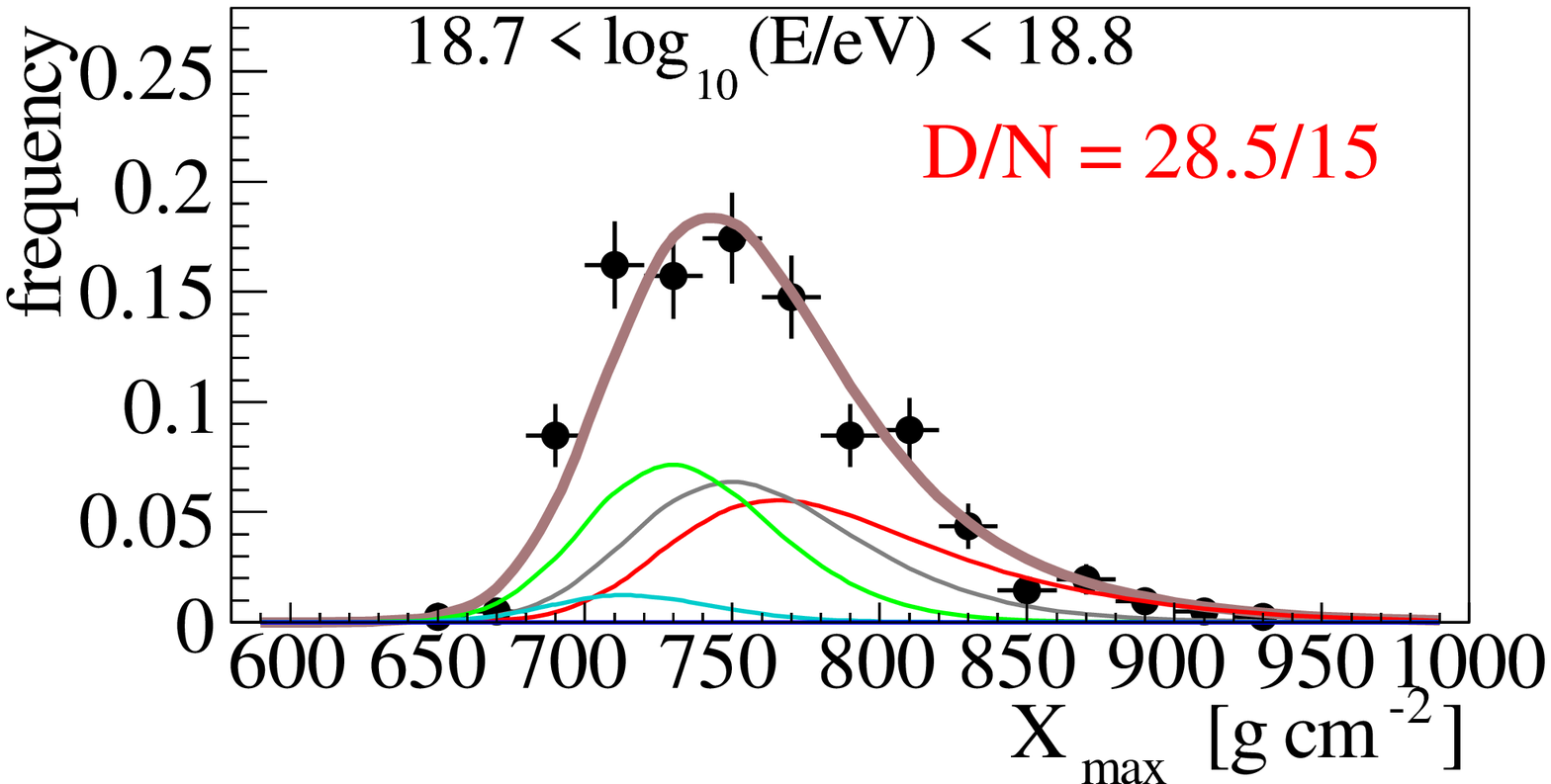}%
\includegraphics[width=0.3\textwidth]{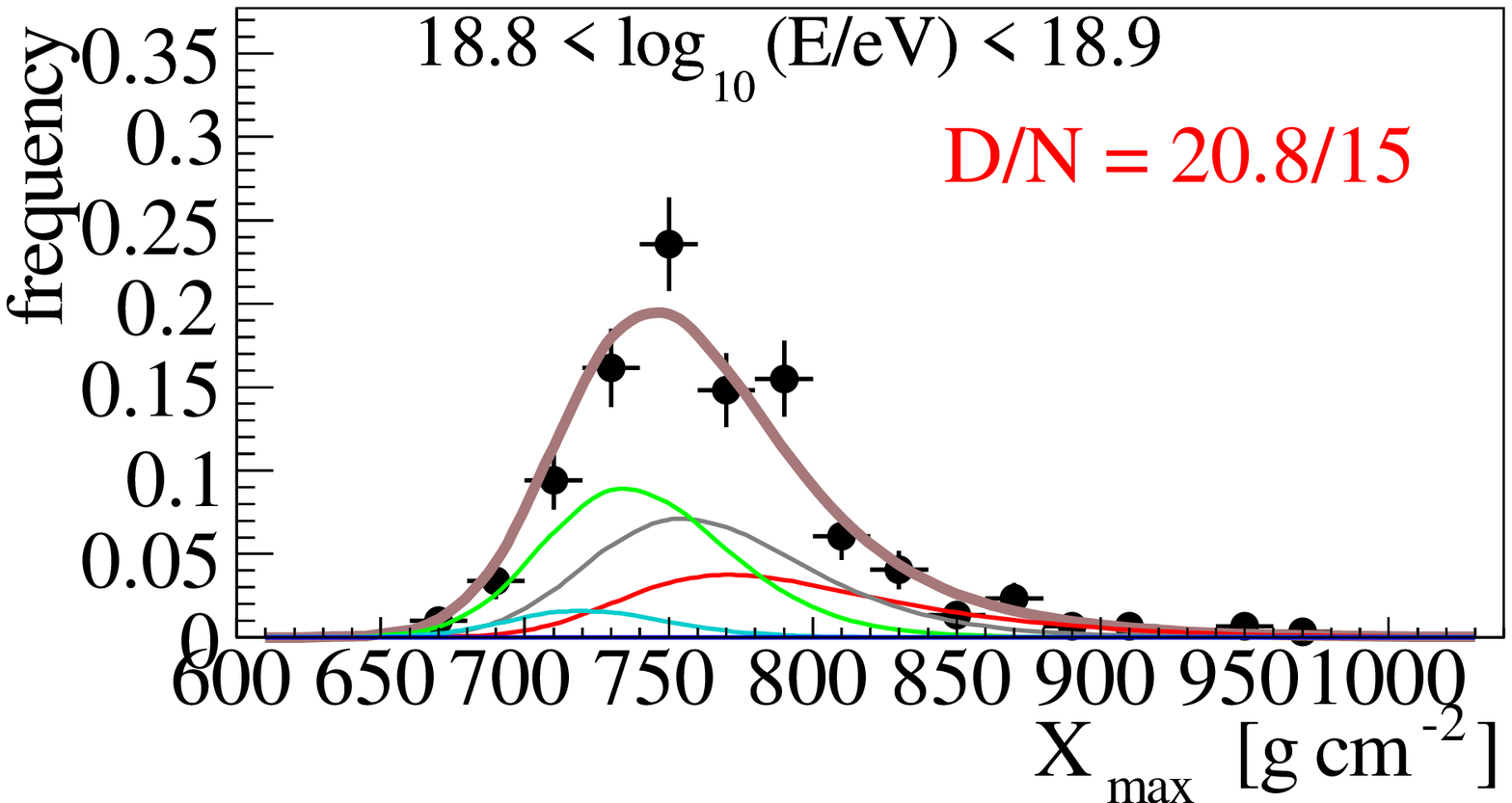}%
\includegraphics[width=0.3\textwidth]{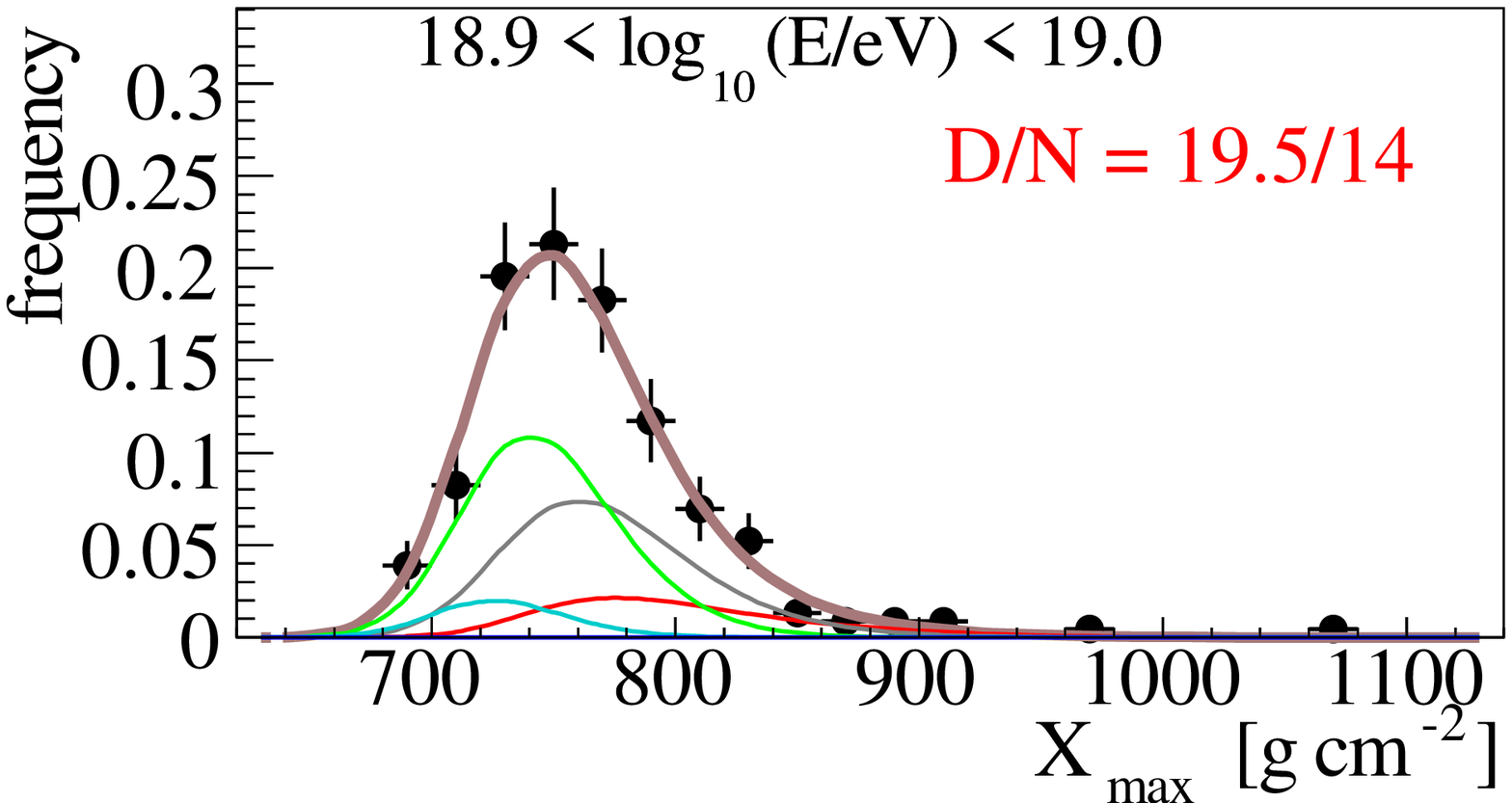}\\
\includegraphics[width=0.3\textwidth]{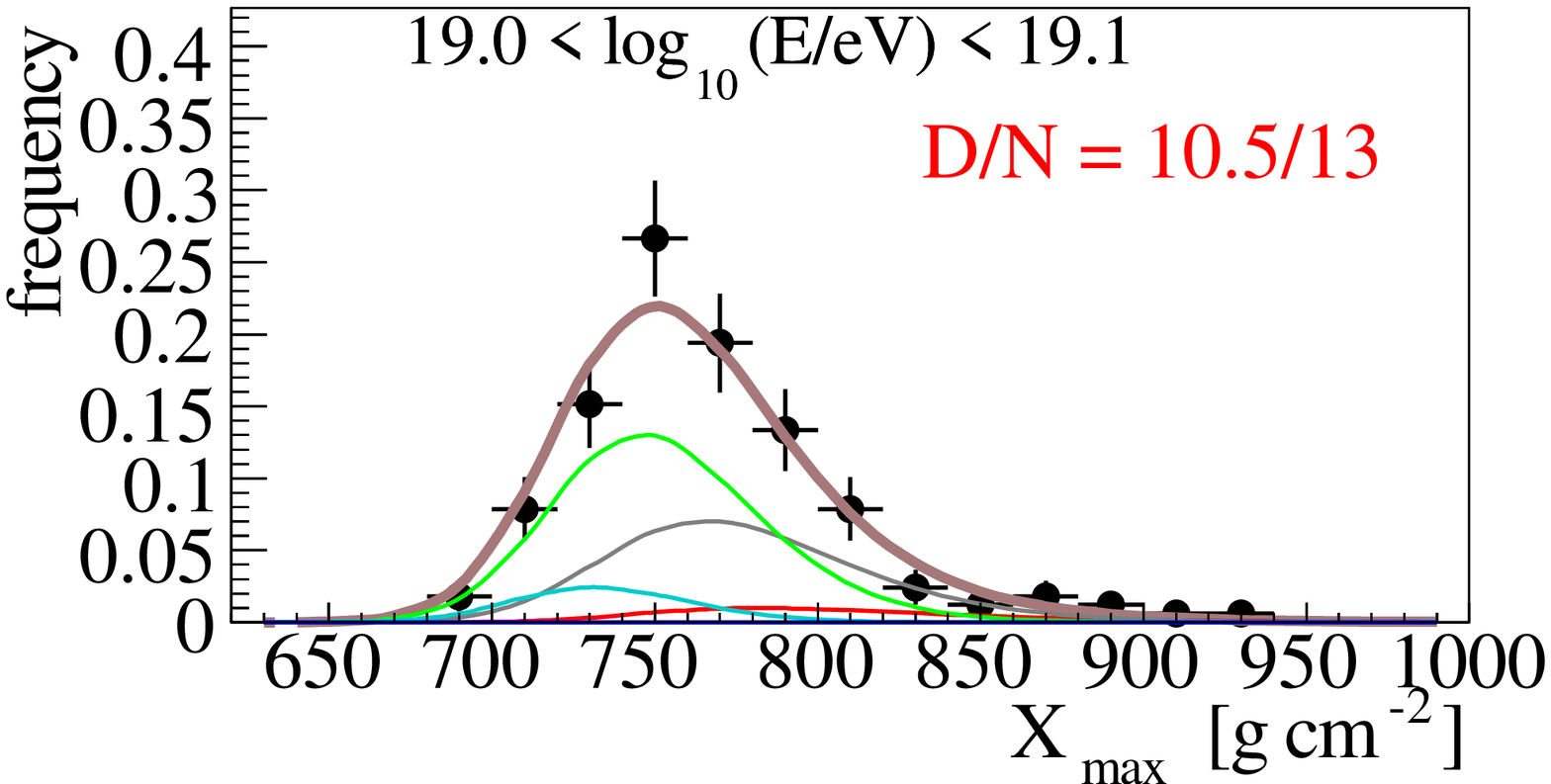}%
\includegraphics[width=0.3\textwidth]{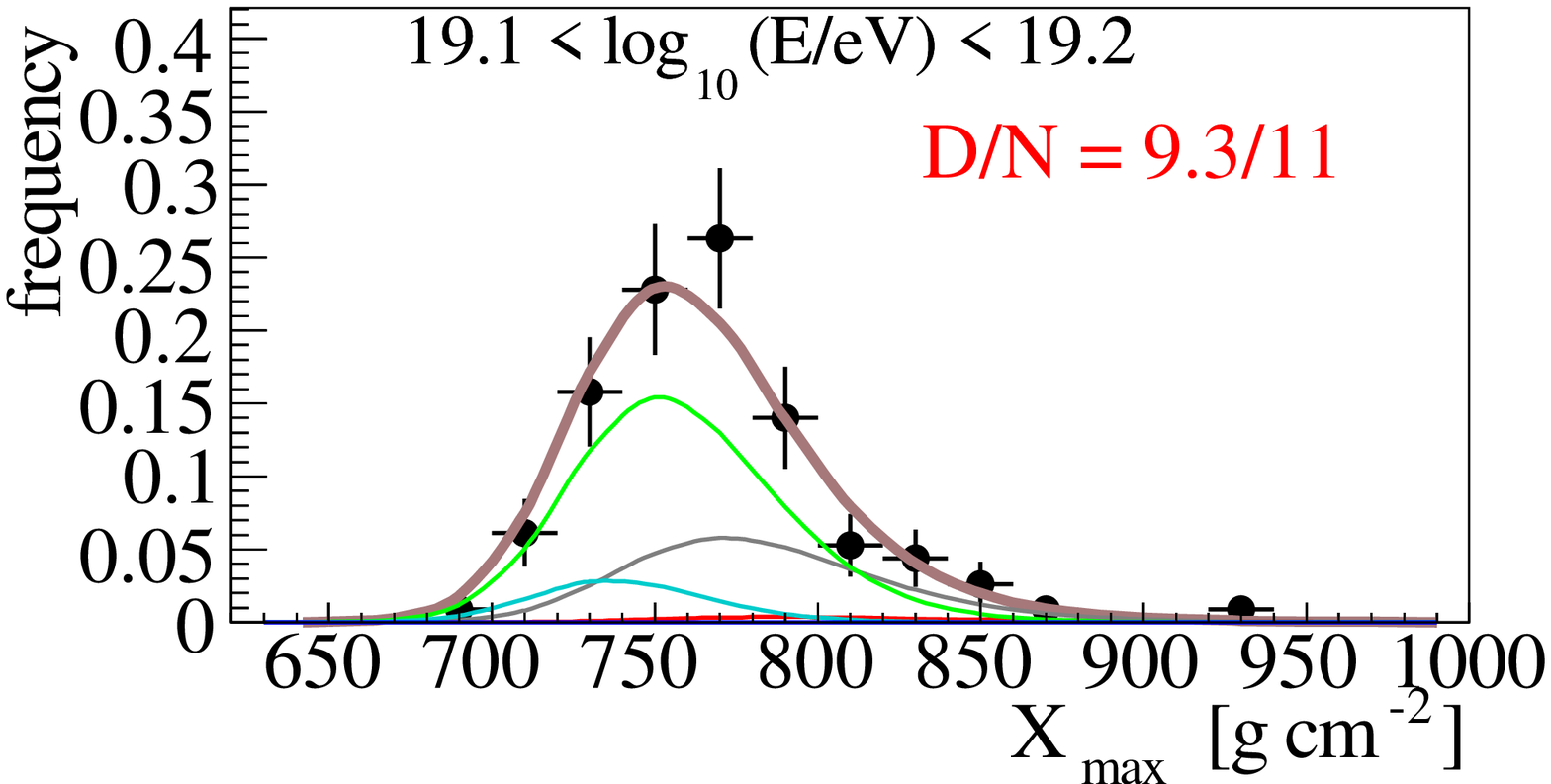}%
\includegraphics[width=0.3\textwidth]{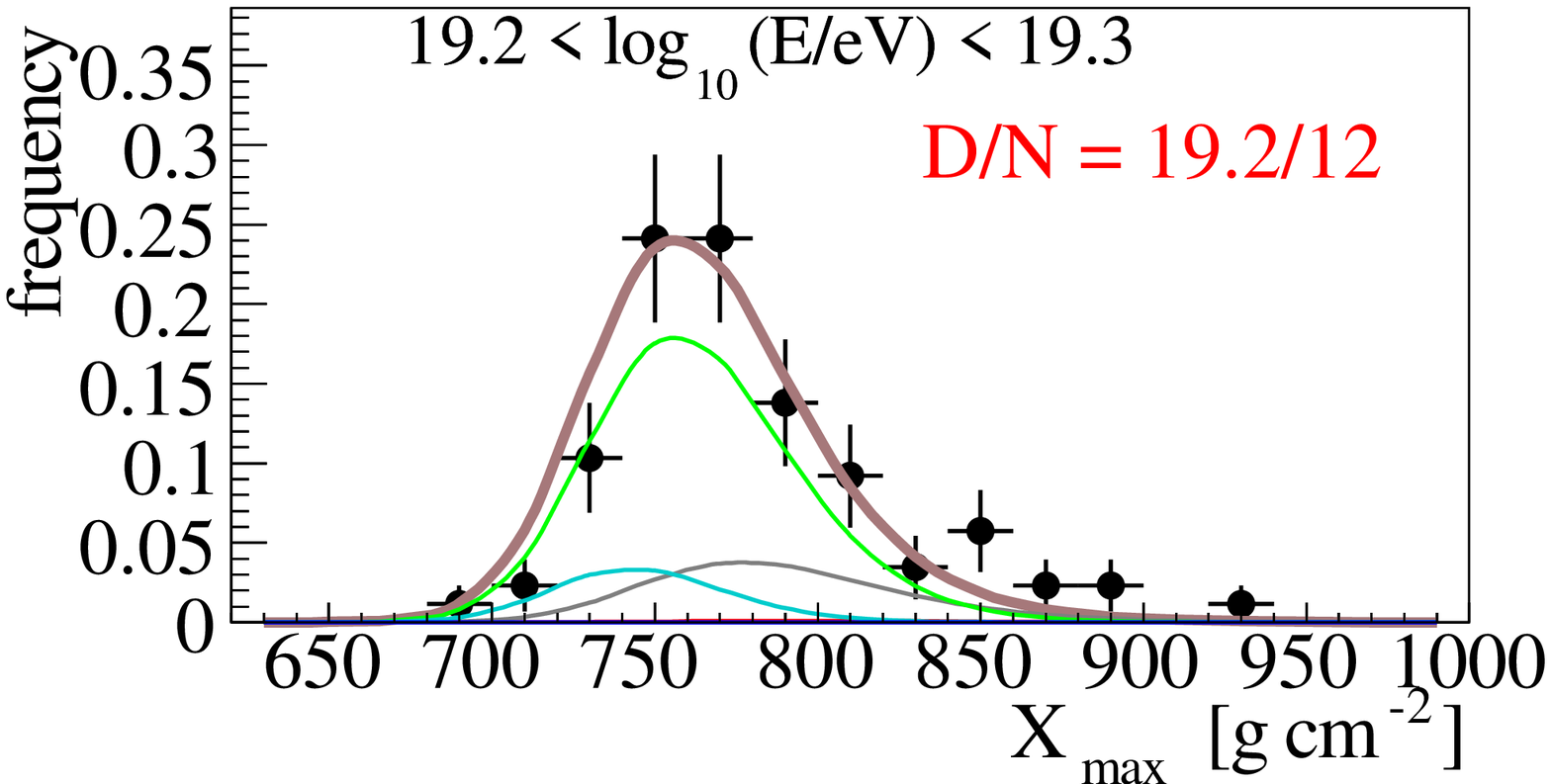}\\
\includegraphics[width=0.3\textwidth]{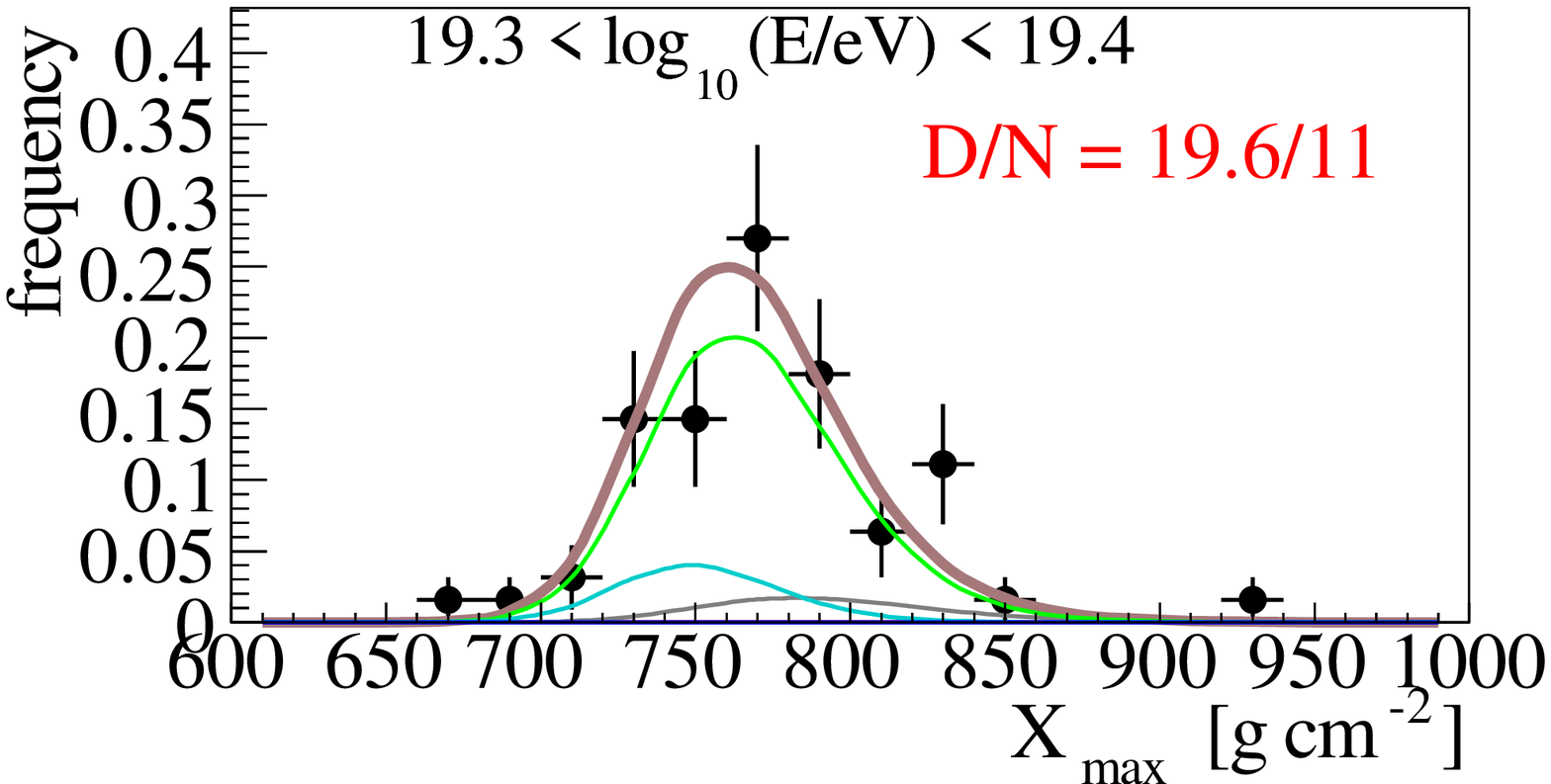}%
\includegraphics[width=0.3\textwidth]{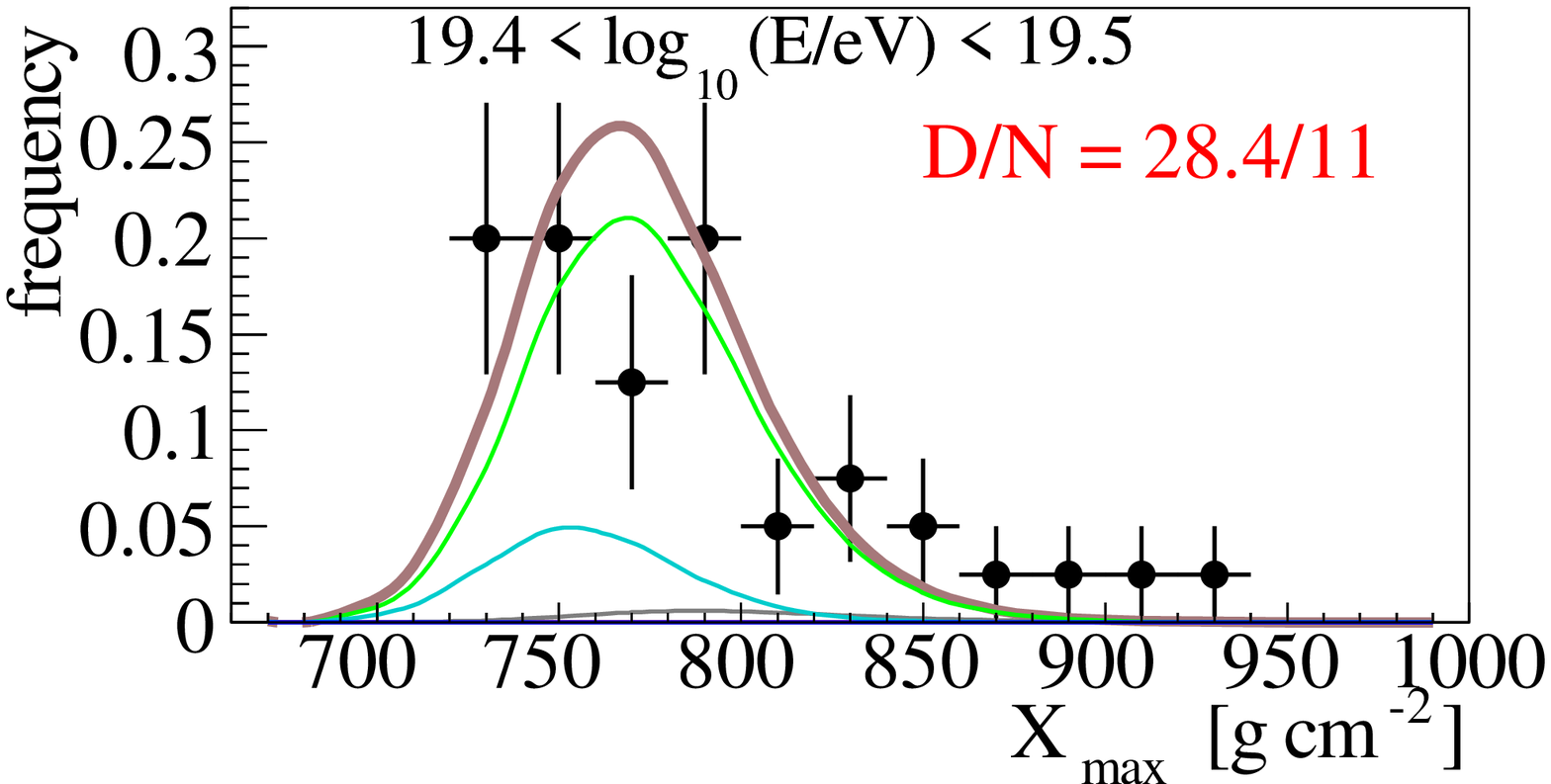}%
\includegraphics[width=0.3\textwidth]{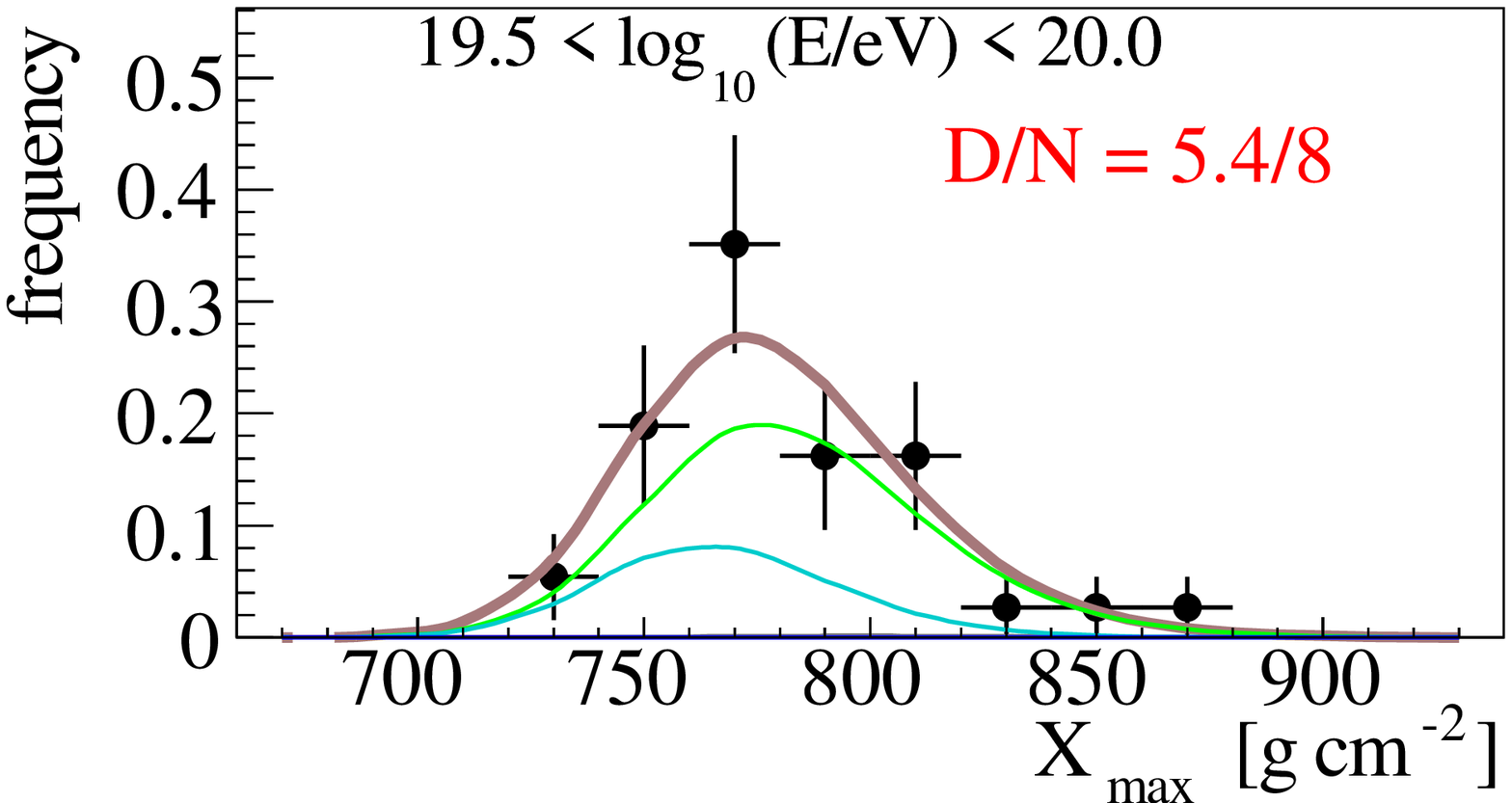}\\
\caption{Top: Fitted spectra, as function of {\it{reconstructed}} energy, compared to experimental counts. The sum of horizontal and vertical counts has been multiplied by $10$ for clarity. Bottom: The distributions of $X_\text{max}$ in the fitted energy bins, best fit minimum, SPG propagation model, EPOS-LHC UHECR-air interactions. Partial distributions are grouped according to the mass number as follows: $A = 1$ (red), $2\le A\le 4$ (grey), $5 \le A \le 22$ (green), $23 \le A \le 38$ (cyan), 
total (brown).}
\label{fig:m5}
\end{figure}

\begin{figure}[!t]
\centering
\includegraphics*[width=0.7\textwidth]{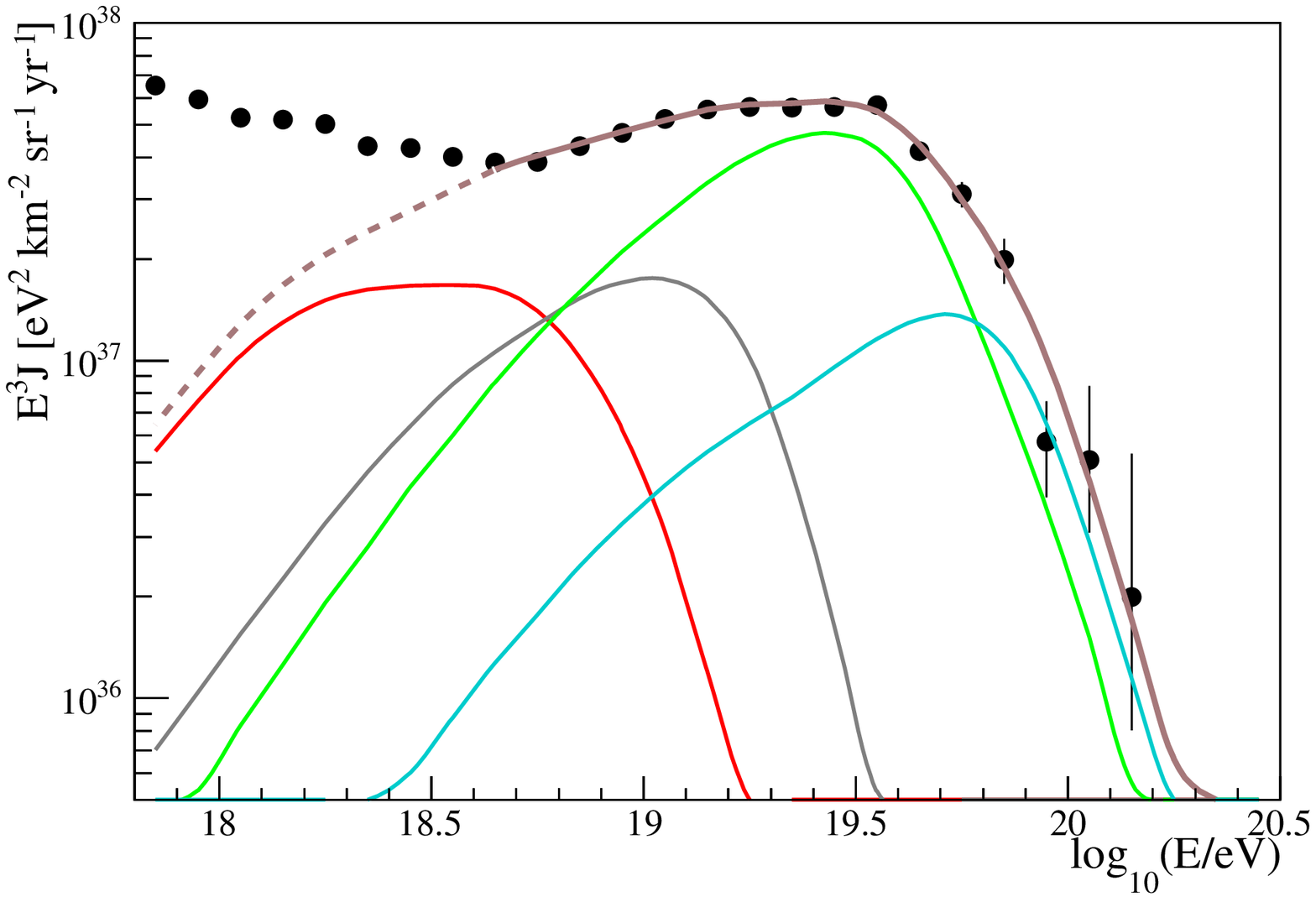}
    \includegraphics[width=0.9\textwidth]{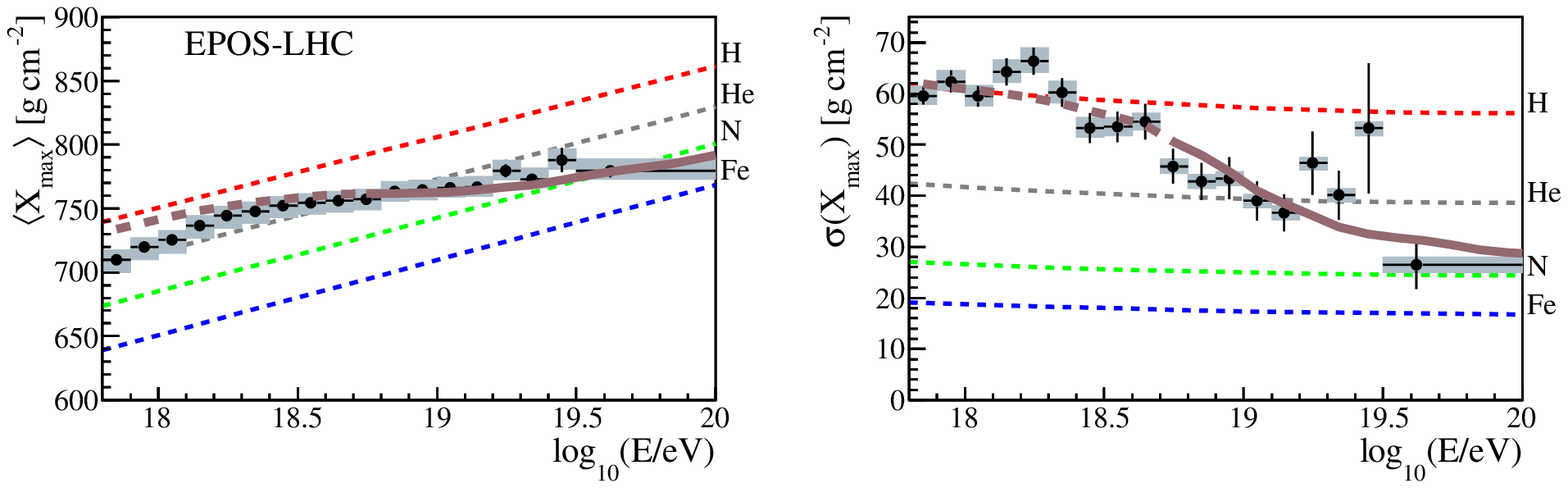}
  \caption{Top: simulated energy spectrum of UHECRs (multiplied by $E^3$) at the top of the Earth's atmosphere, obtained with the best-fit parameters for the reference model using the procedure described in section \ref{sec:fitting}. Partial spectra are grouped as in figure \ref{fig:m5}.
For comparison the fitted spectrum is reported together with the spectrum in \cite{Valino:2015zdi} (filled circles). Bottom: average and standard deviation  of the $X_{\max}$ distribution as predicted (assuming EPOS-LHC UHECR-air interactions) for the model (brown) versus pure $^1$H (red), $^4$He (grey), $^{14}$N (green) and $^{56}$Fe (blue), dashed lines. Only the energy range where the brown lines are solid is included in the fit.}
\label{fig:m4}
\end{figure}

We describe now the results of the fit, taking as reference SimProp propagation with PSB cross sections, and using the Gilmore EBL model (SPG).  The hadronic interaction model used to describe UHECR-air interactions for this fit is EPOS-LHC \cite{Pierog:2013ria}. The choice of this particular set of models will be discussed in section \ref{ssec:propm}. The best fit parameters for this model are reported in table \ref{tab:m2}; errors are calculated as described in section \ref{ssec:minuit}.\\
In figure \ref{fig:m3} we show the value of the pseudo standard deviation $\sqrt{D-D_\text{min}}$ as a function of  ($\gamma, R_\text{cut}$). 
In the inset we show the behaviour of the deviance along the  valley line connecting ($\gamma, \log_{10}(R_\text{cut}/V)$) minima (dashed line in the figure), corresponding in each point to the best fit of the other parameters ($J_0$ and $f_A$).\\
From the figure we see that there is a very definite correlation between $\gamma$ and $R_\text{cut}$: this correlation is a quite general feature of the combined fit, appearing in all the different  variations of the reference fit discussed below. 
\begin{table}[!t]
   \centering
        \bgroup
        \def\arraystretch{1.3}
   \begin{tabular}{l|rrrr}
        reference model &\multicolumn{2}{|c}{main minimum} &   \multicolumn{2}{c}{2nd minimum} \\
        (SPG -- EPOS-LHC)   & best fit & average & best fit & average \\
        \hline
        $\mathcal{L}_0$ 
{\footnotesize [$ 10^{44} ~\rm erg~Mpc^{-3}~yr^{-1}$]} & $4.99 $ & & \multicolumn{2}{c}{$9.46^*$} \\
        $\gamma$ & $0.96_{-0.13}^{+0.08}$ & $0.93 \scriptstyle \pm 0.12$ &$2.04 \scriptstyle \pm 0.01$  &$2.05_{-0.04}^{+0.02}$\\
        $\log_{10}({R_\text{cut}}/{\rm V})$ & $18.68_{-0.04}^{+0.02}$ & $18.66\scriptstyle \pm0.04$& $19.88\scriptstyle \pm 0.02 $&$19.86\scriptstyle \pm 0.06$   \\
        $f_{\rm H}(\%)$ & $0.0 $ &$12.5^{+19.4}_{-12.5}$ & $0.0 $ & $3.3^{+5.2}_{-3.3}$\\
        $f_{\rm He}(\%)$ & $67.3$& $58.6_{-13.5}^{+12.6}$ &$0.0$ & $3.6_{-3.6}^{+6.1}$  \\
        $f_{\rm N}(\%)$ & $28.1$ & $24.6_{-9.1}^{+8.9}$ &$79.8 $ & $72.1_{-10.6\!\!\!}^{+9.3}$  \\
        $f_{\rm Si}$(\%) & $4.6$ & $4.2_{-1.3}^{+1.3}$ & $20.2  $ & $20.9_{-3.9}^{+4.0}$  \\
        $f_{\rm Fe}$(\%) & $0.0$ & & $0.0$ &\\
        \hline
        $D/n$ & $174.4/119$ & & $235.7/119$ &\\
        $D$ ($J$), $D$ ($X_{\max}$) &  $13.3$, $161.1$ & & $19.5$, $216.2$ & \\
        $p$ & $0.026$ & & $5\times10^{-4}$ & \\
        \hline
        \multicolumn{5}{c}{\footnotesize $^*$From $E_{\min}=10^{15}$~eV. } \\
    \end{tabular}
    \egroup
\caption{Main and second local minimum parameters for the reference model. Errors on best-fit spectral parameters are computed from the interval $D \leq D_\text{min}+1$; those on average values are computed using the procedure described in \ref{ssec:minuit}. \protect\label{tab:m2}}
\end{table}
Considering the deviance distribution it is immediate to note that there are two regions of local minima: one, which contains the best minimum, corresponds to a low value of $R_\text{cut}$ and a spectral index $\gamma \approx 1$; this minimum region is quite extended towards smaller values of $\gamma$ at a slowly decreasing $R_\text{cut}$. In figure \ref{fig:m5} we present the spectrum data we actually fit and the $X_\text{max}$ distributions together with the fitted functions, while in figure \ref{fig:m4} the fit results are compared for reference to the all-particle spectrum and  $X_\text{max}$ momenta.~The essential features of such a model have been discussed elsewhere \cite{Aloisio:2013hya,Hooper:2009fd} and, using a similar approach to that of this work, in \cite{bib:fitICRC15}, the general features being a low maximum rigidity around $\log_{10}(R_\text{cut}/\text{V})=18.5$, a hard spectrum and a 
composition dominated by Helium and heavier elements.
\begin{figure}[!t]
\centering
\includegraphics*[width=0.7\textwidth]{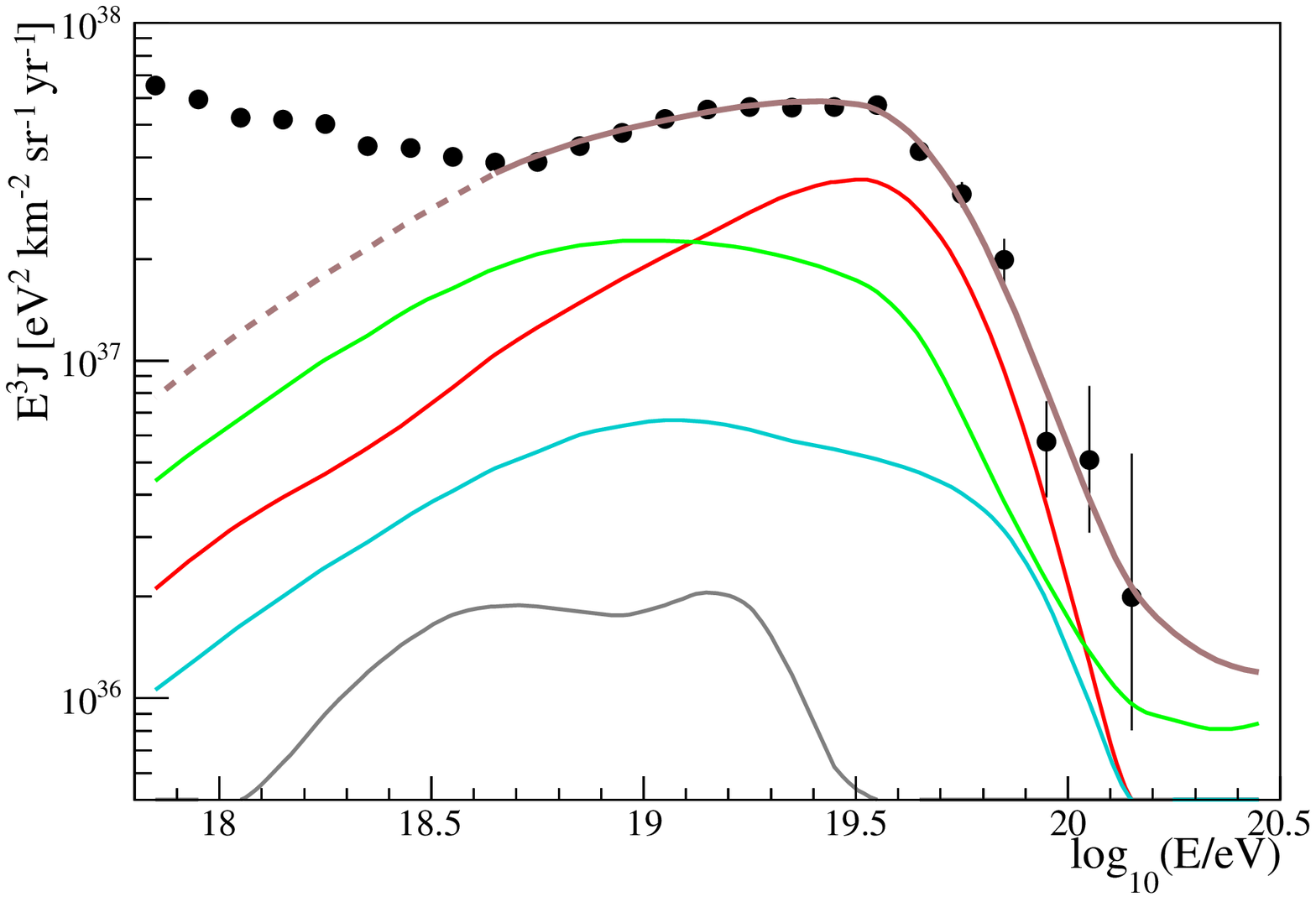}
\includegraphics*[width=0.9\textwidth]{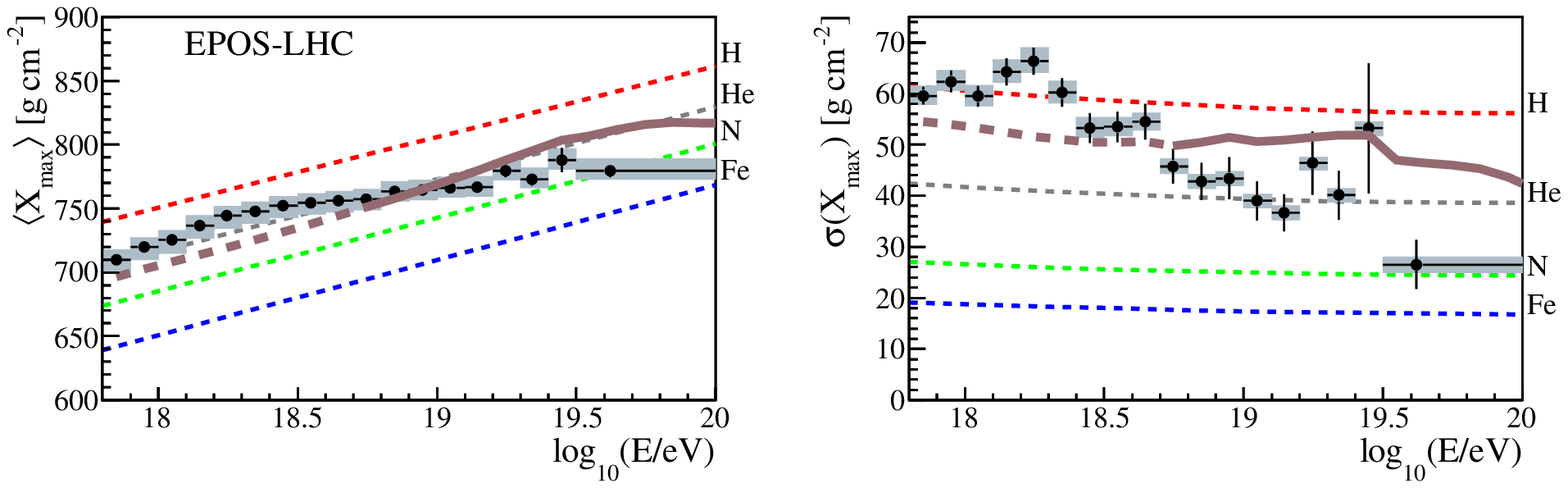}
\caption{Same as figure \ref{fig:m4} at the local minimum at $\gamma=2.04$, SPG propagation model, EPOS-LHC UHECR-air interactions.}
\label{fig:m6}
\end{figure}
\\
There is also a second relative minimum, which appears less extended, around the pair $\gamma = 2.04$ and  $\log_{10}(R_\text{cut}/\text{V})= 19.88$. 
For nuclei injected with these parameters the effects of interactions during propagation are dominant, as it is demonstrated by copious production of high energy secondaries (in particular Hydrogen). This is the reason why in this region the fit to composition is quite bad, as reported in table \ref{tab:m2} and in figure \ref{fig:m6}, with $X_\text{max}$ simulated distributions almost always larger than experimentally observed; 
this solution, in the reference model, can be excluded at the $7.5 \sigma$ level. \\
The low maximum rigidity $ R_\text{cut} \approx 4.9 \cdot 10^{18} $ V in the best fit minimum implies that the maximum energy for Iron nuclei would be $\approx 1.3 \cdot 10^{20}$ eV. This has the very important consequence that 
the shape of the all particle spectrum is likely due to the concurrence of two effects: maximum energy reached at the sources $and$ energy losses during propagation.\\
The injection spectra are very hard, at strong variance with the expectation for the first order Fermi acceleration in shocks, although alternatives are possible, as already mentioned in \ref{sec:astro}. The composition at sources is mixed, and essentially He/N/Si dominated with no contribution from Hydrogen or Iron at the best fit. The value of $J_0$ of the best fit corresponds to a total emissivity $\mathcal{L}_0 = \sum_A \int_{E_\text{min}}^{+\infty} E q_A(E) \mathrm{d}E=4.99 \times 10^{44}$ $\mathrm{erg/Mpc^3/year}$, where $q_A(E)$ is the number of nuclei with mass $A$ injected per unit energy, volume and time, and $\mathcal{L}_\text{He}=0.328 \mathcal{L}_0$, $\mathcal{L}_\text{N}=0.504 \mathcal{L}_0$, $\mathcal{L}_\text{Si}=0.168 \mathcal{L}_0$, with $\mathcal{L}_A / \mathcal{L}_0 = f_A Z_A^{2-\gamma}/ \sum_A (f_A Z_A^{2-\gamma})$.\\
Because of the low value of $R_\text{cut}$, the observed spectra are strongly sensitive to the behaviour of accelerators near the maximum energy and therefore even large differences of injection spectral indices have  little effect on the observable quantities. This is the reason of the large extent of the best minima region, and will be discussed below.\\
Given the deviance reported in table \ref{tab:m2}, the probability of getting a worse fit if the model is correct ($p$-value) is $p=2.6 \%$. Notice however that the effect of experimental systematics is not taken into account here. A discussion of systematics is presented in section \ref{ssec:expsys}. \\
The errors on the parameters are computed as explained in \ref{ssec:minuit}. Those on the elemental fractions are generally large, indicating that different combinations of elemental spectra can give rise to similar observed spectra. This fact is reflected by the presence of large (anti)correlations among the injected nuclear spectra, as shown in table \ref{tab:m2a}.
\begin{table}[!t]
    \centering\begin{tabular}{l|rrrrr}
        & H & He & N &Si & $\gamma$ \\
        \hline
        He & $-0.78$  &   & & &  \\
        N & $-0.61$  &  $-0.01$ & & & \\
        Si & $-0.43$  & $-0.08$  & $+0.75$ & &  \\
        $\gamma$ & $-0.26$ &$ -0.32$&$ +0.80$& $+0.89$ &    \\
        $\log_{10}(R_\text{cut}/\text{V})$ &$-0.59$ &$+0.00$& $+0.93$&$ +0.84$& $+0.86$  \\
        \hline  
    \end{tabular}
\caption{Correlation coefficients among fit parameters (SPG model, EPOS-LHC UHECR-air interactions) as derived from the mock simulated sets.}
\label{tab:m2a}
\end{table}

%
%
%
\subsection{The effect of experimental systematics \label{ssec:expsys}}
The data on which the fit is performed are affected by different experimental systematic uncertainties. In this section we analyze their effect on the fit parameters. \\
The main systematic effects derive from the energy scale in the spectrum \cite{Valino:2015zdi}, and the $X_\text{max}$ scale \cite{Aab:2014kda}. The uncertainty on the former is assumed constant $\Delta E/E = 14 \%$  in the whole energy range considered, while that on composition $\Delta X_\text{max}$ is asymmetric and slightly energy dependent, ranging from about $6$ to
$ 9 $ g/cm$^2$. As described in section \ref{sec:fitting} two approaches are used to take into account the experimental systematics in the fit.\\ 
Including the systematics as nuisance parameters in the fit, we obtain the results in table \ref{tab:m7}.
Here the average value and uncertainty interval of the model parameters include both statistical and systematic uncertainties of the measurement. Also shown are shifts in the energy scale and $X_\text{max}$ scale of the experiment as preferred by the fit. Both remain within one standard deviation of the given uncertainties.
The effect of fixed shifts within the experimental systematics are reported in table \ref{tab:m6}.
\begin{table}[!t] 
   \centering
        \bgroup
        \def\arraystretch{1.3}
        \begin{tabular}{l|rrrr}
        reference model & best fit & average & shortest $68 \%$ int.\\
        \hline
        $\gamma$ & $1.22$ & $1.27$ & $1.20 \div 1.38$ \\
        $\log_{10}({R_\text{cut}}/{\rm V})$ & $18.72 $ &$18.73$ & $18.69 \div 18.77 $ \\
        $f_{\rm H}(\%)$ & $6.4 $ & $15.1$ & $0.0 \div 18.9 $  \\
        $f_{\rm He}(\%) $ & $46.7 $ & $31.6$ & $18.9 \div 47.8 $  \\
        $f_{\rm N}(\%)$ & $37.5 $ & $42.1$ & $ 30.7 \div 51.7 $  \\
        $f_{\rm Si}(\%)$ & $9.4 $ & $11.2$ & $5.4 \div 14.6$  \\
        $\Delta X_\text{max}/\sigma_\text{syst}$ & $-0.63$ & $-0.69 $ & $-0.90 \div -0.48$ \\
        $\Delta E /\sigma_\text{syst}$ & $+0.00$ & $+0.12$ & $-0.57 \div +0.54$\\
        \hline
        $D/n$ & $166.5/117$ &\\
        $D$ ($J$), $D$ ($X_{\max}$) & $12.9$, $153.5$ &\\
         \hline
    \end{tabular} \egroup
\caption{Best-fit parameters for the reference model, including systematic effects as nuisance parameters in the fitting procedure. Errors are computed as described in \ref{ssec:bayes}.  \protect\label{tab:m7}}
\end{table}\\
From the results one can infer that the total deviance of the fit is not strongly sensitive to shifts in the energy scale, though the injection mass fractions are. This is because an increase (or decrease) in the observed position of the energy cutoff can be reproduced by assuming a heavier (lighter) mass composition, as the photo-disintegration threshold energy is roughly proportional to the mass number of the nuclei.\\
On the other hand, a negative  $1 ~\sigma$ change on the $X_\text{max}$ scale does not change $D(J)$ and slightly improves $D(X_{\max})$ and moves $\gamma$ towards somewhat larger values. A positive change dramatically drives $\gamma$ towards negative values outside the fitted interval and moves $R_\text{cut}$ towards lower values, since it implies a lighter composition at all energies, in strong disagreement with the width of the $X_\text{max}$ distributions. Taking into account systematics as in tables \ref{tab:m7} and \ref{tab:m6}, the $p$-value of the best fit becomes $p \approx 6 \%$.
\begin{table}[!t]
	\centering
        \bgroup
        \def\arraystretch{1.3}
	\begin{tabular}{cc|cc|ccc}
		$\Delta X_{\max}$ & $\Delta E/E$ & $\gamma$ & $\log_{10}(R_\text{cut}/\mathrm{V})$ & $D$ & $D(J)$ & $D(X_{\max})$ \\
		\hline
		~ & $-14\%$ & $+1.33\scriptstyle\scriptstyle \pm0.05$ & $18.70\scriptstyle\scriptstyle \pm0.03$ & $167.0$ & $19.0$ & $148.0$ \\
		$-1\sigma_\text{syst}$ & $\phantom{+0}0\phantom{\%}$ & $+1.36\scriptstyle\scriptstyle \pm0.05$ & $18.74_{-0.04}^{+0.03}$ & $166.7$ & $14.7$ & $152.0$ \\
		~ & $+14\%$ & $+1.39_{-0.05}^{+0.03}$ & $18.79_{-0.04}^{+0.03}$ & $169.6$ & $13.0$ & $156.6$ \\
		\hline
		~ & $-14\%$ & $+0.92_{-0.10}^{+0.09}$ & $18.65\scriptstyle\scriptstyle \pm0.02$ & $176.1$ & $18.1$ & $158.0$ \\
		$\phantom{+}0\phantom{\sigma_\text{syst}}$ & $\phantom{+0}0\phantom{\%}$ & $+0.96_{-0.13}^{+0.08}$ & $18.68_{-0.04}^{+0.02}$ & $174.3$ & $13.2$ & $161.1$ \\
		~ & $+14\%$ & $+0.99_{-0.12}^{+0.08}$ & $18.71_{-0.04}^{+0.03}$ & $176.3$ & $11.7$ & $164.4$ \\
		\hline
		~ & $-14\%$ & $-1.50_{* }^{+0.08}$ & $18.22\scriptstyle\scriptstyle \pm0.01$ & $208.1$ & $15.3$ & $192.8$ \\
		$+1\sigma_\text{syst}$ & $\phantom{+0}0\phantom{\%}$ & $-1.49_{* }^{+0.16}$ & $18.25_{-0.01}^{+0.02}$ & $202.6$ & $\phantom{0}9.7$ & $192.8$ \\
		~ & $+14\%$ & $-1.02_{-0.44}^{+0.37}$ & $18.35\scriptstyle\scriptstyle \pm0.05$ & $206.4$ & $11.3$ & $195.1$ \\
		\hline
        \multicolumn{7}{c}{\footnotesize $^*$This interval extends all the way down to $-1.5$,  the lowest value of $\gamma$ we considered. } \\
   \end{tabular}
   \egroup
        \caption{The effect of shifting the data according to the quoted systematics in energy and $X_\text{max}$ scales. In this and all other tables errors are computed from the interval $D \leq D_\text{min}+1$. \protect\label{tab:m6}} 

\end{table}
\begin{figure}
    \centering\includegraphics[width=0.45\textwidth]{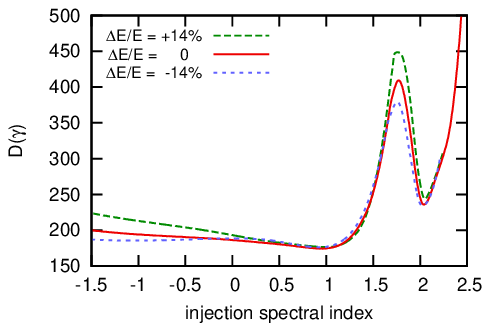}\includegraphics[width=0.45\textwidth]{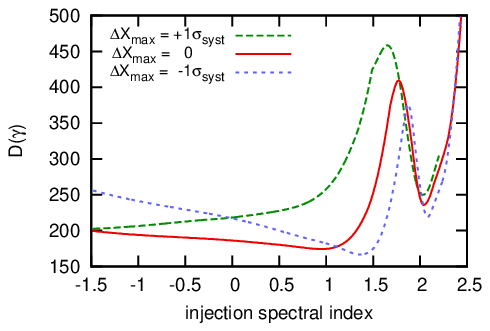}
  \caption{The change in $D_\text{min}$ vs $\gamma$ with respect to change in energy (left) and $X_\text{max}$ scale, at nominal value of the other parameter. \label{fig:m10}}
\end{figure}
\begin{figure}
    \centering\includegraphics[width=0.45\textwidth]{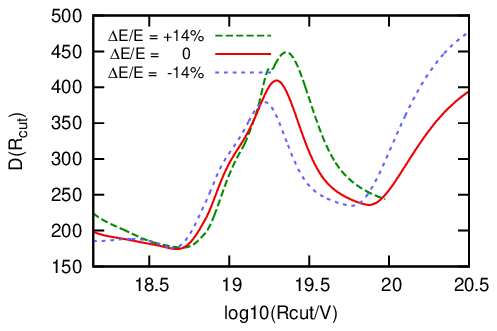}\includegraphics[width=0.45\textwidth]{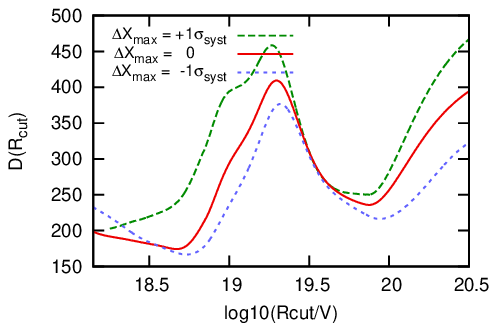}
  \caption{The change in $D_\text{min}$ vs $R_\text{cut}$  with respect to change in energy (left) and $X_\text{max}$ scale, at nominal value of the other parameter. \label{fig:m11}}
\end{figure}
In figures \ref{fig:m10}, \ref{fig:m11} the changes of the $D(\gamma) $
and $D(R_\text{cut})$ relations with systematics are reported. 
\subsection{Effects of physical assumptions \label{ssec:ebl}}
\subsubsection{Air interaction models}
\begin{figure}[t]
    \centering\includegraphics[width=0.45\textwidth]{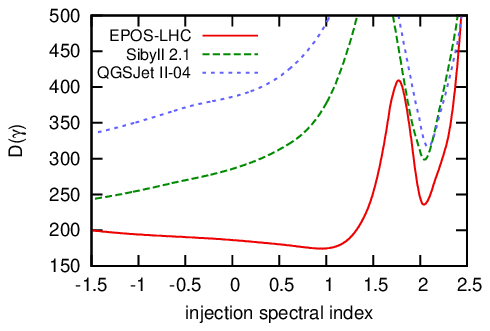}\includegraphics[width=0.45\textwidth]{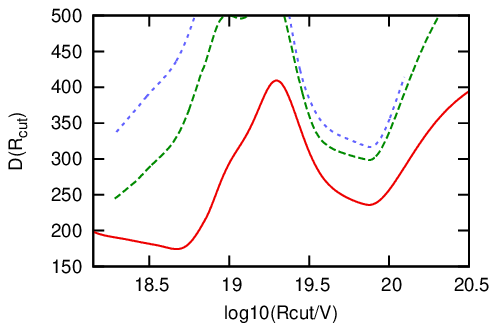}
    \caption{The effect of different hadronic interaction models, using propagation model SPG and various UHECR-air interaction models \label{fig:m8}}
\end{figure}

\begin{figure}[t]
        \centering
        \includegraphics[width=0.90\textwidth]{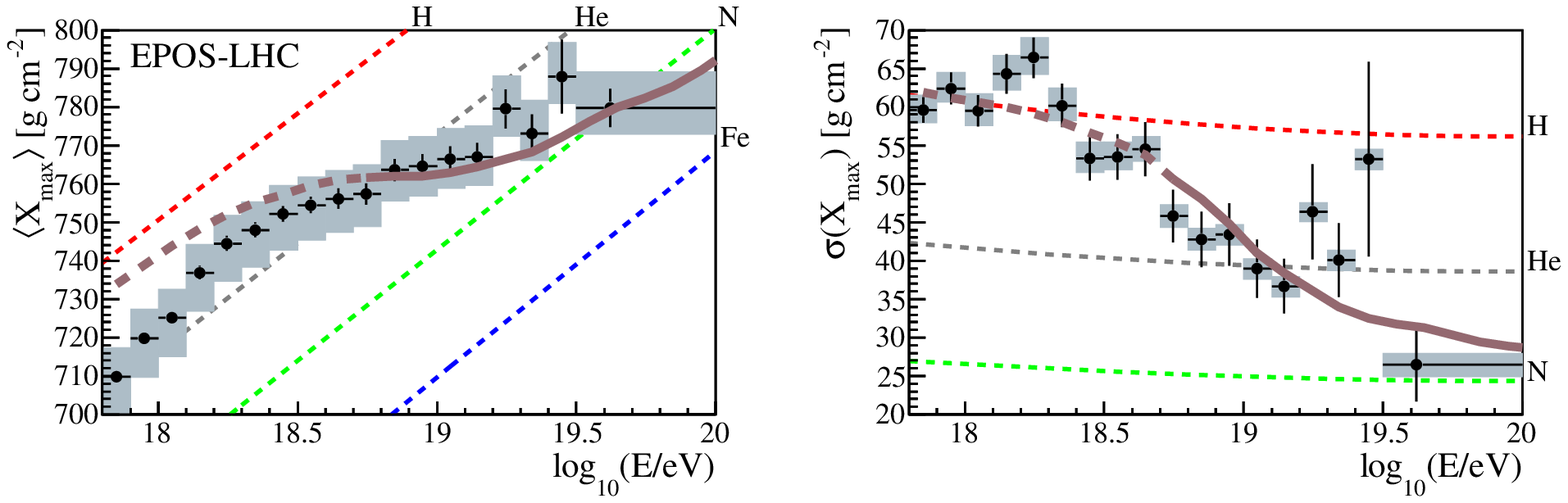}
        \includegraphics[width=0.90\textwidth]{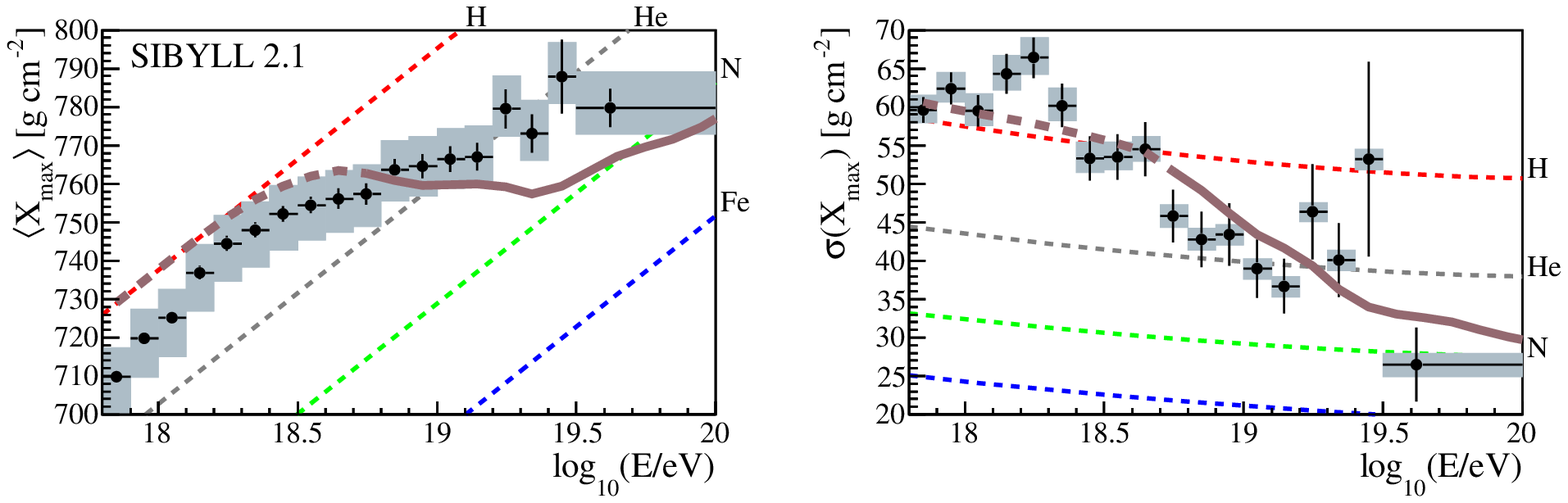}
        \caption{Average (left) and standard deviation (right) of $X_{\max}$ as predicted in the best-fit model assuming EPOS-LHC (top) and Sibyll~2.1 (bottom). The colour code is the same as in the bottom panels of figure~\ref{fig:m4}, but the range of the vertical axis is narrower in order to highlight the differences between the two models. When using QGSJet~II-04, the agreement with the data is even worse than with Sibyll~2.1.}
        \label{fig:SibXmax}
\end{figure}

To derive the results reported above a specific model (EPOS-LHC) of hadronic interactions between UHECR and nuclei in the atmosphere  has been used. It is therefore interesting to consider the influence of this choice on the results. For this reason, we have repeated the fit of the SPG model using  Sibyll 2.1 \cite{Ahn:2009wx} and QGSJet II-04 \cite{Ostapchenko:2010vb}. The results are presented in table~\ref{tab:m5} and in figure~\ref{fig:m8}. The use of these interaction models significantly worsens the goodness of the fit  in the chosen  range of fitted parameters, as shown in figure~\ref{fig:SibXmax} and quantified by the $D(X_{\max})$~values in table~\ref{tab:m5}, pushing towards very low values of $R_\text{cut}$ and consequently extreme negative values of $\gamma$. 
\begin{table}[t]
	\centering
        \bgroup
        \def\arraystretch{1.3}
	\begin{tabular}{c|cc|ccc}
		model & $\gamma$ & $\log_{10}(R_\text{cut}/\mathrm{V})$ & $D$ & $D(J)$ & $D(X_{\max})$ \\
		\hline
		EPOS-LHC & $+0.96_{-0.13}^{+0.08}$ & $18.68_{-0.04}^{+0.02}$ & $174.3$ & $13.2$ & $161.1$ \\
		Sibyll 2.1 & $-1.50^{+0.05}$ & $18.28_{-0.01}^{+0.00}$ & $243.4$ & $19.7$ & $223.7$ \\
		\multirow{2}{*}{QGSJet II-04} & $+2.08_{-0.01}^{+0.02}$ & $19.89_{-0.02}^{+0.01}$ & $316.5$ & $10.5$ & $306.0$ \\
		 & $-1.50_{*}^{+0.02}$ & $18.28_{-0.00}^{+0.01}$ & $334.9$ & $19.6$ & $315.3$ \\
		\hline
\multicolumn{6}{p{0.7\columnwidth}}{\footnotesize $^*$Using QGSJet II-04 the minimum at $\gamma \approx 2$ is better than that at $\gamma \lesssim 1$, which is at the edge of the parameters region we considered.} \\
    \end{tabular} 
    \egroup
    \caption{Same as table~\protect\ref{tab:m2}, using propagation model SPG and various UHECR-air interaction models  \protect\label{tab:m5}}
\end{table}
\subsubsection{Shape of the injection cut-off \label{sec:cutoff}}
\begin{figure}[!t]
   	\centering\includegraphics[width=0.75\textwidth]{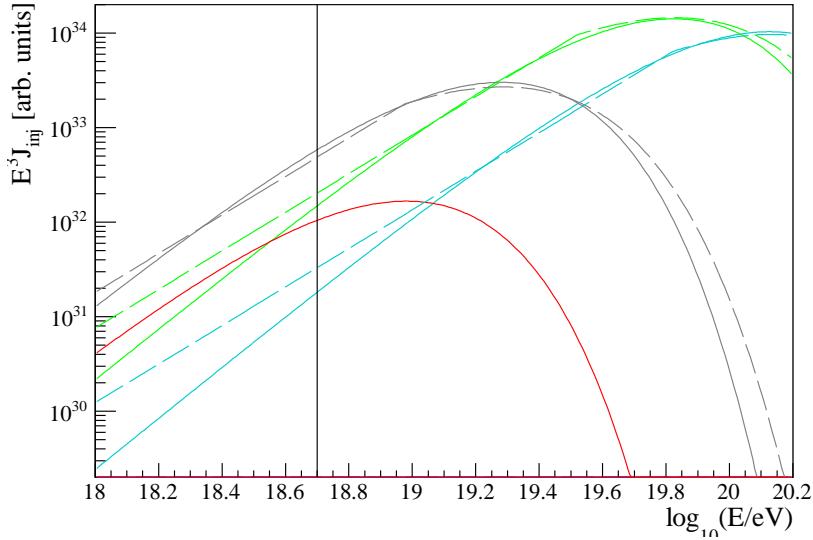}
\caption{Injection spectra corresponding to the two choices of the cutoff function (the continuous lines correspond to the simple exponential and the dashed ones to the broken exponential; the vertical line is the minimum detected energy considered in the fit); $^1$H (red), $^4$He (grey), $^{14}$N (green) and $^{28}$Si (cyan) \label{fig:m11a}}
      \end{figure}

We discuss here the effect of the shape of the cut-off function we have chosen for the reference fit. This choice has been purely instrumental and is not physically motivated. More physical possibilities can be considered, starting from a simple exponential multiplying the power-law flux at all energies, to more complex possibilities (see section \ref{sec:astro}). In table \ref{tab:m10} we present the effect on the fit parameters 
 of the choice of an exponential cut-off function.\\
It has to be noted that the two injection models are not as different as directly comparing the numerical values of the parameters suggests, because the simple exponential cutoff takes over sooner than a broken exponential one with the same nominal cutoff rigidity and makes the spectrum softer ($\gamma_\text{eff} = -\mathrm{d} \ln J/\mathrm{d} \ln E =\gamma + E/(ZR_\text{cut}) > \gamma$, see figure \ref{fig:m11a}).  In any event, the goodness of fit is almost identical in the two cases, so our data are not sensitive to their difference.\\
\begin{table}[!t]
	\centering
        \bgroup
        \def\arraystretch{1.3}
	\begin{tabular}{c|cc|ccc}
		cutoff function & $\gamma$ & $\log_{10}(R_\text{cut}/\mathrm{V})$ & $D$ & $D(J)$ & $D(X_{\max})$ \\
		\hline
		broken exponential & $+0.96_{-0.13}^{+0.08}$ & $18.68_{-0.04}^{+0.02}$ &$174.3$ & $13.2$ & $161.1$ \\
simple exponential & $+0.27_{-0.24}^{+0.21}$ & $18.55_{-0.06}^{+0.09}$ & $178.3$ & $14.4$ & $163.9$ \\
		\hline
	\end{tabular}
        \egroup
\caption{ The effect of the choice of the cut-off function on the fitted parameters, reference model. \label{tab:m10}}
\end{table}
\subsubsection{Propagation models \label{ssec:propm}}
As a consequence of the low maximum rigidity in the best fit, interactions on the CMB are subdominant with respect to those on the EBL, particularly for medium atomic number nuclei (see for instance \cite{Batista:2015mea}).
Therefore, poorly known processes such as photo-disintegration of medium nuclei (e.g. CNO) on EBL can strongly affect extragalactic propagation. In \cite{Batista:2015mea} a detailed discussion of several such effects can be found, as well as a comparison of the two propagation codes used. Particular importance have, for instance, the radiation intensity peak in the far infrared region, and the partial cross sections of photo-disintegration channels in which $\alpha$ particles are ejected. \\
To study the effects of uncertainties in the simulations of UHECR propagation, we repeated the fit using the combinations of Monte Carlo propagation code, photo-disintegration cross sections and EBL spectrum listed in table~\ref{tab:m1}.
In the present analysis EBL spectra and evolution are taken from \cite{Gilmore:2011ks} (Gilmore 2012) and \cite{Dominguez:2010bv} (Dom\'inguez 2011). 
As for photo-disintegration, here we use the cross sections from \cite{Stecker:2006eh,Stecker:2005qs} (PSB), \cite{Koning20122841} (TALYS, as described in \cite{Batista:2015mea}), and Geant4 \cite{Allison:2006ve} total cross sections with TALYS branching ratios (\cite{Batista:2015mea}).  \\
In table \ref{tab:m3} we present the spectrum parameters of the best fit at the principal minimum, while table \ref{tab:m4} contains the elemental fractions. We have verified that the different fitting procedures outlined in section \ref{ssec:minuit}, \ref{ssec:bayes} have no influence on the fit results.\\
The difference among models with different physical assumptions are generally much larger than the statistical errors on the parameters, implying that they really correspond to different physical cases, at least in the best minimum region.
\begin{table}[!t]
\centering
    \begin{tabular}{ l | l l l }
        ~ & MC code & $\sigma_\text{photodisint.}$ & EBL model \\
        \hline
        SPG & SimProp & PSB & Gilmore 2012 \\
        STG & SimProp & TALYS & Gilmore 2012 \\
        SPD & SimProp & PSB & Dom\'inguez 2011 \\
        CTG & CRPropa & TALYS & Gilmore 2012 \\
        CTD & CRPropa & TALYS & Dom\'inguez 2011 \\
        CGD & CRPropa & Geant4 & Dom\'inguez 2011 \\
        \hline
    \end{tabular}

    \caption{The  propagation models  used (see~Ref.~\cite{Batista:2015mea} and references therein for details)  \protect\label{tab:m1}}
\end{table}
\begin{table}[!t]
	\centering
        \bgroup
        \def\arraystretch{1.3}
	\begin{tabular}{c|cc|ccc}
		model & $\gamma$ & $\log_{10}(R_\text{cut}/\mathrm{V})$ & $D$ & $D(J)$ & $D(X_{\max})$ \\
		\hline
		SPG & $+0.96_{-0.13}^{+0.08}$ & $18.68_{-0.04}^{+0.02}$ & $174.3$ & $13.2$ & $161.1$ \\
		STG & $+0.77_{-0.13}^{+0.07}$ & $18.62_{-0.04}^{+0.02}$ & $175.9$ & $18.8$ & $157.1$ \\
		SPD & $-1.02_{-0.26}^{+0.31}$ & $18.19_{-0.03}^{+0.04}$ & $187.0$ & $\phantom{0}8.4$ & $178.6$ \\
		\multirow{2}{*}{CTG} & $-1.03_{-0.30}^{+0.35}$ & $18.21_{-0.04}^{+0.05}$ & $189.7$ & $\phantom{0}8.3$ & $181.4$ \\
		 & $+0.87_{-0.06}^{+0.08}$ & $18.62\scriptstyle\pm0.02$ & $191.9$ & $29.2$ & $162.7$ \\
		CTD & $-1.47_{*}^{+0.28}$ & $18.15_{-0.01}^{+0.03}$ & $187.3$ & $\phantom{0}8.8$ & $178.5$ \\
		CGD & $-1.01_{-0.28}^{+0.26}$ & $18.21\scriptstyle\pm0.03$ & $179.5$ & $\phantom{0}7.9$ & $171.6$ \\
		\hline
    \end{tabular}
    \egroup
    \\ {\footnotesize $^*$This interval extends all the way down to $-1.5$, the lowest value of $\gamma$ we considered.}
\caption{Best-fit parameters and 68\% uncertainties for the various propagation models we used (see table \ref{tab:m1}). For the CTG model we report the two main local minima, whose total deviances differ by $2.2$. 
\protect\label{tab:m3}}
\end{table}
  \begin{table}[!t]
    \centering
    \begin{tabular}{l|llll|llll}
      model & $f_\mathrm{H}$ & $f_\mathrm{He}$ & $f_\mathrm{N}$ & $f_\mathrm{Si}$ & $\frac{\mathcal{L}_\mathrm{H}}{\mathcal{L}_0}$ & $\frac{\mathcal{L}_\mathrm{He}}{\mathcal{L}_0}$ & $\frac{\mathcal{L}_\mathrm{N}}{\mathcal{L}_0}$ & $\frac{\mathcal{L}_\mathrm{Si}}{\mathcal{L}_0}$ \\
      \hline
      SPG & $\phantom{0}0\%$ & $67\%$ & $28\%$ & $\phantom{0}5\%$ & $\phantom{0}0\%$ & $33\%$ & $50\%$ & $17\%$ \\
      STG & $\phantom{0}0\%$ & $\phantom{0}7\%$ & $85\%$ & $\phantom{0}8\%$ & $\phantom{0}0\%$ & $\phantom{0}1\%$ & $81\%$ & $17\%$ \\
      SPD & $63\%$ & $37\%$ & $\phantom{0}0.6\%$ & $\phantom{0}0.03\%$ & $\phantom{0}9\%$ & $45\%$ & $30\%$ & $15\%$ \\
      CTG ($\gamma = -1.03$) & $68\%$ & $31\%$ & $\phantom{0}1\%$ & $\phantom{0}0.06\%$ & $\phantom{0}7\%$ & $26\%$ & $50\%$ & $18\%$ \\
      CTG ($\gamma = +0.87$) & $\phantom{0}0\%$ & $\phantom{0}0\%$ & $88\%$ & $12\%$ & $\phantom{0}0\%$ & $\phantom{0}0\%$ & $77\%$ & $23\%$ \\
      CTD & $45\%$ & $52\%$ & $\phantom{0}3\%$ & $\phantom{0}0.06\%$ & $\phantom{0}1\%$ & $15\%$ & $70\%$ & $14\%$ \\
      CGD & $90\%$ & $\phantom{0}5\%$ & $\phantom{0}4\%$ & $\phantom{0}0.09\%$ & $\phantom{0}5\%$ & $\phantom{0}2\%$ & $79\%$ & $14\%$ \\
      \hline
    \end{tabular}
    \caption{Element fractions at injection (at $E_0=10^{18}$~eV and in terms of total emissivity) at the best fit for the various propagation models we used.  \protect\label{tab:m4}}
  \end{table}
\begin{figure}[t]
	\includegraphics[width=0.45\columnwidth]{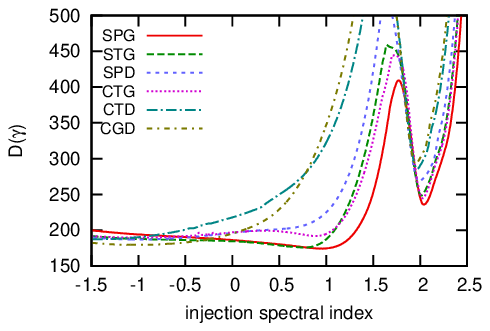}
	\includegraphics[width=0.45\columnwidth]{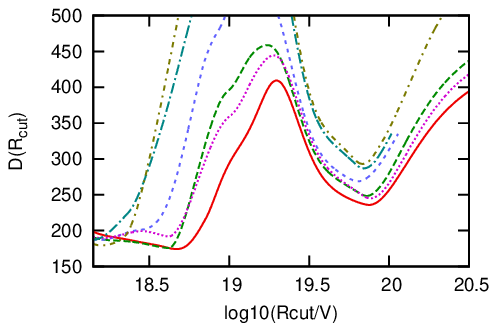}
        \caption{Deviance $D$ versus $\gamma$ (left) and $R_\text{cut}$ (right) along the valley line connecting ($\gamma, \log_{10}(R_\text{cut}/V)$) minima, in the propagation models considered.
}
	\label{fig:m7}
\end{figure}  \\
In figure \ref{fig:m7} we show the dependence of the deviance $D$ on $\gamma$ and $R_\text{cut}$ (with all the other parameters fixed at their best fit values) for the models we used in the fit.
From the parameters in tables \ref{tab:m3}, \ref{tab:m4} and the behaviours in figure \ref{fig:m7} some considerations can be drawn. Concerning EBL models it is clear that the Dom\'inguez model of EBL, having a stronger peak in the far infrared affects more the propagation and would result in too many low-energy secondary protons unless the cutoff rigidity is lowered, with a resulting negative spectral index and lighter composition. \\ 
The strength of interactions has a similar effect: PSB \cite{Puget:1976nz, Stecker:1998ib} cross sections (which altogether neglect $\alpha$ production) imply generally larger maximum rigidity than TALYS \cite{Koning20122841} cross sections (which largely overestimate $\alpha$ production), with a similar effect on spectral index and elemental fractions at injection.\\
As a consequence, the reference model exhibits the lowest deviance $D$ as a function of $\gamma$, while the models using Dom\'inguez EBL have a much less defined minimum in the region considered in the fits. \\
%
In the local minimum at $\gamma \approx 2$ the difference among models is greatly reduced; this reflects the fact that here interactions happen dominantly on CMB, which is  known to much higher precision than EBL 
and interaction lengths are so short that almost all nuclei fully photo-disintegrate regardless of the choice of cross sections. The larger  $D$ however in all cases  disfavours this minimum with respect to the best one. 

\subsubsection{Sensitivity of the fit to the source parameters}

\begin{figure}[!t]
	\centering
	\includegraphics[width=0.85\columnwidth]{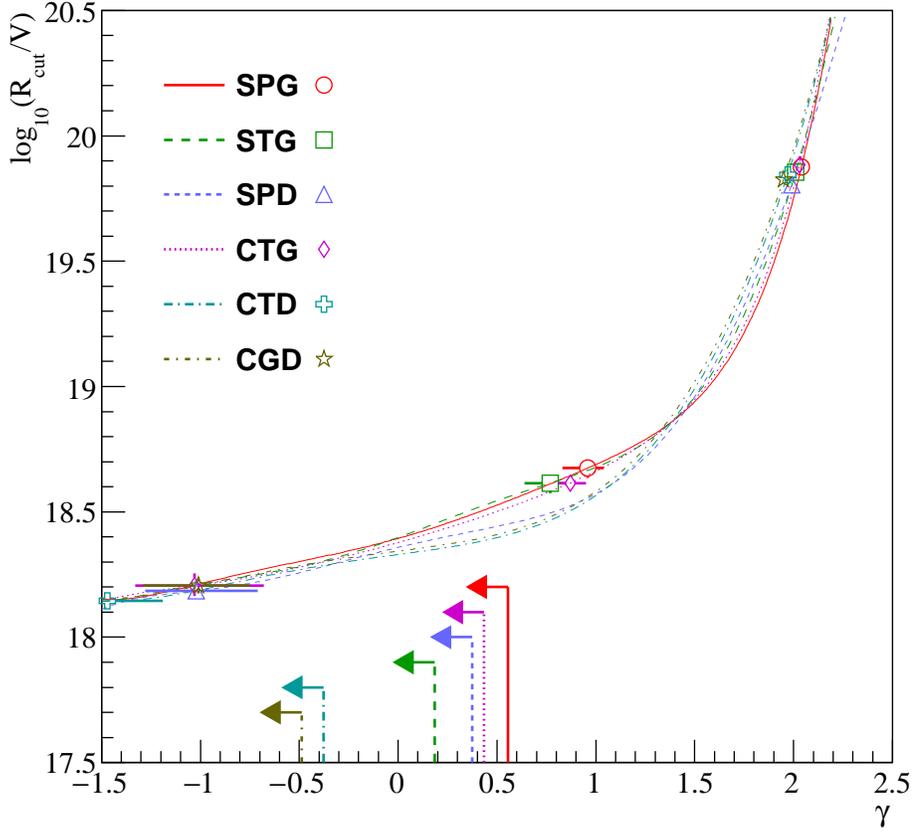}
        \caption{The lines connecting the local minima for the six
models given in table \ref{tab:m1}.
The lines and arrows at the bottom of the figure indicate the $\gamma$ regions where the light elements are dominant ($f_\text{H} + f_\text{He} > 90 \%$). Symbols indicate the position of the minima of each model. Both the best fit at $\gamma \lesssim 1$ and the second local minimum at $\gamma \approx 2$ are shown. For the CTG model both the $\gamma \lesssim 1$ minima reported in table \ref{tab:m3} are presented.}
    	\label{fig:leones}
\end{figure}

The fitting procedure has different sensitivity on the parameters that all together characterize the sources.
The spectral index $\gamma$ and the rigidity cutoff $\log_{10}(R_\text{cut}/\text{V})$ are well fitted in a wide region of astrophysical interest, whereas the fractions of injected nuclei are always poorly determined, mainly because of the sizable correlations among them, as shown in table \ref{tab:m2a} for the  SPG model.\footnote{This correlation is present in all propagation models, at least for lighter nuclei. For the CTG model, for instance, the correlation coefficient among H and He is $\approx -1.0$.}
Moreover, the detected observables, the all-particle spectrum and the longitudinal shower profiles, are either weakly dependent on the nuclei that reach the Earth or, as for the $X_\mathrm{max}$ distributions, depend logarithmically on their mass number. For this reason, different combinations of  injected nuclear species can  produce similar observables.
In figure \ref{fig:leones} we show the ($\gamma,~\log_{10}(R_\text{cut}/\text{V})$) region considered in our fits (tables \ref{tab:m1}, \ref{tab:m3}, \ref{tab:m4}). 
The valley lines of the fit, corresponding to the values of $R_\text{cut}$, $f_A$ and spectrum normalization that minimize $D$ for each value of $\gamma$, show a slightly increasing region of low spectral indexes below $\gamma \approx 0.5$, with a logarithmic rigidity cut between $18$ and $18.5$, and one with a steeply increasing spectral index region between $2$ and $2.5$, and corresponding large rigidity; these lines are a common feature of the models with small differences among them.\\
A noticeable fact is the variation of mean mass at injection along the valley lines. 
The arrows at the bottom of figure \ref{fig:leones} show for each propagation model the $\gamma$ region where the injection is dominated by light elements ($f_\text{H} + f_\text{He} > 90 \%$). 
This happens when the spectral index is below some value ranging from about $-0.5$ to $+0.5$ depending on the propagation model used (tables \ref{tab:m3}, \ref{tab:m4}).
\begin{figure}[!t]
	\centering
	\includegraphics[width=0.95\columnwidth]{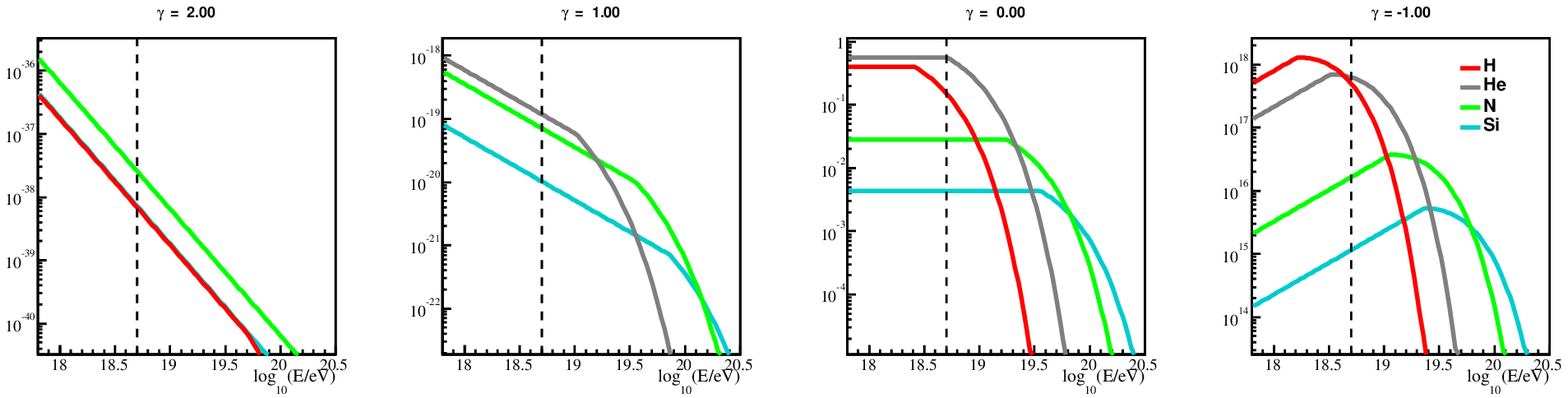}
	\includegraphics[width=0.95\columnwidth]{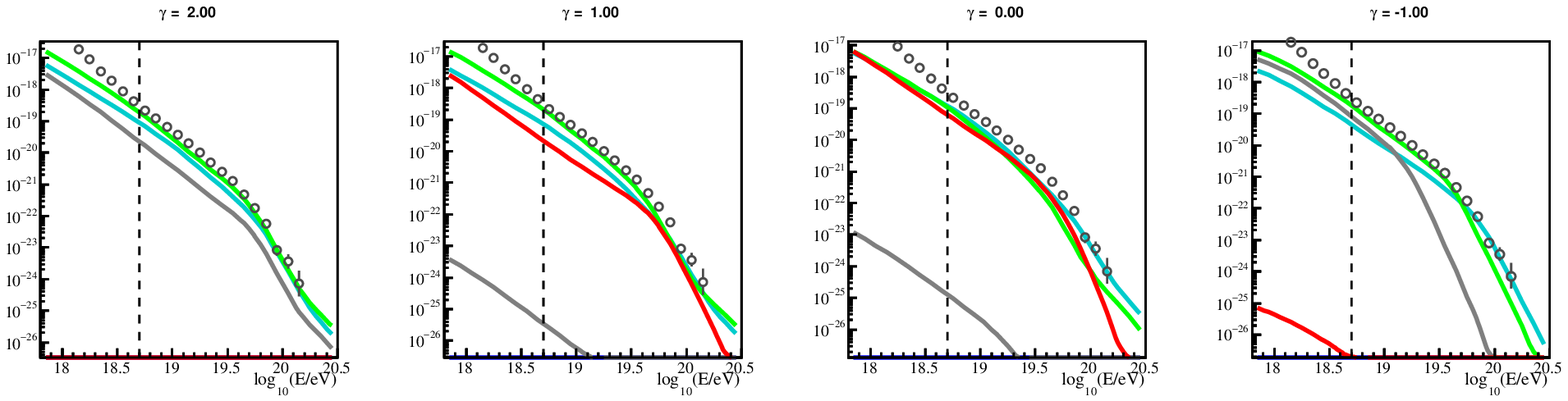}
        \caption{Top: injected spectra (in arbitrary units) as function of $\gamma$ along the valley line for the reference model. Bottom: spectra (in arbitrary units) at detection as function of $\gamma$ along the same line, compared with experimental points 
(open circles). The partial fluxes at detection represent the total propagated flux originating from a given primary nucleus grouping together all detected nuclei (primary and secondaries).}
	\label{fig:mla}
\end{figure}\\
%
In figure \ref{fig:mla} we show the injected spectra (top) as a function of the spectral index along the valley line and the corresponding fluxes at detection (bottom) for the SPG propagation model.
It is clear that for values of the spectral index sufficiently small the form of the overall observed spectrum loses almost every dependence on the injection spectrum of single elements; it is rather the tuning of elemental fractions that determines the final overall injection spectrum. 
In the negative $\gamma$ region fractions effectively substitute the spectral parameters to shape the overall flux: this is the reason why here the sensitivity to the spectral index becomes poor. \\
The  values of $R_\mathrm{cut}$ along the valley line for   $\gamma \le +0.5$ correspond to a propagation regime dominated by EBL photons, with energy loss lengths 
from hundreds of Mpc to Gpc (at the cutoff energy). 
The propagation with ($\gamma$, $R_\mathrm{cut}$) in this region depends strongly on the photo-disintegration cross sections and EBL parameterization, and, in negative-$\gamma$ region, the sensitivity to the propagation details becomes so extreme to make sub-dominant channels to play a major role~\cite{Batista:2015mea}.
This fact can explain the change of regime along the valley lines.\\

\section{Possible extensions of the basic fit \label{sec:improve}}
 \subsection{Homogeneity of source distribution and evolution \label{sec:hom}}
 \begin{table}[!t]
	\centering 
        \bgroup
        \def\arraystretch{1.3}
	\begin{tabular}{cc|cc|ccc}
		\multicolumn{2}{c|}{source evolution} & $\gamma$ & $\log_{10}(R_\text{cut}/\mathrm{V})$ & $D$ & $D(J)$ & $D(X_{\max})$ \\
		\hline
		\multirow{5}{*}{$(1+z)^m$} & $m=+3$ & $-1.40_{-0.09}^{+0.35}$ & $18.22_{-0.02}^{+0.05}$ & $179.1$ & $\phantom{0}7.5$ & $171.7$ \\
		& $m=0$ & $+0.96_{-0.13}^{+0.08}$ & $18.68_{-0.04}^{+0.02}$ & $174.3$ & $13.2$ & $161.1$ \\
& $m=-3$ & $+1.42_{-0.07}^{+0.06}$ & $18.85_{-0.07}^{+0.04}$  & $173.9$ & $19.3$ & $154.6$ \\
		& $m=-6$ & $+1.56_{-0.07}^{+0.06}$ & $18.74\scriptstyle\pm0.03$ & $182.4$ & $19.1$ & $163.3$ \\
		& $m=-12$ & $+1.79\scriptstyle\pm0.06$ & $18.73\scriptstyle\pm0.03$ & $182.1$ & $18.1$ & $164.0$ \\
		\hline
		&  {$z\le 0.02\phantom{0}$} & $+2.69\scriptstyle\pm0.01$ & $19.50_{-0.07}^{+0.08}$ & $178.6$ & $15.3$ & $163.3$ \\
                 \hline
     \end{tabular}\egroup
 \caption{Best fit parameters (reference model) corresponding to different assumptions on the evolution or spatial distribution of sources.     \protect\label{tab:i1}}
 \end{table}
As indicated in section \ref{sec:astro} we have assumed homogeneity (and isotropy) in the distribution of the sources. It is clear that nearby sources are not distributed homogeneously (nor isotropically). 
In \cite{Taylor:2011ta} the effect of nearby sources on the description of the data has been discussed, and, more recently \cite{Taylor:2015rla}, the effect of the evolution of the sources (with redshift). \\
Only particles originating from $z \lesssim 0.5$ are able to reach the Earth with $E > 10^{18.7}$~eV, so only the source evolution at small redshifts is relevant. At such small redshifts, most proposed parameterizations of the evolution of the emissivity (luminosity times density) of sources with redshift are of the form  $(1+z)^m$ \cite{Gelmini:2011kg}, although more detailed dependences are possible in selected cases.\\
In the simple case above, positive $m$ implies more luminous 
 and/or dense far sources, with increased importance of interactions on the photon backgrounds, and the contrary for negative evolution. 
To evaluate the possible effect on the fitted parameters, we have repeated the fit, using the reference SPG model, for several values of $m$, and (for $m=0$) assuming  a maximal source redshift at $z_\text{max}=0.02$, corresponding to a distance of $\approx 80$  Mpc.
The results of the fit are summarized in table \ref{tab:i1}.\\
The changing of the evolution has a strong effect on the spectral index: negative $m$  allow values of $\gamma$ nearer to the expected for the standard Fermi mechanism, and a corresponding slight increase of  $R_\text{cut}$. 
Limiting the distances has a similar effect.\footnote{Also propagation in extragalactic magnetic fields may induce similar changes \cite{Mollerach:2013dza} since in presence of a turbulent magnetic field the distance from which UHECRs reach detection is effectively limited.}
On the other hand, in the cases considered in the table, the deviance of the best solution does not change much, so we cannot conclude that these scenarios are required by the data.


 \subsection{The ankle and the need for an additional component \label{sec:proton}}
\begin{figure}[!t]
\centering
\includegraphics*[width=0.7\textwidth]{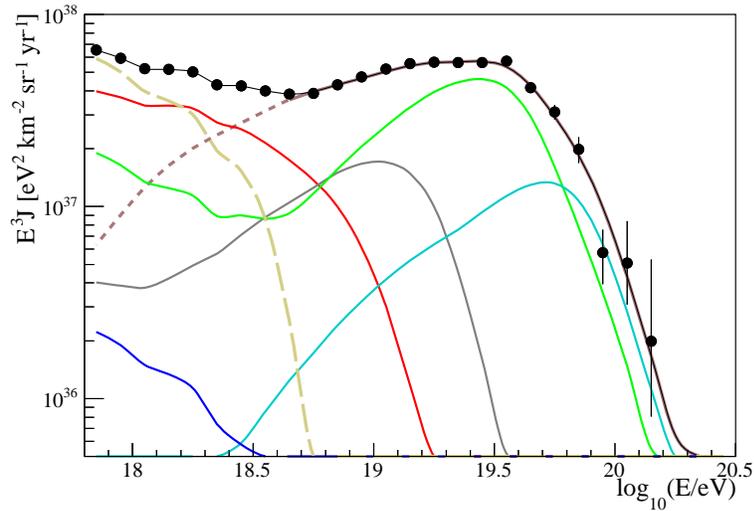}
  \caption{Simulated energy spectrum of UHECRs (multiplied by
$E^3$) at the top of the Earth's atmosphere, best-fit parameters for
model SPG, along with Auger data points. The dashed (yellowish) line shows the sub-ankle component obtained by subtracting from the experimental data the continuation of the all-particle spectrum (figure \ref{fig:m4}) below the fitted energy region. 
The composition below the ankle is derived from the mass fractions obtained in  \cite{Aab:2014aea} averaged in the same energy range (see text).}
%
\label{fig:addFermilab}
\end{figure}
\begin{figure}[!t]
\centering
\includegraphics*[width=0.7\textwidth]{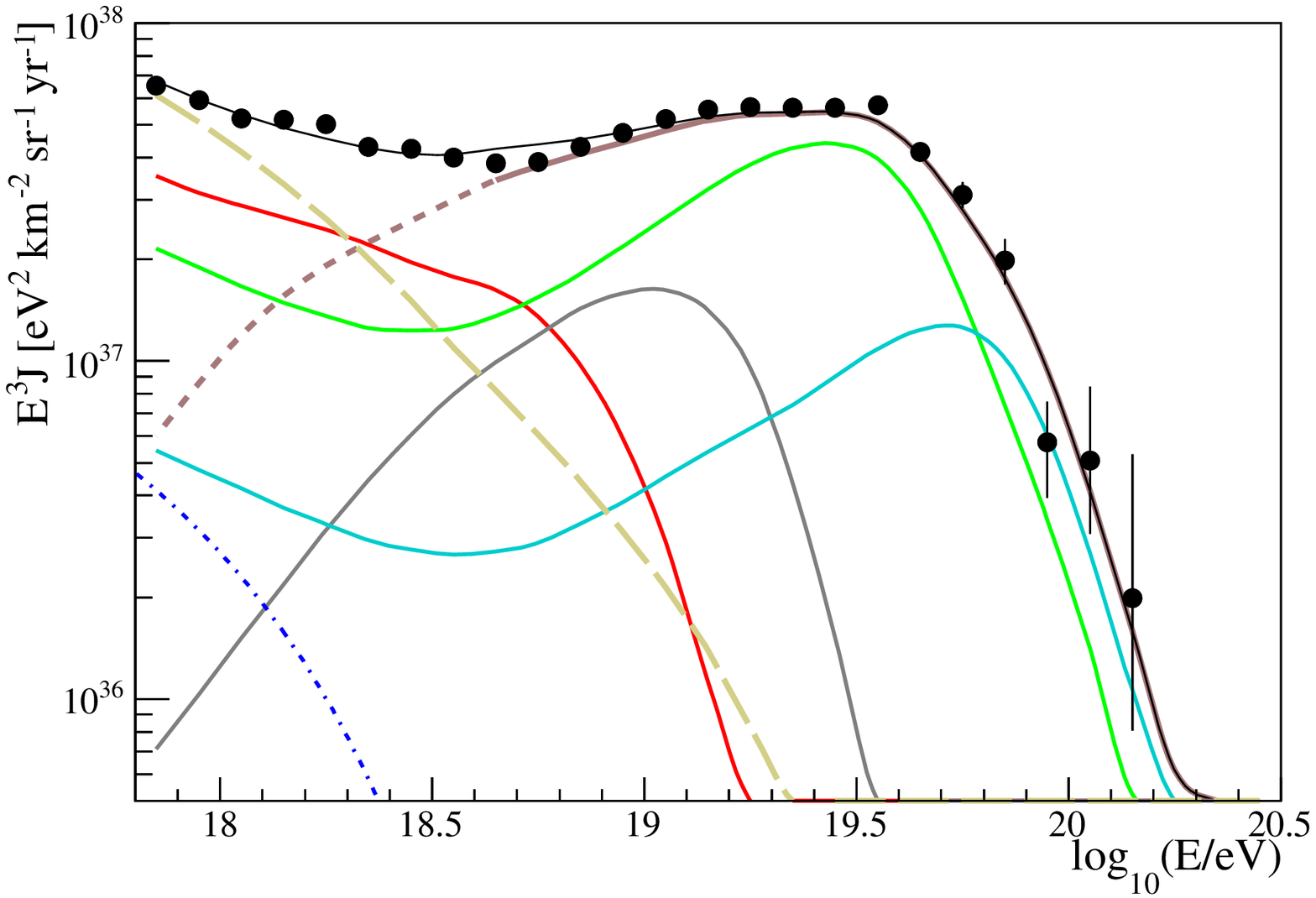}
    \includegraphics[width=0.9\textwidth]{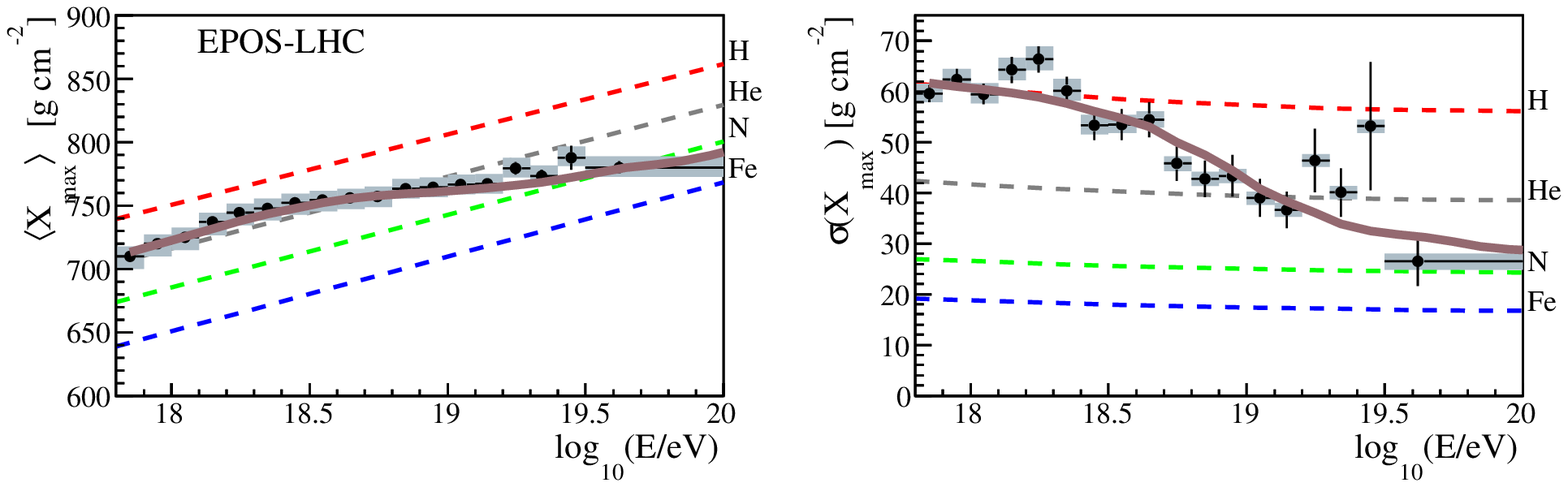}
  \caption{Top: 
simulated energy spectrum of UHECRs (multiplied by
$E^3$) at the top of the Earth's atmosphere, best-fit parameters for
model SPG, along with Auger data points. Partial spectra are grouped
as in figure \ref{fig:m5}. The dashed (yellowish) line shows the sub-ankle component
obtained as described in the text. The dot-dashed (blue) shows the 
KASCADE-Grande electron-poor flux, here assumed to be only iron.
Bottom: 
average and standard deviation  of the $X_{\max}$ distribution
as predicted (assuming EPOS-LHC UHECR-air interactions). Markers and colours as 
in figures \ref{fig:m5},\ref{fig:m4}.
}
\label{fig:addSub}
\end{figure}
%
In section \ref{sec:ref} we have shown the results of the combined fit of spectrum and composition data in the energy region above the ankle. 
This choice was motivated by the mixed nature of the measured composition and the impossibility to generate the ankle feature with the basic scenario outlined in section \ref{sec:astro}. 
As a consequence, we have implicitly assumed that the flux below the ankle has to be  explained as due to the superposition of additional component(s). 
This component can be originated by different sources and mechanisms. An exclusive  galactic origin  is difficult to accomodate, up to the ankle energy, in the standard paradigm of acceleration in SNRs.
Therefore extragalactic CR sources are expected to contribute to the low energy component that generate the ankle feature. These sources should reasonably  belong to a different class with respect to the one used in our fit above the ankle. 
The study of the flux and composition below the ankle  is beyond the scope of this paper and  shall be more effectively addressed using in the combined fit the data from the Auger detectors specially dedicated to low energy showers \cite{ThePierreAuger:2015rma}: the Infill 750 m-spacing array and HEAT (High Elevation Auger Telescopes) \cite{Valino:2015zdi,Porcelli:2015jli}.
Here we limit the discussion on the possible effects of a sub-ankle component on the solutions found in the section \ref{sec:ref}, taking as a reference the best fit solution for the reference propagation model.\\ 
For this purpose, we first obtain the  flux for $\log_{10}(E/\text{eV})<18.7$ by subtracting from the experimental  data the lower energy continuation of the all-particle flux fitted above this value. 
We assume below the ankle the elemental fractions in \cite{Aab:2014aea} where they are obtained from the same $X_\mathrm{max}$ distributions used in this work, independently in each energy bin; to reduce fluctuations from bin to bin we use their averages for $\log_{10}(E/\mathrm{eV}) \leq 18.6$ (for EPOS-LHC, 59\% H, 5.6\% He, 32\% N and 3.8\% Fe).
The resultant fluxes are presented in figure \ref{fig:addFermilab}.   
This approach by construction gives a description of the spectrum and composition at lower energy fully consistent with the data, but cannot give any indication on the nature of the sources. \\
In order to obtain more information, we then  fit  the sub-ankle flux obtained as described above, assuming UHECRs in this energy range are injected by a class of sources similar to those described in  section \ref{sec:astro} (although with different physical parameters).
To account for the possible presence of a sub-dominant component of the galactic flux, we assume an iron flux modeled as the one obtained by the KASCADE-Grande (KG) collaboration assuming EPOS as hadronic interaction model
\cite{Apel:2014uka}.\footnote {An exponential cutoff is applied to the KG flux with $E_\mathrm{cut} = 10^{18}$ eV, above the energy range of measured data.}
Under this assumption we find that a reasonable description of the sub-ankle component is obtained for a spectral index  $\gamma = 3.6$, a rigidity cutoff $\log_{10}(R_\text{cut}/\mathrm{V}) = 18.4$ and a mix of about $56 \%$ H, $35 \%$ N and $9\%$ Si. \\
The two components are summed each multiplied by adjustable weight factors to account for the superposition effects, and the spectrum errors below the ankle are increased adding in quadrature a 3\% offset to account for  possible related uncertainties. The limited interaction between the two components is reflected by the values, close to 1, of the two weight factors, 1.01 (0.97) for the sub-ankle (super-ankle) component.\\
The result of this procedure is shown in figure \ref{fig:addSub}. The sub-ankle
spectrum and composition merge with the fitted spectrum giving rise to a comprehensive description in the whole energy range. It has to be stressed that this does not come out from an overall minimization of a two-component source model and therefore it simply shows that there are possible sub-ankle components consistent with our  fitted spectrum and composition. \\
The result of this procedure gives a similar description to  that discussed above (figure \ref{fig:addFermilab}); moreover, the sub-ankle component used here is mixed as well as the one describing the super-ankle region, containing elements heavier than protons, as consequence of the $X_\mathrm{max}$ behaviour below the ankle, where the mean value is close to the one generated by pure He, but the dispersion close to that of H. 
%
 This is the main reason of the excess of simulated flux in the ankle
region, since the presence of nuclei heavier than protons does not allow us to 
reproduce a steep flux like the one obtained by subtraction. 
Although the procedure used does not allow us to draw firm conclusions,  it appears difficult that a population of sources with a rigidity dependent cutoff can reproduce a sharp ankle as in the Auger data.\\
As already stated the approach followed here is partial and cannot provide a full description of the data from the lowest energies; however it suggests that a description of the sub-ankle data does not necessarily spoil the main features of the fit as discussed in section \ref{sec:ref}.\\

\section{Discussion\label{sec:discussion}}
We have presented in this paper a fit of the experimental measurements (spectrum and mass composition) performed by the Pierre Auger Observatory at UHECRs energies above the ankle, assuming an extragalactic origin.
Although the best fit obtained depends to considerable extent on the models used for propagation in the extragalactic space and interactions in the atmosphere, we have found some general features characterizing the parameters of the astrophysical model chosen ($\gamma$, $R_\text{cut}$, the elemental fractions $f_A$ and total emissivity $\mathcal{L}_0$). \\
Referring to figures \ref{fig:m4} and \ref{fig:leones}, it is evident that the best fit solutions present a marked correlation between $\gamma$ and $\log_{10}(R_\text{cut}/\text{V})$, and two local minima regions:
\begin{itemize}
\item  An elongated region at $R_\text{cut} \lessapprox 5 \times 10^{18} $ V, $\gamma \lessapprox 1$, where the best minimum falls. In this region both the spectrum and the $X_\text{max}$ data are reproduced reasonably well, but the precise location of the best fit strongly depends on the propagation model (i.e.~Monte Carlo code, EBL spectrum, photo-disintegration cross sections and air interaction models).
\item A smaller region at $R_\text{cut} \approxeq 7 \times 10^{19}$ V, $\gamma \approxeq 2$, where the spectrum is well reproduced but there are too many high-energy protons at strong variance with the  $X_\text{max}$ data, while the position of the local minimum does not vary much among the various propagation models.
\end{itemize}
For large values of the maximum energy at the source, $Z R_\text{cut}$, the observed drop of the spectrum is a consequence of interactions during propagation in the background radiation. 
However, the copious secondary production implies a very mixed composition at odds with observations.  In this region interactions occur predominantly on the CMB, and almost all nuclei fully photo-disintegrate into nucleons, which explains the little dependence on details of propagation.\\
Decreasing  $R_\text{cut}$, the propagated  fluxes  start to show  the effect of the cutoff at the sources with the consequence that the maximum energy of secondary protons is pushed to low values, which in turn 
produces a less mixed composition in better agreement with data. 
In this region the observed spectrum starts to be reproduced by the envelope of hard elemental fluxes ($\gamma \approx 1$), cut by a decrease that is caused by both the source cutoff (for the secondary nucleons) and the photo-disintegration (for the surviving primary medium and heavy nuclei). This is the region of parameters in which the best fit of the reference case resides. Since the cutoff rigidity corresponds to an energy per nucleon way below the threshold for pion production on the CMB, the resulting flux of cosmogenic neutrinos at EeV~energies is negligible. Also, particles with magnetic rigidity~$E/Z \lesssim 5$~EV can be deflected by intergalactic and galactic magnetic fields by several tens of degrees\footnote{Indeed, the conclusion that the highest-energy CRs include many light and medium-mass nuclei but few
protons was independently reached by other authors \cite{Fargion:2008sp,Fargion:2009rb,Fargion:2009ki,Fargion:2014jma,Fargion:2016cep} from
the observation of a few excesses in the angular distribution of
UHECR arrival directions in regions of $\sim 20^{\circ}$~radius and the lack
of excesses on smaller scales.} even when originating from relatively nearby sources~\cite{Smida:2015kga}, making it very hard to infer source positions. \\
At even lower values of $R_\text{cut}$ interactions on EBL begin to dominate, and are in any case relatively weak. Primary Hydrogen and Helium become then dominant in order to reproduce composition data, and the observed spectra are the product of fine tuning of the elemental fluxes at injection. \\
This interplay between astrophysical source properties and effects of propagation then explains the general trend observed in section \ref{sec:ref}: copious interactions, both depending on the choice of background and of cross sections require a small $R_\text{cut}$ and possibly negative $\gamma$ in a very flat minima region. This partly explains why the position of the best fit for low $\gamma$ is so strongly model-dependent and why the models with lower best-fit values of~$\gamma$ tend to have larger uncertainty intervals on it (see figure \ref{fig:mla}). \\
The use of the hadronic interaction models Sibyll 2.1 or QGSJet II-04 in place of EPOS-LHC worsens the fit, pushing the best fit to the lowest considered values of spectral index and requiring lighter mass compositions. This is because the widths of measured $X_\text{max}$ distributions, which depend both on shower-to-shower fluctuations and the amount of superposition of different masses, are relatively narrow. Sibyll 2.1 and QGSJet II-04 predict very broad $X_\text{max}$ distributions which are hard to reconcile with Auger data even assuming a pure mass composition. Therefore it is not surprising that in this situation the fit seeks to keep the propagated energy spectra for individual mass groups as separated as possible, corresponding to negative $\gamma$, and even then cannot reasonably reproduce Auger data.\\
The fit results are also somewhat dependent on the Auger energy and $X_\text{max}$ scales, on which there are sizeable systematic uncertainties. The effect of shifting the energy scale is compensated by a change of composition at the sources, with very small effects in the other parameters or the overall goodness of fit. On the other hand, shifting the $X_\text{max}$  distributions downwards by their systematic uncertainty requires a somewhat higher spectral index and heavier composition and improves the fit;  shifting them upwards requires a much lower spectral index and lighter composition and worsens the fit. \\
Some departures from the simple astrophysical model used are considered in section \ref{sec:hom}, where we discuss the effect of modifications of the hypothesis of constant emissivity of the sources. We find that an evolution of source emissivity $\propto (1+z)^m$ with $m>0$ would make the fit worse due to the increased level of interactions that produce abundant secondary protons, whereas with $m<0$ the goodness of fit is not affected much and a higher injection spectral index is required. Limiting the source distance has a similar effect.
It has been noted that diffusion in extragalactic magnetic fields \cite{Mollerach:2013dza} may effectively limit the distance from which particles reach detection, and produce softer injection spectra: to better evaluate the importance of this mechanism, three-dimensional propagation simulations are needed.\\ 
Finally, in section \ref{sec:proton} we have discussed how the results obtained fitting spectrum and composition above the ankle can be affected by the sub-ankle flux. An additional component of extragalactic nuclei, mostly H and N,  with a generation spectrum much steeper than the one obtained by the fit above the ankle can be introduced to provide a reasonable description of the data in the whole energy range. The new component appears not to interfere much with the general picture discussed above. However it appears difficult to reproduce a sharp ankle as in the Auger data once a rigidity cutoff is assumed for the sub-ankle component. One possible way out is to assume from the beginning a model generating the ankle feature as a consequence of interactions in the source photon environment, such as  for example \cite{Unger:2015laa,Globus:2015xga} and compare with experimental data: we did not follow this strategy in this paper. 
\section{Conclusions}
In this paper we have shown that, within given hypotheses on propagation and interaction at Earth, Auger data can bind the physical parameters of the sources in the simple astrophysical model considered. However several different hypotheses (i.e.\ atmospheric interaction and  EBL models, choices of photo-disintegration cross sections) can be made with resulting source parameters well outside the statistical uncertainties of the fit. Better models of UHECR-air hadronic interactions, EBL spectrum and evolution, or
photo-disintegration cross sections and branching ratios would help reduce
these uncertainties.\\
The results obtained show some sensitivity to experimental systematics, in particular to that on $X_\text{max}$. About that, the new operation of Auger (AugerPrime) \cite{arXiv:1604.03637} will produce more composition sensitive observables,  in particular connected to muons in showers, in an extended energy range, including the highest energies, and with reduced systematics. 
Also the planned extended FD operation will increase the statistics of shower development measurements at the highest energies.\\

\appendix
\section{Generation of simulated propagated spectra \label{app-a}}
In order to compute the simulated spectrum that the measured data points are compared to, we use either SimProp \cite{Aloisio:2012wj,Aloisio:2015sga,Aloisio:2016tqp} or CRPropa  \cite{Armengaud:2006fx,Batista:2014xza,Batista:2016yrx} simulations.
Both SimProp and CRPropa runs simulate events with a uniform distribution of $\log_{10}(E_\text{inj}/\mathrm{eV})$, but the injection points are uniform in $z_\text{inj}$ in SimProp and in $t_\text{inj}$ in CRPropa. A uniform distribution of sources per unit comoving volume corresponds to a uniform distribution of injection times for the events arriving at Earth (if counting as one event the arrival of all the particles originating from the same primary), so in the case of SimProp each event is weighed by a factor $w(z_\text{inj}) \propto \left.\mathrm{d}t/\mathrm{d}z\right|_{z={z_\text{inj}}}$, whereas no such weighing is required for CRPropa events.\\
When using SimProp, we used seven redshift intervals $[0, 0.01)$, $[0.01, 0.05)$, $[0.05, 0.10)$, $[0.10, 0.20)$, $[0.20, 0.30)$, $[0.30, 0.50)$, and $[0.50, 2.50)$, simulating $5 \cdot 10^5$ events with\linebreak $\log_{10}(E_\text{inj}/\mathrm{eV})$ from 17.5 to 22.5 for each primary mass and each redshift interval, for a total of $3.5 \cdot 10^6$ events per primary mass. (Each redshift interval is weighed by its width.) When using CRPropa, we simulated $4 \cdot 10^6$ primary H events, $2 \cdot 10^6$ primary He events, $8 \cdot 10^6$ primary N events, $2 \cdot 10^6$ primary Si events, and $4 \cdot 10^6$ primary Fe events, with $\log_{10}(E_\text{inj}/\mathrm{eV})$ from 17.5 to $21.5+\log_{10}Z_\text{inj}$ and $t_\text{inj}$ from $t(z=1)$ to the present time $t(z=0)$. We verified that these numbers of events result in squared statistical uncertainties on the simulations less than $10\%$ of those on the Auger data in each of the energy bin of the fit.
We then bin all the particles arriving at Earth in bins of both $\log_{10}(E_\text{inj}/\mathrm{eV})$ and $\log_{10}(E_\text{Earth}/\mathrm{eV})$ of width 0.01, obtaining a four-dimensional matrix giving the average number of nuclei arriving at Earth with a given mass number $A_\text{Earth}$ in a given $E_\text{Earth}$ bin for each primary injected with a given mass number $A_\text{inj}$ in a given $E_\text{inj}$ bin. This matrix can be multiplied by a vector representing the injection spectrum to obtain a vector representing the true spectrum at Earth, or also by an analogous matrix representing the detector properties (see appendix B) to obtain the folded spectrum at Earth, which when multiplied by the detector exposure and integrated over the energy bins results in the expected number of events $\mu_m$ which enters the deviance function.

\section{Treatment of detector effects \label{app}}
The astrophysical models, combined with propagation in extragalactic photon backgrounds, predict elemental fluxes at the top of the atmosphere. The signal generated on the detectors, after interactions in the atmosphere, is then reconstructed in terms of physical observables and, to do so, experimental uncertainty and biases have to be taken into account. For each generated true flux 
$ J(E_\text{true})$ (which depends on source parameters) we have a corresponding reconstructed (folded) flux
\begin{equation}
    J_\text{fold}(E_\text{rec}) = \int_{0}^{+\infty} p(E_\text{rec}|E_\text{true}) J(E_\text{true}) \,\mathrm{d}E_\text{true}
\end{equation}
    where 
\begin{equation} p(E_\text{rec}|E_\text{true})=\mathcal{T}\operatorname{Gauss}\left(E_\text{rec}|bE_\text{true},\frac{\sigma_E}{E}E_\text{true}\right)
\end{equation}
where $\mathcal{T}$ is the trigger efficiency, $b$ the energy bias, $\frac{\sigma_E}{E}$ is the SD energy resolution, which are all functions of $E_\text{true}$.
In terms of these function the expected counts are: 
\begin{equation}
\mu_m = \int_{\text{bin}~m}  \mathcal{E}J_\text{fold}(E_\text{rec})\,\mathrm{d}E_\text{rec}
\end{equation}
 where $\mathcal{E}$ is the exposure of the surface detector(s).
$\mathcal{T}$ , $b$, ${\sigma_E}$ are obtained through detailed detector simulations or from the data themselves (see below).\\
For the vertical SD spectrum, we used:
\begin{align}
\sigma_E/E &= \sqrt{\sigma_\text{det}^2 + \sigma_\text{sh}^2}; &
\sigma_\text{det} &= A + \frac{B}{\sqrt{E_\text{true}}} + \frac{C}{E_\text{true}}; &
\sigma_\text{sh} &= p_0 +p_1 y
\label{eq:reso}
\end{align}
where $A = 1.3\times 10^{-3}$, $B = 0.18~\mathrm{EeV}^{1/2}$, $C = 0.052~\mathrm{EeV}$ and $y = \log_{10}(E_\text{true}/\mathrm{EeV})$, $p_0 =
0.154$, $p_1=-0.030$.
In the range of energies we are considering $\mathcal{T}=1$ and the energy bias $b$ is (in this energy range there is no zenith angle dependence for an isotropic UHECR distribution):
\begin{align}
b = 1 + P_0 + P_1 y + P_2 y^2;~~
\end{align}
with $P_0 = 0.0566$, $P_1 = -0.0720$, $P_2 =0.0227$.\\
%
For the inclined spectrum \cite{ThePierreAuger:2015rha} we used:
\begin{align}
{\sigma_E}/{E} &= \sqrt{\sigma_\text{det}^2 + \sigma_\text{sh}^2}/c_1; &
\sigma_\text{det} &= p_0 + \frac{p_1}{\sqrt{c_0 (E/10~\mathrm{EeV})^{c_1}}}; &
\sigma_\text{sh} &= 0.143,
\end{align}
where $p_0 = 0.03896$, $p_1 = 0.1128$ and $c_0 = 1.746$, $c_1 = 0.9938$, and $\mathcal{T} = b = 1$ at all energies.
\\
It is worthwhile noting that the model fit gives a flux that is lower than the observed flux below the energy threshold of the fit. As a consequence, the migration of events from low energies into the fitting range is underestimated by $\lesssim 3\%$ in the first energy bin%
\footnote{In the best-fit model, 34\% of the events with $E_\text{rec}/\text{eV}$ in $[10^{18.7}, 10^{18.8})$ have $E_\text{true}/\text{eV}$ in $[10^{18.6}, 10^{18.7})$, and 5\% have $E_\text{true}/\text{eV}$ in $[10^{18.5}, 10^{18.6})$. Compared to the data in ref.~\cite{Valino:2015zdi}, this model underestimates the total flux with $E_\text{true}/\text{eV} \in [10^{18.6}, 10^{18.7})$ by 5\% and that in $[10^{18.5}, 10^{18.6})$ by 25\%. }
 and negligible at higher energies.\\
In each energy bin, we only fit the total number of observed SD events~$n_m = n_m^\text{v} + n_m^\text{i}$ to the total
model prediction~$\mu_m = \mu_m^\text{v} + \mu_m^\text{i}$ rather than the vertical and inclined counts separately.
It can be shown in the latter case that the total deviance
\begin{equation}
 D^\text{v} + D^\text{i} = -2\sum_m \left( \mu_m^\text{v} - n_m^\text{v} + n_m^\text{v} \ln \frac{n_m^\text{v}}{\mu_m^\text{v}} + \mu_m^\text{i} - n_m^\text{i} + n_m^\text{i} \ln \frac{n_m^\text{i}}{\mu_m^\text{i}} \right) 
\label{eq:sumdev}
\end{equation}
would be equal to that computed from the summed spectra (\ref{eq:devsum}) plus a term given by
\begin{equation}
D^\text{rel} = -2\sum_m \left(n_m^\text{v} \ln \left(\frac{n_m}{n_m^\text{v}} \frac{\mu_m^\text{v}}{\mu_m} \right) + n_m^\text{i} \ln \left(\frac{n_m}{n_m^\text{i}} \frac{\mu_m^\text{i}}{\mu_m} \right) \right),
\end{equation}
which quantifies 
possible differences between the two observed spectra and only depends on the model predictions through the ratios ${\mu_m^\text{v}}/{\mu_m}, {\mu_m^\text{i}}/{\mu_m}$. Since ${\mu_m^\text{v}}/{\mu_m}$ (${\mu_m^\text{i}}/{\mu_m}$) is almost\footnote{Except for very small effects due to the two datasets having different $p(E_\text{rec}|E_\text{true})$ functions.} equal to the ratio of the vertical (inclined) exposure to the total SD exposure, $D^\text{rel}$ does not depend on the astrophysical model but only on the data, and including or excluding it makes no difference on the best-fit parameter values or their uncertainty intervals. 
We chose to exclude it from the values of~$D(J)$ we mention in this work (by using Eq. \ref{eq:devsum} rather than \ref{eq:sumdev}) because any 
difference between the observed SD vertical and inclined spectra cannot be due to the astrophysical models.
%
%
%
For what concerns the composition, we apply to the Gumbel parameterization of $X_\text{max}$ (see section \ref{sec:fitting}), the parameterization for the resolution, $\mathcal{R}$, and 
acceptance, $\mathcal{A}$, as given in \cite{Aab:2014kda}, in order to account for the detector response:
\begin{equation}
g(X_\mathrm{max}^\mathrm{rec} | E,A) = (g(X_\mathrm{max} | E,A) \cdot 
\mathcal{A}(X_\mathrm{max} | E)) \otimes \mathcal{R}(X_\mathrm{max}^\mathrm{rec} |X_\mathrm{max},E)
\label{eq:GumbelRec}
\end{equation}
The model probability $G_m(X_\mathrm{max}^\mathrm{rec})$, evaluated at the logarithmic energy bin centre $m$, for a given mass distribution at detection $\{p_A\}$ is then given by: 
\begin{equation}
G_{m}^\text{model}(X_\mathrm{max}^\mathrm{rec}) = \sum_A p_A \cdot g(X_\mathrm{max}^\mathrm{rec} | E_m ,A) 
\label{eq:GumbelMix}
\end{equation}
Finally, since the energy resolution of the Fluorescence Detector is narrower than the width of the energy bins we use, its effect has been neglected in this analysis.
\\
Not using the forward-folding procedure and directly fitting the values of $J$,
$\langle X_{\max} \rangle$ and $\sigma(X_{\max})$ presented in refs.~\cite{Valino:2015zdi, Aab:2014kda}
assuming a Gaussian likelihood (so that the deviance to be minimized is the $\chi^2$~statistic)
would yield qualitatively similar results, but with slightly lower best-fit values for $\gamma$
and $R_\text{cut}$ and somewhat larger uncertainty intervals
($\gamma = 0.68_{-0.17}^{+0.12}$,
$\log_{10}(R_\text{cut}/\mathrm{V}) = 18.59_{-0.04}^{+0.03}$ in the reference
model).

\subsection{Effect of the choice of the SD energy resolution}
The energy resolution of the SD, required for the forward-folding
procedure, can be estimated either from shower and detector simulations
(but with possibly strongly model-dependent results) or directly from
the calibration data (but with large statistical uncertainties
especially at the highest energies). Throughout this work we used the
data-based parameterization of eq.~\eqref{eq:reso}, but in
order to assess how sensitive our fit is to this choice, we also tried
using the QGSJet~II-03 simulation-based parameterization
\begin{equation}
  \frac{\sigma_E} {E} = 0.109 + 0.435\times 10^{-17}
  \left(\frac{E}{\mathrm{eV}}\right)^{-1/2}
\end{equation}
for the vertical SD resolution, which exceeds the data-based one by about twice the statistical standard deviation of the latter. (We did not change the inclined SD
resolution.) We found the effects of this to be negligible, with the
best fit at the same $(\gamma, \log_{10}R_\text{cut})$ pair to within
our grid spacing (but with slightly narrower lower uncertainty
intervals, $0.96_{-0.11}^{+0.08}$ and $18.68_{-0.03}^{+0.02}$), with
the same mass fractions to within 0.4\%, and with the same total
deviance to within $3\times 10^{-3}$ (though with higher spectrum
deviance and lower $X_{\max}$ deviance by about 0.5).


\section*{Acknowledgements}

\begin{sloppypar}
The successful installation, commissioning, and operation of the Pierre Auger Observatory would not have been possible without the strong commitment and effort from the technical and administrative staff in Malarg\"ue. We are very grateful to the following agencies and organizations for financial support:
\end{sloppypar}

\begin{sloppypar}
Argentina -- Comisi\'on Nacional de Energ\'\i{}a At\'omica; Agencia Nacional de Promoci\'on Cient\'\i{}fica y Tecnol\'ogica (ANPCyT); Consejo Nacional de Investigaciones Cient\'\i{}ficas y T\'ecnicas (CONICET); Gobierno de la Provincia de Mendoza; Municipalidad de Malarg\"ue; NDM Holdings and Valle Las Le\~nas; in gratitude for their continuing cooperation over land access; Australia -- the Australian Research Council; Brazil -- Conselho Nacional de Desenvolvimento Cient\'\i{}fico e Tecnol\'ogico (CNPq); Financiadora de Estudos e Projetos (FINEP); Funda\c{c}\~ao de Amparo \`a Pesquisa do Estado de Rio de Janeiro (FAPERJ); S\~ao Paulo Research Foundation (FAPESP) Grants No.\ 2010/07359-6 and No.\ 1999/05404-3; Minist\'erio de Ci\^encia e Tecnologia (MCT); Czech Republic -- Grant No.\ MSMT CR LG15014, LO1305 and LM2015038 and the Czech Science Foundation Grant No.\ 14-17501S; France -- Centre de Calcul IN2P3/CNRS; Centre National de la Recherche Scientifique (CNRS); Conseil R\'egional Ile-de-France; D\'epartement Physique Nucl\'eaire et Corpusculaire (PNC-IN2P3/CNRS); D\'epartement Sciences de l'Univers (SDU-INSU/CNRS); Institut Lagrange de Paris (ILP) Grant No.\ LABEX ANR-10-LABX-63 within the Investissements d'Avenir Programme Grant No.\ ANR-11-IDEX-0004-02; Germany -- Bundesministerium f\"ur Bildung und Forschung (BMBF); Deutsche Forschungsgemeinschaft (DFG); Finanzministerium Baden-W\"urttemberg; Helmholtz Alliance for Astroparticle Physics (HAP); Helmholtz-Gemeinschaft Deutscher Forschungszentren (HGF); Ministerium f\"ur Innovation, Wissenschaft und Forschung des Landes Nordrhein-Westfalen; Ministerium f\"ur Wissenschaft, Forschung und Kunst des Landes Baden-W\"urttemberg; Italy -- Istituto Nazionale di Fisica Nucleare (INFN); Istituto Nazionale di Astrofisica (INAF); Ministero dell'Istruzione, dell'Universit\'a e della Ricerca (MIUR); CETEMPS Center of Excellence; Ministero degli Affari Esteri (MAE); Mexico -- Consejo Nacional de Ciencia y Tecnolog\'\i{}a (CONACYT) No.\ 167733; Universidad Nacional Aut\'onoma de M\'exico (UNAM); PAPIIT DGAPA-UNAM; The Netherlands -- Ministerie van Onderwijs, Cultuur en Wetenschap; Nederlandse Organisatie voor Wetenschappelijk Onderzoek (NWO); Stichting voor Fundamenteel Onderzoek der Materie (FOM); Poland -- National Centre for Research and Development, Grants No.\ ERA-NET-ASPERA/01/11 and No.\ ERA-NET-ASPERA/02/11; National Science Centre, Grants No.\ 2013/08/M/ST9/00322, No.\ 2013/08/M/ST9/00728 and No.\ HARMONIA 5 -- 2013/10/M/ST9/00062; Portugal -- Portuguese national funds and FEDER funds within Programa Operacional Factores de Competitividade through Funda\c{c}\~ao para a Ci\^encia e a Tecnologia (COMPETE); Romania -- Romanian Authority for Scientific Research ANCS; CNDI-UEFISCDI partnership projects Grants No.\ 20/2012 and No.194/2012 and PN 16 42 01 02; Slovenia -- Slovenian Research Agency; Spain -- Comunidad de Madrid; Fondo Europeo de Desarrollo Regional (FEDER) funds; Ministerio de Econom\'\i{}a y Competitividad; Xunta de Galicia; European Community 7th Framework Program Grant No.\ FP7-PEOPLE-2012-IEF-328826; USA -- Department of Energy, Contracts No.\ DE-AC02-07CH11359, No.\ DE-FR02-04ER41300, No.\ DE-FG02-99ER41107 and No.\ DE-SC0011689; National Science Foundation, Grant No.\ 0450696; The Grainger Foundation; Marie Curie-IRSES/EPLANET; European Particle Physics Latin American Network; European Union 7th Framework Program, Grant No.\ PIRSES-2009-GA-246806; European Union's Horizon 2020 research and innovation programme (Grant No. 646623);
and UNESCO.
\end{sloppypar}

\bibliography{FALmain}

\providecommand{\href}[2]{#2}\begingroup\raggedright\begin{thebibliography}{10}

\bibitem{Auger:2012an}
{\scshape Pierre Auger} collaboration, P.~Abreu et~al., \emph{{Large scale
  distribution of arrival directions of cosmic rays detected above $10^{18}$ eV
  at the Pierre Auger Observatory}},
  \href{http://dx.doi.org/10.1088/0067-0049/203/2/34}{\emph{Astrophys. J.
  Suppl.} {\bf 203} (2012) 34}, [\href{https://arxiv.org/abs/1210.3736}{{\tt
  1210.3736}}].

\bibitem{ThePierreAuger:2014nja}
{\scshape Pierre Auger} collaboration, A.~Aab et~al., \emph{{Large Scale
  Distribution of Ultra High Energy Cosmic Rays Detected at the Pierre Auger
  Observatory With Zenith Angles up to 80$^{\circ}$}},
  \href{http://dx.doi.org/10.1088/0004-637X/802/2/111}{\emph{Astrophys. J.}
  {\bf 802} (2015) 111}, [\href{https://arxiv.org/abs/1411.6953}{{\tt
  1411.6953}}].

\bibitem{ThePierreAuger:2015rma}
{\scshape Pierre Auger} collaboration, A.~Aab et~al., \emph{{The Pierre Auger
  Cosmic Ray Observatory}},
  \href{http://dx.doi.org/10.1016/j.nima.2015.06.058}{\emph{Nucl. Instrum.
  Meth.} {\bf A798} (2015) 172--213},
  [\href{https://arxiv.org/abs/1502.01323}{{\tt 1502.01323}}].

\bibitem{Valino:2015zdi}
{\scshape Pierre Auger} collaboration, I.~Valino, \emph{{The flux of ultra-high
  energy cosmic rays after ten years of operation of the Pierre Auger
  Observatory}}, {\emph{PoS} {\bf ICRC2015} (2016) 271},
  [\href{https://arxiv.org/abs/1509.03732}{{\tt 1509.03732}}].

\bibitem{Aab:2014kda}
{\scshape Pierre Auger} collaboration, A.~Aab et~al., \emph{{Depth of maximum
  of air-shower profiles at the Pierre Auger Observatory. I. Measurements at
  energies above $10^{17.8}$  eV}},
  \href{http://dx.doi.org/10.1103/PhysRevD.90.122005}{\emph{Phys.Rev.} {\bf
  D90} (2014) 122005}, [\href{https://arxiv.org/abs/1409.4809}{{\tt
  1409.4809}}].

\bibitem{TheTelescopeArray:2015mgw}
{\scshape Telescope Array} collaboration, R.~U. Abbasi et~al., \emph{{The
  energy spectrum of cosmic rays above 10$^{17.2}$ eV measured by the
  fluorescence detectors of the Telescope Array experiment in seven years}},
  \href{http://dx.doi.org/10.1016/j.astropartphys.2016.04.002}{\emph{Astropart.
  Phys.} {\bf 80} (2016) 131--140},
  [\href{https://arxiv.org/abs/1511.07510}{{\tt 1511.07510}}].

\bibitem{Greisen:1966jv}
K.~Greisen, \emph{{End to the cosmic ray spectrum?}},
  \href{http://dx.doi.org/10.1103/PhysRevLett.16.748}{\emph{Phys.Rev.Lett.}
  {\bf 16} (1966) 748--750}.

\bibitem{Zatsepin:1966jv}
G.~Zatsepin and V.~Kuzmin, \emph{{Upper limit of the spectrum of cosmic rays}},
  {\emph{JETP Lett.} {\bf 4} (1966) 78--80}.

\bibitem{Aab:2014aea}
{\scshape Pierre Auger} collaboration, A.~Aab et~al., \emph{{Depth of maximum
  of air-shower profiles at the Pierre Auger Observatory. II. Composition
  implications}},
  \href{http://dx.doi.org/10.1103/PhysRevD.90.122006}{\emph{Phys.Rev.} {\bf
  D90} (2014) 122006}, [\href{https://arxiv.org/abs/1409.5083}{{\tt
  1409.5083}}].

\bibitem{Abreu:2013env}
{\scshape Pierre Auger} collaboration, P.~Abreu et~al., \emph{{Interpretation
  of the Depths of Maximum of Extensive Air Showers Measured by the Pierre
  Auger Observatory}},
  \href{http://dx.doi.org/10.1088/1475-7516/2013/02/026}{\emph{JCAP} {\bf 1302}
  (2013) 026}, [\href{https://arxiv.org/abs/1301.6637}{{\tt 1301.6637}}].

\bibitem{Abbasi:2014sfa}
R.~U. Abbasi et~al., \emph{{Study of Ultra-High Energy Cosmic Ray composition
  using Telescope Array’s Middle Drum detector and surface array in hybrid
  mode}},
  \href{http://dx.doi.org/10.1016/j.astropartphys.2014.11.004}{\emph{Astropart.
  Phys.} {\bf 64} (2015) 49--62}, [\href{https://arxiv.org/abs/1408.1726}{{\tt
  1408.1726}}].

\bibitem{Unger:2015rzh}
{\scshape Pierre Auger and Telescope Array} collaboration, M.~Unger,
  \emph{{Report of the Working Group on the Composition of Ultra-High Energy
  Cosmic Rays}}, {\emph{PoS} {\bf ICRC2015} (2016) 307},
  [\href{https://arxiv.org/abs/1511.02103}{{\tt 1511.02103}}].

\bibitem{Armengaud:2006fx}
E.~Armengaud, G.~Sigl, T.~Beau and F.~Miniati, \emph{{Crpropa: a numerical tool
  for the propagation of uhe cosmic rays, gamma-rays and neutrinos}},
  \href{http://dx.doi.org/10.1016/j.astropartphys.2007.09.004}{\emph{Astropart.Phys.}
  {\bf 28} (2007) 463--471},
  [\href{https://arxiv.org/abs/astro-ph/0603675}{{\tt astro-ph/0603675}}].

\bibitem{Batista:2014xza}
R.~Alves~Batista and G.~Sigl, \emph{{Diffusion of cosmic rays at EeV energies
  in inhomogeneous extragalactic magnetic fields}},
  \href{http://dx.doi.org/10.1088/1475-7516/2014/11/031}{\emph{JCAP} {\bf 1411}
  (2014) 031}, [\href{https://arxiv.org/abs/1407.6150}{{\tt 1407.6150}}].

\bibitem{Batista:2016yrx}
R.~Alves~Batista, A.~Dundovic, M.~Erdmann, K.-H. Kampert, D.~Kuempel,
  G.~Mueller et~al., \emph{{CRPropa 3 - a Public Astrophysical Simulation
  Framework for Propagating Extraterrestrial Ultra-High Energy Particles}},
  \href{http://dx.doi.org/10.1088/1475-7516/2016/05/038}{\emph{JCAP} {\bf 1605}
  (2016) 038}, [\href{https://arxiv.org/abs/1603.07142}{{\tt 1603.07142}}].

\bibitem{Aloisio:2012wj}
R.~Aloisio, D.~Boncioli, A.~Grillo, S.~Petrera and F.~Salamida, \emph{{SimProp:
  a Simulation Code for Ultra High Energy Cosmic Ray Propagation}},
  \href{http://dx.doi.org/10.1088/1475-7516/2012/10/007}{\emph{JCAP} {\bf 1210}
  (2012) 007}, [\href{https://arxiv.org/abs/1204.2970}{{\tt 1204.2970}}].

\bibitem{Aloisio:2015sga}
R.~Aloisio, D.~Boncioli, A.~di~Matteo, A.~Grillo, S.~Petrera and F.~Salamida,
  \emph{{SimProp v2r2: a Monte Carlo simulation to compute cosmogenic neutrino
  fluxes}},  \href{https://arxiv.org/abs/1505.01347}{{\tt 1505.01347}}.

\bibitem{Aloisio:2016tqp}
R.~Aloisio, D.~Boncioli, A.~di~Matteo, A.~Grillo, S.~Petrera and F.~Salamida,
  \emph{{SimProp v2r3: Monte Carlo simulation code of UHECR propagation}},
  \href{https://arxiv.org/abs/1602.01239}{{\tt 1602.01239}}.

\bibitem{Aloisio:2013hya}
R.~Aloisio, V.~Berezinsky and P.~Blasi, \emph{{Ultra high energy cosmic rays:
  implications of Auger data for source spectra and chemical composition}},
  \href{http://dx.doi.org/10.1088/1475-7516/2014/10/020}{\emph{JCAP} {\bf 1410}
  (2014) 020}, [\href{https://arxiv.org/abs/1312.7459}{{\tt 1312.7459}}].

\bibitem{Hooper:2009fd}
D.~Hooper and A.~M. Taylor, \emph{{On The Heavy Chemical Composition of the
  Ultra-High Energy Cosmic Rays}},
  \href{http://dx.doi.org/10.1016/j.astropartphys.2010.01.003}{\emph{Astropart.Phys.}
  {\bf 33} (2010) 151--159}, [\href{https://arxiv.org/abs/0910.1842}{{\tt
  0910.1842}}].

\bibitem{bib:fitICRC15}
{\scshape Pierre Auger} collaboration, A.~di~Matteo, \emph{{Combined fit of
  spectrum and composition data as measured by the Pierre Auger Observatory}},
  {\emph{PoS} {\bf ICRC2015} (2016) 249},
  [\href{https://arxiv.org/abs/1509.03732}{{\tt 1509.03732}}].

\bibitem{Boncioli:2015pds}
{\scshape Pierre Auger} collaboration, D.~Boncioli, A.~di~Matteo and A.~Grillo,
  \emph{{Surprises from extragalactic propagation of UHECRs}},
  \href{http://dx.doi.org/10.1016/j.nuclphysbps.2016.10.020}{\emph{Nucl. Part.
  Phys. Proc.} {\bf 279-281} (2016) 139--143},
  [\href{https://arxiv.org/abs/1512.02314}{{\tt 1512.02314}}].

\bibitem{Abraham:2009qb}
{\scshape Pierre Auger} collaboration, J.~Abraham et~al., \emph{{Upper limit on
  the cosmic-ray photon fraction at EeV energies from the Pierre Auger
  Observatory}},
  \href{http://dx.doi.org/10.1016/j.astropartphys.2009.04.003}{\emph{Astropart.Phys.}
  {\bf 31} (2009) 399--406}, [\href{https://arxiv.org/abs/0903.1127}{{\tt
  0903.1127}}].

\bibitem{Aglietta:2007yx}
{\scshape Pierre Auger} collaboration, J.~Abraham et~al., \emph{{Upper limit on
  the cosmic-ray photon flux above 10$^{19}$ eV using the surface detector of
  the Pierre Auger Observatory}},
  \href{http://dx.doi.org/10.1016/j.astropartphys.2008.01.003}{\emph{Astropart.Phys.}
  {\bf 29} (2008) 243--256}, [\href{https://arxiv.org/abs/0712.1147}{{\tt
  0712.1147}}].

\bibitem{Abraham:2006ar}
{\scshape Pierre Auger} collaboration, J.~Abraham et~al., \emph{{An upper limit
  to the photon fraction in cosmic rays above $10^{19}$-eV from the Pierre
  Auger Observatory}},
  \href{http://dx.doi.org/10.1016/j.astropartphys.2006.10.004}{\emph{Astropart.Phys.}
  {\bf 27} (2007) 155--168},
  [\href{https://arxiv.org/abs/astro-ph/0606619}{{\tt astro-ph/0606619}}].

\bibitem{Abraham:2007rj}
{\scshape Pierre Auger} collaboration, J.~Abraham et~al., \emph{{Upper limit on
  the diffuse flux of UHE tau neutrinos from the Pierre Auger Observatory}},
  \href{http://dx.doi.org/10.1103/PhysRevLett.100.211101}{\emph{Phys.Rev.Lett.}
  {\bf 100} (2008) 211101}, [\href{https://arxiv.org/abs/0712.1909}{{\tt
  0712.1909}}].

\bibitem{Abreu:2011zze}
{\scshape Pierre Auger} collaboration, P.~Abreu et~al., \emph{{A Search for
  Ultra-High Energy Neutrinos in Highly Inclined Events at the Pierre Auger
  Observatory}}, \href{http://dx.doi.org/10.1103/PhysRevD.85.029902,
  10.1103/PhysRevD.84.122005}{\emph{Phys.Rev.} {\bf D84} (2011) 122005},
  [\href{https://arxiv.org/abs/1202.1493}{{\tt 1202.1493}}].

\bibitem{Abreu:2012zz}
{\scshape Pierre Auger} collaboration, P.~Abreu et~al., \emph{{Search for
  point-like sources of ultra-high energy neutrinos at the Pierre Auger
  Observatory and improved limit on the diffuse flux of tau neutrinos}},
  \href{http://dx.doi.org/10.1088/2041-8205/755/1/L4}{\emph{Astrophys.J.} {\bf
  755} (2012) L4}, [\href{https://arxiv.org/abs/1210.3143}{{\tt 1210.3143}}].

\bibitem{Abreu:2013zbq}
{\scshape Pierre Auger} collaboration, P.~Abreu et~al., \emph{{Ultrahigh Energy
  Neutrinos at the Pierre Auger Observatory}}, {\emph{Adv.High Energy Phys.}
  {\bf 2013} (2013) 708680}, [\href{https://arxiv.org/abs/1304.1630}{{\tt
  1304.1630}}].

\bibitem{Bykov:2012ca}
A.~Bykov, N.~Gehrels, H.~Krawczynski, M.~Lemoine, G.~Pelletier et~al.,
  \emph{{Particle acceleration in relativistic outflows}},
  \href{http://dx.doi.org/10.1007/s11214-012-9896-y}{\emph{Space Sci.Rev.} {\bf
  173} (2012) 309--339}, [\href{https://arxiv.org/abs/1205.2208}{{\tt
  1205.2208}}].

\bibitem{Chen:2002nd}
P.~Chen, T.~Tajima and Y.~Takahashi, \emph{{Plasma wakefield acceleration for
  ultrahigh-energy cosmic rays}},
  \href{http://dx.doi.org/10.1103/PhysRevLett.89.161101}{\emph{Phys.Rev.Lett.}
  {\bf 89} (2002) 161101}, [\href{https://arxiv.org/abs/astro-ph/0205287}{{\tt
  astro-ph/0205287}}].

\bibitem{Guo:2014via}
F.~Guo, H.~Li, W.~Daughton and Y.-H. Liu, \emph{{Formation of Hard Power-laws
  in the Energetic Particle Spectra Resulting from Relativistic Magnetic
  Reconnection}},
  \href{http://dx.doi.org/10.1103/PhysRevLett.113.155005}{\emph{Phys.Rev.Lett.}
  {\bf 113} (2014) 155005}, [\href{https://arxiv.org/abs/1405.4040}{{\tt
  1405.4040}}].

\bibitem{Blasi:2000xm}
P.~Blasi, R.~I. Epstein and A.~V. Olinto, \emph{{Ultrahigh-energy cosmic rays
  from young neutron star winds}},
  \href{http://dx.doi.org/10.1086/312626}{\emph{Astrophys.J.} {\bf 533} (2000)
  L123}, [\href{https://arxiv.org/abs/astro-ph/9912240}{{\tt
  astro-ph/9912240}}].

\bibitem{Kotera:2015pya}
K.~Kotera, E.~Amato and P.~Blasi, \emph{{The fate of ultrahigh energy nuclei in
  the immediate environment of young fast-rotating pulsars}},
  \href{http://dx.doi.org/10.1088/1475-7516/2015/08/026}{\emph{JCAP} {\bf 1508}
  (2015) 026}, [\href{https://arxiv.org/abs/1503.07907}{{\tt 1503.07907}}].

\bibitem{Ptitsyna:2015nta}
K.~Ptitsyna and A.~Neronov, \emph{{Particle acceleration in the vacuum gaps in
  black hole magnetospheres}},
  \href{http://dx.doi.org/10.1051/0004-6361/201527549}{\emph{Astron.
  Astrophys.} {\bf 593} (2016) A8},
  [\href{https://arxiv.org/abs/1510.04023}{{\tt 1510.04023}}].

\bibitem{Winchen:2016koj}
T.~Winchen and S.~Buitink, \emph{{Efficient Second Order Fermi Accelerators as
  Sources of Ultra-High-Energy Cosmic Rays}},
  \href{https://arxiv.org/abs/1612.03675}{{\tt 1612.03675}}.

\bibitem{Niemiec:2006zd}
J.~Niemiec, M.~Ostrowski and M.~Pohl, \emph{{Cosmic-ray acceleration at
  ultrarelativistic shock waves: effects of downstream short-wave turbulence}},
  \href{http://dx.doi.org/10.1086/506901}{\emph{Astrophys.J.} {\bf 650} (2006)
  1020--1027}, [\href{https://arxiv.org/abs/astro-ph/0603363}{{\tt
  astro-ph/0603363}}].

\bibitem{Arons:2002yj}
J.~Arons, \emph{{Magnetars in the metagalaxy: an origin for ultrahigh-energy
  cosmic rays in the nearby universe}},
  \href{http://dx.doi.org/10.1086/374776}{\emph{Astrophys.J.} {\bf 589} (2003)
  871--892}, [\href{https://arxiv.org/abs/astro-ph/0208444}{{\tt
  astro-ph/0208444}}].

\bibitem{Fang:2013cba}
K.~Fang, K.~Kotera and A.~V. Olinto, \emph{{Ultrahigh Energy Cosmic Ray Nuclei
  from Extragalactic Pulsars and the effect of their Galactic counterparts}},
  \href{http://dx.doi.org/10.1088/1475-7516/2013/03/010}{\emph{JCAP} {\bf 1303}
  (2013) 010}, [\href{https://arxiv.org/abs/1302.4482}{{\tt 1302.4482}}].

\bibitem{Neronov:2007mh}
A.~Y. Neronov, D.~Semikoz and I.~Tkachev, \emph{{Ultra-High Energy Cosmic Ray
  production in the polar cap regions of black hole magnetospheres}},
  \href{http://dx.doi.org/10.1088/1367-2630/11/6/065015}{\emph{New J.Phys.}
  {\bf 11} (2009) 065015}, [\href{https://arxiv.org/abs/0712.1737}{{\tt
  0712.1737}}].

\bibitem{Kachelriess:2005xh}
M.~Kachelriess and D.~V. Semikoz, \emph{{Reconciling the ultra-high energy
  cosmic ray spectrum with Fermi shock acceleration}},
  \href{http://dx.doi.org/10.1016/j.physletb.2006.01.009}{\emph{Phys.Lett.}
  {\bf B634} (2006) 143--147},
  [\href{https://arxiv.org/abs/astro-ph/0510188}{{\tt astro-ph/0510188}}].

\bibitem{Blaksley:2011kw}
C.~Blaksley and E.~Parizot, \emph{{Enhancing the Relative Fe-to-Proton
  Abundance in Ultra-High-Energy Cosmic Rays}},
  \href{http://dx.doi.org/10.1016/j.astropartphys.2011.10.006}{\emph{Astropart.
  Phys.} {\bf 35} (2012) 342--345},
  [\href{https://arxiv.org/abs/1111.1607}{{\tt 1111.1607}}].

\bibitem{Aab:2016htd}
{\scshape Pierre Auger} collaboration, A.~Aab et~al., \emph{{Evidence for a
  mixed mass composition at the `ankle' in the cosmic-ray spectrum}},
  \href{http://dx.doi.org/10.1016/j.physletb.2016.09.039}{\emph{Phys. Lett. B}
  (2016) }, [\href{https://arxiv.org/abs/1609.08567}{{\tt 1609.08567}}].

\bibitem{Aloisio:2015ega}
R.~Aloisio, D.~Boncioli, A.~di~Matteo, A.~F. Grillo, S.~Petrera and
  F.~Salamida, \emph{{Cosmogenic neutrinos and ultra-high energy cosmic ray
  models}}, \href{http://dx.doi.org/10.1088/1475-7516/2015/10/006}{\emph{JCAP}
  {\bf 1510} (2015) 006}, [\href{https://arxiv.org/abs/1505.04020}{{\tt
  1505.04020}}].

\bibitem{Heinze:2015hhp}
J.~Heinze, D.~Boncioli, M.~Bustamante and W.~Winter, \emph{{Cosmogenic
  Neutrinos Challenge the Cosmic Ray Proton Dip Model}},
  \href{http://dx.doi.org/10.3847/0004-637X/825/2/122}{\emph{Astrophys. J.}
  {\bf 825} (2016) 122}, [\href{https://arxiv.org/abs/1512.05988}{{\tt
  1512.05988}}].

\bibitem{Aartsen:2016ngq}
{\scshape IceCube} collaboration, M.~G. Aartsen et~al., \emph{{Constraints on
  ultra-high-energy cosmic ray sources from a search for neutrinos above 10 PeV
  with IceCube}},  \href{https://arxiv.org/abs/1607.05886}{{\tt 1607.05886}}.

\bibitem{Unger:2015laa}
M.~Unger, G.~R. Farrar and L.~A. Anchordoqui, \emph{{Origin of the ankle in the
  ultrahigh energy cosmic ray spectrum, and of the extragalactic protons below
  it}}, \href{http://dx.doi.org/10.1103/PhysRevD.92.123001}{\emph{Phys. Rev.}
  {\bf D92} (2015) 123001}, [\href{https://arxiv.org/abs/1505.02153}{{\tt
  1505.02153}}].

\bibitem{Globus:2015xga}
N.~Globus, D.~Allard and E.~Parizot, \emph{{A complete model of the cosmic ray
  spectrum and composition across the Galactic to extragalactic transition}},
  \href{http://dx.doi.org/10.1103/PhysRevD.92.021302}{\emph{Phys. Rev.} {\bf
  D92} (2015) 021302}, [\href{https://arxiv.org/abs/1505.01377}{{\tt
  1505.01377}}].

\bibitem{Batista:2015mea}
R.~Alves~Batista, D.~Boncioli, A.~di~Matteo, A.~van Vliet and D.~Walz,
  \emph{{Effects of uncertainties in simulations of extragalactic UHECR
  propagation, using CRPropa and SimProp}},
  \href{http://dx.doi.org/10.1088/1475-7516/2015/10/063}{\emph{JCAP} {\bf 1510}
  (2015) 063}, [\href{https://arxiv.org/abs/1508.01824}{{\tt 1508.01824}}].

\bibitem{Stecker:2005qs}
F.~W. Stecker, M.~Malkan and S.~Scully, \emph{{Intergalactic photon spectra
  from the far ir to the uv lyman limit for $0 < z < 6$ and the optical depth
  of the universe to high energy gamma-rays}},
  \href{http://dx.doi.org/10.1086/506188}{\emph{Astrophys.J.} {\bf 648} (2006)
  774--783}, [\href{https://arxiv.org/abs/astro-ph/0510449}{{\tt
  astro-ph/0510449}}].

\bibitem{Stecker:2006eh}
F.~W. Stecker, M.~Malkan and S.~Scully, \emph{{Corrected Table for the
  Parametric Coefficients for the Optical Depth of the Universe to Gamma-rays
  at Various Redshifts}},
  \href{http://dx.doi.org/10.1086/511738}{\emph{Astrophys.J.} {\bf 658} (2007)
  1392}, [\href{https://arxiv.org/abs/astro-ph/0612048}{{\tt
  astro-ph/0612048}}].

\bibitem{Kneiske:2003tx}
T.~M. Kneiske, T.~Bretz, K.~Mannheim and D.~Hartmann, \emph{{Implications of
  cosmological gamma-ray absorption. 2. Modification of gamma-ray spectra}},
  \href{http://dx.doi.org/10.1051/0004-6361:20031542}{\emph{Astron.Astrophys.}
  {\bf 413} (2004) 807--815},
  [\href{https://arxiv.org/abs/astro-ph/0309141}{{\tt astro-ph/0309141}}].

\bibitem{Gilmore:2011ks}
R.~Gilmore, R.~Somerville, J.~Primack and A.~Dominguez, \emph{{Semi-analytic
  modeling of the EBL and consequences for extragalactic gamma-ray spectra}},
  \href{http://dx.doi.org/10.1111/j.1365-2966.2012.20841.x}{\emph{Mon.Not.Roy.Astron.Soc.}
  {\bf 422} (2012) 3189}, [\href{https://arxiv.org/abs/1104.0671}{{\tt
  1104.0671}}].

\bibitem{Dominguez:2010bv}
A.~Dominguez, J.~Primack, D.~Rosario, F.~Prada, R.~Gilmore et~al.,
  \emph{{Extragalactic Background Light Inferred from AEGIS Galaxy SED-type
  Fractions}},
  \href{http://dx.doi.org/10.1111/j.1365-2966.2010.17631.x}{\emph{Mon.Not.Roy.Astron.Soc.}
  {\bf 410} (2011) 2556}, [\href{https://arxiv.org/abs/1007.1459}{{\tt
  1007.1459}}].

\bibitem{Franceschini:2008tp}
A.~Franceschini, G.~Rodighiero and M.~Vaccari, \emph{{The extragalactic
  optical-infrared background radiations, their time evolution and the cosmic
  photon-photon opacity}},
  \href{http://dx.doi.org/10.1051/0004-6361:200809691}{\emph{Astron.Astrophys.}
  {\bf 487} (2008) 837}, [\href{https://arxiv.org/abs/0805.1841}{{\tt
  0805.1841}}].

\bibitem{Inoue:2012bk}
Y.~Inoue, S.~Inoue, M.~A.~R. Kobayashi, R.~Makiya, Y.~Niino and T.~Totani,
  \emph{{Extragalactic Background Light from Hierarchical Galaxy Formation:
  Gamma-ray Attenuation up to the Epoch of Cosmic Reionization and the First
  Stars}},
  \href{http://dx.doi.org/10.1088/0004-637X/768/2/197}{\emph{Astrophys. J.}
  {\bf 768} (2013) 197}, [\href{https://arxiv.org/abs/1212.1683}{{\tt
  1212.1683}}].

\bibitem{Mucke:1999yb}
A.~Mucke, R.~Engel, J.~Rachen, R.~Protheroe and T.~Stanev, \emph{{SOPHIA: Monte
  Carlo simulations of photohadronic processes in astrophysics}},
  \href{http://dx.doi.org/10.1016/S0010-4655(99)00446-4}{\emph{Comput.Phys.Commun.}
  {\bf 124} (2000) 290--314},
  [\href{https://arxiv.org/abs/astro-ph/9903478}{{\tt astro-ph/9903478}}].

\bibitem{Puget:1976nz}
J.~Puget, F.~Stecker and J.~Bredekamp, \emph{{Photonuclear Interactions of
  Ultrahigh-Energy Cosmic Rays and their Astrophysical Consequences}},
  \href{http://dx.doi.org/10.1086/154321}{\emph{Astrophys.J.} {\bf 205} (1976)
  638--654}.

\bibitem{Stecker:1998ib}
F.~Stecker and M.~Salamon, \emph{{Photodisintegration of ultrahigh-energy
  cosmic rays: A New determination}},
  \href{http://dx.doi.org/10.1086/306816}{\emph{Astrophys.J.} {\bf 512} (1999)
  521--526}, [\href{https://arxiv.org/abs/astro-ph/9808110}{{\tt
  astro-ph/9808110}}].

\bibitem{talys}
A.~J. {Koning}, S.~{Hilaire} and M.~C. {Duijvestijn}, \emph{{TALYS:
  Comprehensive Nuclear Reaction Modeling}},  in \emph{International Conference
  on Nuclear Data for Science and Technology} (R.~C. {Haight}, M.~B.
  {Chadwick}, T.~{Kawano} and P.~{Talou}, eds.), vol.~769 of \emph{American
  Institute of Physics Conference Series}, pp.~1154--1159, May, 2005.
\newblock \href{http://dx.doi.org/10.1063/1.1945212}{DOI}.

\bibitem{Koning20122841}
A.~Koning and D.~Rochman, \emph{{Modern Nuclear Data Evaluation with the TALYS
  Code System }},
  \href{http://dx.doi.org/http://dx.doi.org/10.1016/j.nds.2012.11.002}{\emph{Nuclear
  Data Sheets} {\bf 113} (2012) 2841 -- 2934}.

\bibitem{talys1.6-manual}
A.~Koning, S.~Hilaire and S.~Goriely, \emph{TALYS 1.6 User Manual}.

\bibitem{Allison:2006ve}
J.~Allison et~al., \emph{{Geant4 developments and applications}},
  \href{http://dx.doi.org/10.1109/TNS.2006.869826}{\emph{IEEE Trans. Nucl.
  Sci.} {\bf 53} (2006) 270}.

\bibitem{Pierog:2013ria}
T.~Pierog, I.~Karpenko, J.~M. Katzy, E.~Yatsenko and K.~Werner, \emph{{EPOS
  LHC: Test of collective hadronization with data measured at the CERN Large
  Hadron Collider}},
  \href{http://dx.doi.org/10.1103/PhysRevC.92.034906}{\emph{Phys. Rev.} {\bf
  C92} (2015) 034906}, [\href{https://arxiv.org/abs/1306.0121}{{\tt
  1306.0121}}].

\bibitem{Ostapchenko:2010vb}
S.~Ostapchenko, \emph{{Monte Carlo treatment of hadronic interactions in
  enhanced Pomeron scheme: I. QGSJET-II model}},
  \href{http://dx.doi.org/10.1103/PhysRevD.83.014018}{\emph{Phys.Rev.} {\bf
  D83} (2011) 014018}, [\href{https://arxiv.org/abs/1010.1869}{{\tt
  1010.1869}}].

\bibitem{Ahn:2009wx}
E.-J. Ahn, R.~Engel, T.~K. Gaisser, P.~Lipari and T.~Stanev, \emph{{Cosmic ray
  interaction event generator SIBYLL 2.1}},
  \href{http://dx.doi.org/10.1103/PhysRevD.80.094003}{\emph{Phys.Rev.} {\bf
  D80} (2009) 094003}, [\href{https://arxiv.org/abs/0906.4113}{{\tt
  0906.4113}}].

\bibitem{Ghia:2015kfz}
{\scshape Pierre Auger} collaboration, P.~L. Ghia, \emph{{Highlights from the
  Pierre Auger Observatory}}, {\emph{PoS} {\bf ICRC2015} (2016) 034}.

\bibitem{ThePierreAuger:2015rha}
{\scshape Pierre Auger} collaboration, A.~Aab et~al., \emph{{Measurement of the
  cosmic ray spectrum above 4 $\times$ 10$^{18}$ eV using inclined events
  detected with the Pierre Auger Observatory}},
  \href{http://dx.doi.org/10.1088/1475-7516/2015/08/049}{\emph{JCAP} {\bf 1508}
  (2015) 049}, [\href{https://arxiv.org/abs/1503.07786}{{\tt 1503.07786}}].

\bibitem{DeDomenico:2013wwa}
M.~De~Domenico, M.~Settimo, S.~Riggi and E.~Bertin, \emph{{Reinterpreting the
  development of extensive air showers initiated by nuclei and photons}},
  \href{http://dx.doi.org/10.1088/1475-7516/2013/07/050}{\emph{JCAP} {\bf 1307}
  (2013) 050}, [\href{https://arxiv.org/abs/1305.2331}{{\tt 1305.2331}}].

\bibitem{Pierog:2004re}
T.~Pierog, M.~Alekseeva, T.~Bergmann, V.~Chernatkin, R.~Engel et~al.,
  \emph{{First results of fast one-dimensional hybrid simulation of EAS using
  CONEX}},
  \href{http://dx.doi.org/10.1016/j.nuclphysbps.2005.07.029}{\emph{Nucl.Phys.Proc.Suppl.}
  {\bf 151} (2006) 159--162},
  [\href{https://arxiv.org/abs/astro-ph/0411260}{{\tt astro-ph/0411260}}].

\bibitem{Minuit}
F.~James and M.~Winkler, \emph{{Minuit User's Guide, (CERN)}}, .

\bibitem{Rolke:2004mj}
W.~A. Rolke, A.~M. Lopez and J.~Conrad, \emph{{Limits and confidence intervals
  in the presence of nuisance parameters}},
  \href{http://dx.doi.org/10.1016/j.nima.2005.05.068}{\emph{Nucl. Instrum.
  Meth.} {\bf A551} (2005) 493--503},
  [\href{https://arxiv.org/abs/physics/0403059}{{\tt physics/0403059}}].

\bibitem{Onn2011}
S.~Onn and I.~Weissman, \emph{Generating uniform random vectors over a simplex
  with implications to the volume of a certain polytope and to multivariate
  extremes}, \href{http://dx.doi.org/10.1007/s10479-009-0567-7}{\emph{Annals of
  Operations Research} {\bf 189} (2011) 331--342}.

\bibitem{Patil:2010}
A.~Patil, D.~Huard and C.~J. Fonnesbeck, \emph{Pymc: Bayesian stochastic
  modelling in python}, {\emph{Journal of Statistical Software} {\bf 35} (7,
  2010) 1--81}.

\bibitem{gelman1992}
A.~Gelman and D.~B. Rubin, \emph{Inference from iterative simulation using
  multiple sequences},
  \href{http://dx.doi.org/10.1214/ss/1177011136}{\emph{Statist. Sci.} {\bf 7}
  (11, 1992) 457--472}.

\bibitem{Barlow:2003sg}
R.~Barlow, \emph{{Asymmetric systematic errors}},
  \href{https://arxiv.org/abs/physics/0306138}{{\tt physics/0306138}}.

\bibitem{Taylor:2011ta}
A.~M. Taylor, M.~Ahlers and F.~A. Aharonian, \emph{{The need for a local source
  of UHE CR nuclei}},
  \href{http://dx.doi.org/10.1103/PhysRevD.84.105007}{\emph{Phys.Rev.} {\bf
  D84} (2011) 105007}, [\href{https://arxiv.org/abs/1107.2055}{{\tt
  1107.2055}}].

\bibitem{Taylor:2015rla}
A.~M. Taylor, M.~Ahlers and D.~Hooper, \emph{{Indications of Negative Evolution
  for the Sources of the Highest Energy Cosmic Rays}},
  \href{http://dx.doi.org/10.1103/PhysRevD.92.063011}{\emph{Phys. Rev.} {\bf
  D92} (2015) 063011}, [\href{https://arxiv.org/abs/1505.06090}{{\tt
  1505.06090}}].

\bibitem{Gelmini:2011kg}
G.~B. Gelmini, O.~Kalashev and D.~V. Semikoz, \emph{{Gamma-Ray Constraints on
  Maximum Cosmogenic Neutrino Fluxes and UHECR Source Evolution Models}},
  \href{http://dx.doi.org/10.1088/1475-7516/2012/01/044}{\emph{JCAP} {\bf 1201}
  (2012) 044}, [\href{https://arxiv.org/abs/1107.1672}{{\tt 1107.1672}}].

\bibitem{Mollerach:2013dza}
S.~Mollerach and E.~Roulet, \emph{{Magnetic diffusion effects on the ultra-high
  energy cosmic ray spectrum and composition}},
  \href{http://dx.doi.org/10.1088/1475-7516/2013/10/013}{\emph{JCAP} {\bf 1310}
  (2013) 013}, [\href{https://arxiv.org/abs/1305.6519}{{\tt 1305.6519}}].

\bibitem{Porcelli:2015jli}
{\scshape Pierre Auger} collaboration, A.~Porcelli, \emph{{Measurements of the
  first two moments of the depth of shower maximum over nearly three decades of
  energy, combining data from}}, {\emph{PoS} {\bf ICRC2015} (2016) 420},
  [\href{https://arxiv.org/abs/1509.03732}{{\tt 1509.03732}}].

\bibitem{Apel:2014uka}
W.~D. Apel et~al., \emph{{The KASCADE-Grande energy spectrum of cosmic rays and
  the role of hadronic interaction models}},
  \href{http://dx.doi.org/10.1016/j.asr.2013.05.008}{\emph{Adv. Space Res.}
  {\bf 53} (2014) 1456--1469}.

\bibitem{Fargion:2008sp}
D.~Fargion, \emph{{Light Nuclei solving Auger puzzles?}},
  \href{http://dx.doi.org/10.1088/0031-8949/78/04/045901}{\emph{Phys. Scripta}
  {\bf 78} (2008) 045901}, [\href{https://arxiv.org/abs/0801.0227}{{\tt
  0801.0227}}].

\bibitem{Fargion:2009rb}
D.~Fargion, D.~D'Armiento, P.~Paggi and S.~Patri', \emph{{Lightest Nuclei in
  UHECR versus Tau Neutrino Astronomy}},
  \href{http://dx.doi.org/10.1016/j.nuclphysbps.2009.03.083}{\emph{Nucl. Phys.
  Proc. Suppl.} {\bf 190} (2009) 162--166},
  [\href{https://arxiv.org/abs/0902.3290}{{\tt 0902.3290}}].

\bibitem{Fargion:2009ki}
D.~Fargion, \emph{{Coherent and random UHECR Spectroscopy of Lightest Nuclei
  along CenA: Shadows on GZK Tau Neutrinos spread in a near sky and time}},
  \href{http://dx.doi.org/10.1016/j.nima.2010.06.040}{\emph{Nucl. Instrum.
  Meth.} {\bf A630} (2011) 111--114},
  [\href{https://arxiv.org/abs/0908.2650}{{\tt 0908.2650}}].

\bibitem{Fargion:2014jma}
D.~Fargion, G.~Ucci, P.~Oliva and P.~G. De~Sanctis~Lucentini, \emph{{The
  meaning of the UHECR Hot Spots: A Light Nuclei Nearby Astronomy}},
  \href{http://dx.doi.org/10.1051/epjconf/20159908002}{\emph{EPJ Web Conf.}
  {\bf 99} (2015) 08002}, [\href{https://arxiv.org/abs/1412.1573}{{\tt
  1412.1573}}].

\bibitem{Fargion:2016cep}
D.~Fargion, P.~Oliva, P.~G. De~Sanctis~Lucentini, D.~D. Armiento and P.~Paggi,
  \emph{{UHECR narrow clustering correlating IceCube through-going muons}},
  2016.
\newblock \href{https://arxiv.org/abs/1611.00079}{{\tt 1611.00079}}.

\bibitem{Smida:2015kga}
R.~Smida and R.~Engel, \emph{{The ultra-high energy cosmic rays image of Virgo
  A}}, {\emph{PoS} {\bf ICRC2015} (2016) 470},
  [\href{https://arxiv.org/abs/1509.09033}{{\tt 1509.09033}}].

\bibitem{arXiv:1604.03637}
{\scshape Pierre Auger} collaboration, A.~Aab et~al., \emph{{The Pierre Auger
  Observatory Upgrade}},  \href{https://arxiv.org/abs/arXiv:1604.03637}{{\tt
  arXiv:1604.03637}}.

\end{thebibliography}\endgroup
\bibliographystyle{JHEP}


\section*{The Pierre Auger Collaboration}

\end{document}